\documentclass[useAMS,usenatbib]{mn2e}
\usepackage{longtable}
\usepackage[pdftex]{graphicx}

\def\ltsima{$\; \buildrel < \over \sim \;$}
\def\simlt{\lower.5ex\hbox{\ltsima}}
\def\gtsima{$\; \buildrel > \over \sim \;$}
\def\simgt{\lower.5ex\hbox{\gtsima}}
\def\gsimeq
{\hbox{\raise0.5ex\hbox{$>\lower1.06ex\hbox{$\kern-1.07em{\sim}$}$}}}
\def\lsimeq
{\hbox{\raise0.5ex\hbox{$<\lower1.06ex\hbox{$\kern-1.07em{\sim}$}$}}}
\def\pn{\par\noindent}
\def\ss{\smallskip\pn}

\def\asca{{\it ASCA}}

\def\xmm{{\it XMM-Newton}}
\def\chandra{{\it Chandra}}
\def\suzaku{{\it Suzaku}}

\def\fermi{{\it Fermi}}
\def\swift{{\it Swift}}
\def\nustar{{\it NuSTAR}}
\def\integral{{\it INTEGRAL}}

\def\apj{ApJ}
\def\aj{AJ}
\def\mnras{MNRAS}
\def\aap{A\&A}
\def\aaps{A\&AS}
\def\apjl{ApJ}
\def\apjs{ApJS}
\def\araa{ARA\&A}
\def\pasj{PASJ}
\def\nat{Nature}

\def\procspie{Proc. SPIE}
\def\memsai{Memorie della Societa Astronomica Italiana}

\def\sgr{Sgr~A$^{\star}$}

\def\sgras{Sgr~A$^\star$}

\def\xis{XIS}
\def\xis1{XIS1}
\def\xis2{XIS2}
\def\xis3{XIS3}

\title[] 
 {{The \xmm\ view of the central degrees of the Milky Way}}

 \author[G.\ Ponti et al. ]
 {G.~Ponti$^{1}$\thanks{ponti@mpe.mpg.de},  M. R. Morris$^{2}$,  
 R. Terrier$^{3}$, F. Haberl$^{1}$, R. Sturm$^{1}$, M. Clavel$^{3,4}$, S. Soldi$^{3,4}$, 
\newauthor
A. Goldwurm$^{3,4}$, P. Predehl$^{1}$, K. Nandra$^{1}$,
G. Belanger$^{5}$, R.~S.~Warwick$^{6}$ and V. Tatischeff$^{7}$ 
\\ \\
   $^1$ Max Planck Institut f\"{u}r Extraterrestrische Physik, 85748, Garching, Germany\\
   $^2$ Department of Physics and Astronomy, University of California, Los Angeles, 
   CA 90095-1547, USA\\
   $^3$Unit\'e mixte de recherche Astroparticule et Cosmologie, 10 rue Alice Domon 
   et L\'eonie Duquet, 75205 Paris, France \\
   $^4$Service d'Astrophysique (SAp), IRFU/DSM/CEA-Saclay, 91191 
   Gif-sur-Yvette Cedex, France\\
   $^5$ESA/ESAC, PO Box 78, 28691 Villanueva de la Ca\~{n}ada, Spain\\
   $^6$Department of Physics and Astronomy, University of Leicester, University Road, 
   Leicester, LE1 7RH, UK\\
   $^7$Centre de Sciences Nucl\'eaires et de Sciences de la Mati\`ere, 
   IN2P3-CNRS and Univ Paris-Sud, F-91405 Orsay Cedex, France\\
}

\pagerange{\pageref{firstpage}--\pageref{lastpage}}

\usepackage{times}

\begin{document}

\label{firstpage}

\maketitle

\begin{abstract}
The deepest \xmm\ mosaic map of the central $1.5^\circ$ of the Galaxy is presented, 
including a total of about $1.5$~Ms of EPIC-pn cleaned exposures in the central 
$15"$ and about $200$~ks outside. 
This compendium presents broad-band X-ray continuum maps, 
soft X-ray intensity maps, a decomposition into spectral components and a comparison 
of the X-ray maps with emission at other wavelengths. 
Newly-discovered extended features, such as supernova remnants (SNRs), superbubbles 
and X-ray filaments are reported. We provide an atlas of extended features within $\pm$1 
degree of \sgras. We discover the presence of a coherent X-ray 
emitting region peaking around G0.1-0.1 and surrounded by the ring of cold,
mid-IR-emitting material known from previous work as the "Radio Arc Bubble"
and with the addition of the X-ray data now appears to be a candidate superbubble.
Sgr~A's bipolar lobes show sharp edges, suggesting that they could be the remnant, 
collimated by the circumnuclear disc, of a SN explosion that created the recently 
discovered magnetar, SGR~J1745-2900. 
Soft X-ray features, most probably from SNRs, are observed to fill holes in the 
dust distribution, and to indicate a direct interaction between SN explosions and 
Galactic center (GC) molecular clouds. 
We also discover warm plasma at high Galactic latitude, showing a sharp edge to 
its distribution that correlates with the location of known radio/mid-IR features such 
as the "GC Lobe". These features might be associated with an 
inhomogeneous hot "atmosphere" over the GC, perhaps fed 
by continuous or episodic outflows of mass and energy from the GC region.
\end{abstract}


\section{Introduction} 

At a distance of only $\sim8$ kpc, the center of the Milky Way is the 
closest Galactic nucleus, allowing us to directly image, with incomparable 
spatial resolution, the physical processes typical of galactic nuclei. 
The central region of the Galaxy is one of the richest laboratories for astrophysics 
(Genzel et al. 2010; Morris et al. 2012; Ponti et al. 2013). 
Within the inner $\sim200$~pc about $3-5\times10^{7}$~M$_\odot$ 
of molecular material are concentrated, the so called Central Molecular Zone (CMZ). 
This corresponds to about 1~\% of the molecular mass of the entire Galaxy 
and it is concentrated in a region of about $\sim10^{-6}$ of its volume (Morris \& Serabyn 1996). 
In this region many thousands of persistent and transient point-like 
X-ray sources are embedded, such as active stars, bright accreting binary 
systems (and many more quiescent massive bodies) and cataclysmic variables, 
which have been beautifully imaged thanks to the superior spatial resolution 
of \chandra\ (Wang et al.\ 2002; Muno et 
al.\ 2003; 2009).
One of the best jewels in the GC is Sgr~A$^{\star}$, the electromagnetic counterpart of 
the closest supermassive black hole (BH; Genzel et al. 2010). 
In addition to this large population of point sources, extended X-ray sources, 
such as supernova remnants, non-thermal filaments, pulsar wind nebulae, and 
massive star clusters, populate the GC (Wang et al.\ 2002). 
The GC is considered a mini-starburst environment, giving us the possibility to study 
the interaction between supernova remnants (SNRs) and molecular clouds and 
the impact of massive-and-young star clusters on their surroundings. It allows us 
to image, in superb detail, the creation and evolution of bubbles and superbubbles 
and the generation of Galactic outflows, 
powered by past starbursts and/or accretion events onto \sgras, and their impact 
on the GC environment. 

Warm ($kT\sim1$~keV) and hot ($kT\sim6.5$~keV) thermal plasma emission  
plus non-thermal hard X-ray emission associated with X-ray reflection nebulae 
(see Ponti et al. 2013 for a review) pervade the central region, 
producing a high background of soft and hard X-ray radiation. 
About $90$\% (Ebisawa et al. 2001; Wang et al. 2002) of the soft X-ray emission 
appears to be due to a diffuse, patchy and thermal component (Bamba et al. 2002) 
with a temperature $kT\sim1$~keV, most probably associated with supernova remnants. 
The origin of the hot component is, instead, still highly debated. 
At $\sim1.5^\circ$ from the GC, $\sim80$~\% of this emission has 
been resolved into point sources (e.g., accreting white dwarfs and coronally 
active stars) by a deep \chandra\ observation (Revnivtsev et al. 2009).
Although the intensity of the hot plasma emission increases rapidly towards the GC,
point sources continue to make a substantial contribution to the observed 
hard emission (Muno et al.\ 2004; Heard \& Warwick 2013a). 
Additionally, some of the emission may arise due to scattering of the radiation 
from bright X-ray binaries by the dense interstellar medium (Sunyaev et al.\ 1993; 
Molaro et al.\ 2014). 
Nevertheless, it is not excluded that a truly diffuse hot-plasma 
component is also present in the GC (Koyama et al.\ 2009; 
Uchiyama et al.\ 2013). Such hot plasma would be unbound to the Galaxy and 
it would require a huge energy ($E\sim10^{55}$ erg) and energy loss rate 
of the mass outflow of $\sim10^{43}$~erg~s$^{-1}$, corresponding to a rate of 
1 supernova/yr, to continuously replenish it (Tanaka 2002). 
However, it has recently been proposed that such hot plasma might be trapped 
by the GC magnetic field (Nishiyama et al.\ 2013). 

Indeed, the magnetic field is thought to be an important ingredient of the GC 
environment. The first high-resolution radio images of the Milky Way center 
(see bottom panel of Fig.\ 6), revealed the presence of many straight, 
long (up to $\sim20-30$~pc) and thin (with width $\lsimeq0.1$~pc), linearly 
polarised vertical filaments with spectral index consistent with synchrotron 
radiation (Yusef-Zadeh et al. 1984; 1987a,b; Anantharamaiah 
et al. 1991; Lang et al. 1999; LaRosa et al. 2000). 
These filaments are hypothesized to be magnetic flux tubes trapping energetic electrons 
and therefore tracing the diffuse interstellar GC poloidal magnetic field (Morris \& 
Yusef-Zadeh 1985; Lang et al. 1999). 
A staggeringly powerful poloidal magnetic field pervading the GC, with a
field strength of $B~\gsimeq~50~\mu G$, and very possibly $B\sim~1~mG$, has 
been inferred (Morris 1990; Crocker et al. 2010; Ferriere et al. 2011). 
The details of the physical process creating the filaments and energising the 
magnetic field are still debated; however, it appears clear that the magnetic 
filaments are interacting with the ionised surfaces of massive molecular clouds. 

Recent far-infrared/sub-millimeter polarization studies of thermal dust emission 
made it possible to probe the direction of the interstellar magnetic field inside 
dense molecular clouds. The magnetic field threading GC molecular clouds 
is measured to be parallel to the Galactic plane (Novak et al. 2003; 
Chuss et al. 2003; Nishiyama et al. 2009). 
Therefore, it appears that the large-scale GC magnetic field is poloidal in the 
diffuse interstellar medium and toroidal in dense regions in the plane. 
If the strength of the diffuse magnetic field is on the high side ($B\sim1~mG$) 
a huge amount of magnetic energy, $E\sim10^{55}$~erg, would be stored 
in the central $\sim300$~pc. This is comparable to the kinetic energy associated 
with the rotation of the gas in the CMZ. Therefore, it is thought to be a key player for the 
GC physics and phenomenology. 

A large scale structure with a possible magnetic origin and appearing to be interacting 
with massive clouds of the CMZ (similar to the non-thermal filaments) is the Galactic 
center lobe (GCL). 
The GCL has a limb brightened shell structure in the 10.5~GHz map, defined 
primarily by two spurs (see Fig.\ 1 and 2 of Law et al. 2009). The eastern one arises 
from the location of the GC Radio Arc\footnote{This is a well known radio feature 
(see Fig.\ \ref{FC2}) composed of an array of straight, long, thin and linearly 
polarised vertical filaments, indicating the importance of the GC magnetic field. }, 
while the second starts from the Sgr~C thread. 
It was proposed that the GCL is produced by channelling of plasma from the 
Galactic plane, induced by energetic GC activity (e.g.\ episode of AGN activity, 
or a large mass outflow due to the high star formation rate, etc.; 
see Law et al.\ 2011) or from twisting of poloidal magnetic field 
lines by Galactic rotation (Sofue et al.\ 1984; 1985; Uchida 1985; 1990; Shibata 1989). 
Located at the western limb of the GCL is an interesting feature, AFGL~5376 
(Uchida et al.\ 1994), an unusually warm, shock heated and extended IR source,
thought to be associated with the GCL. 

All major X-ray telescopes devoted a significant fraction of their time to 
the study of the GC. \chandra\ invested several Ms to monitor both \sgras's 
activity (Baganoff et al.\ 2001; 2003; Neilsen et al.\ 2013) as well as diffuse soft 
and hard X-ray emission (Wang et al. 2002; Park et al. 2004). 
\suzaku\ and \swift\ also performed large observational campaigns to scan 
the Milky Way center (Koyama et al. 2007; Degenaar \& Wijnands 2010) and 
monitor the transients in the region (Degenaar et al. 2012). The study of the GC is one of 
the key programs of the \nustar\ mission (Harrison et al.\ 2013; Barriere et al.\ 2014; 
Mori et al.\ 2013). 
\xmm\ completed a first shallow ($\sim30$~ks total cleaned exposure in each 
point) scan of the CMZ within a couple of years after launch (see the conference 
proceedings: Sakano et al. 2003; 2004; Decourchelle et al. 2003). 
A larger amount of time (more than $\sim1.5$~Ms) has been invested 
by \xmm\ on studying the emission properties of Sgr~A$^\star$ (Goldwurm et al.\ 2003; 
Porquet et al.\ 2003; 2008; Belanger et al.\ 2005; Trap et al.\ 2011; 
Mossoux et al.\ 2014), focussing on the 
central $\sim15$~arcmin, only. Using the \xmm\ observations from the shallow
scan of the CMZ together with a number of the \sgras\ pointings, Heard \& 
Warwick (2013a,b) have investigated the distribution of the X-ray emission 
within the central region of the Galaxy. With the aim of studying the propagation
of echoes of the past GC activity within the CMZ (Sunyaev et al. 1993; Koyama 
et al. 1996; 2008; Revnivtsev et al. 2004; Muno et al. 2007; Inui et al. 2009; 
Ponti et al. 2010; 2013; Terrier et al. 2010; Nobukawa et al. 2011; Capelli et al. 2011; 2012;
Clavel et al. 2013; 2014; Krivonos et al. 2014), recently, a new deep (with $\sim100$~ks 
exposure at each location) \xmm\ scan of the CMZ has been completed (in fall 2012). 
We present here the combined images of both the new and old \xmm\ scans, as well
as all the \xmm\ observations within the central degree of the Galaxy. 

In \S \ref{datared} we present the data reduction process and the key steps 
to produce the GC EPIC mosaic maps. 
Section 3 introduces the broad band X-ray images, discussing 
the (transient) emission from the brightest point sources, 
the contribution from the foreground emission, as well as the soft 
and hard GC diffuse emission.
In section~4, the narrow band images at the energies of the soft X-ray lines 
are displayed. Section~5 presents a new technique of spectral-imaging 
decomposition of the soft X-ray emission into three physical components.
Section 6 presents an atlas of all the new and known diffuse features 
within the surveyed area. 
Section 7 presents the comparison with the distribution of column density 
of intervening matter. 
Discussion and conclusions are in \S~8 and 9, respectively. 
Hereinafter, unless otherwise stated, we will state all locations and 
positions in Galactic coordinates and Galactic cardinal points.
Errors are given at 90~\% confidence for one interesting parameter. 

\section{Data reduction and cleaning}
\label{datared}

The new \xmm\ CMZ scan has been performed 
in 2012 starting on August $30^{th}$ and ending on October $10^{th}$. 
It comprises of 16 \xmm\ observations all performed with all the EPIC 
instruments in full-frame CCD readout mode with the medium optical 
blocking filters applied (we refer to Tab. \ref{TabObs1} and \ref{TabObs2} 
for more details on the instruments set-ups). 

This paper is not limited to the use of the 2012 \xmm\ scan of the CMZ. 
Instead it is using all \xmm\ observations pointed within 1 degree from 
\sgr. Therefore we combined the 16 observations of the new CMZ scan 
with the 14 observations of the previous CMZ scan accumulated between 
2000 and 2002. We also include the 30 observations pointing at \sgr\ and 
other 49 observations aimed at studies of different sources in the vicinity 
of \sgr\ (see Tab. \ref{TabObs1} and \ref{TabObs2}).

We performed the analysis of the EPIC data with the version 13.0.0 of 
the \xmm\ Science Analysis System (SAS). Periods of increased particle 
background have been removed from the data. To perform this, we first 
selected the Good Time Intervals (GTI) starting from the 7-15 keV background 
light curves, then we applied a threshold of 8 and 2.5 cts ks$^{-1}$ 
arcmin$^{-2}$ for EPIC pn and EPIC MOS, respectively (see e.g. Haberl 
et al. 2012). 
The chosen thresholds efficiently cut out all the periods of most extreme 
activity of soft proton flares. We noted, however, that an enhanced, but weak, 
background activity was still present in the data during several observations. 
Because of the non-uniform distribution of the GC diffuse 
emission, lowering the threshold uniformly in all observations, would 
result in cutting truly good time intervals in observations with higher GC 
diffuse emission. Thus we decided to visually inspect the background 
light curve of all data-sets and select a different threshold for each 
observation (see Tab. \ref{TabObs1} and \ref{TabObs2}). 
Such as in Haberl et al. (2012), when the data from several detectors were 
available, we combined the GTIs using only common time intervals, 
otherwise we included GTIs of the single detector. 
Most of the 2012 CMZ scan data were affected by negligible particle flaring 
activity. On the other hand, many of the previous observations have been 
severely affected by soft proton flares (see the reduction in exposure 
in Tab. \ref{TabObs1} and \ref{TabObs2}). 

To prevent infrared, optical and UV photons from bright sources 
in the field of view that would increase the noise and degrade the CCD 
energy scale, the different \xmm\ observations have been performed 
with different filters applied, according to the optical-UV brightness of 
the sources in the field of view (see Tab. \ref{TabObs1} and \ref{TabObs2}). 
In particular, we used the filter wheel closed observations to remove 
the internal EPIC background. 

\subsection{Images and exposure maps}

Images and exposure maps, corrected for vignetting, have been 
produced with an image pixel size of $2"\times2"$ for each 
energy band (for the definition of all bands, see \S~\ref{ParEbands}). 
To increase the sky coverage, we selected EPIC-pn 
events requiring (FLAG \& 0xfa0000) = 0, which also includes events 
in pixels next to bad pixels or bad columns. Moreover, we 
used single to double pixel events. EPIC-MOS events were required 
to have FLAG = 0 and single to quadruple-pixel events were allowed. 

\begin{figure*} 
\centering
\includegraphics[width=0.8\textwidth,angle=0]{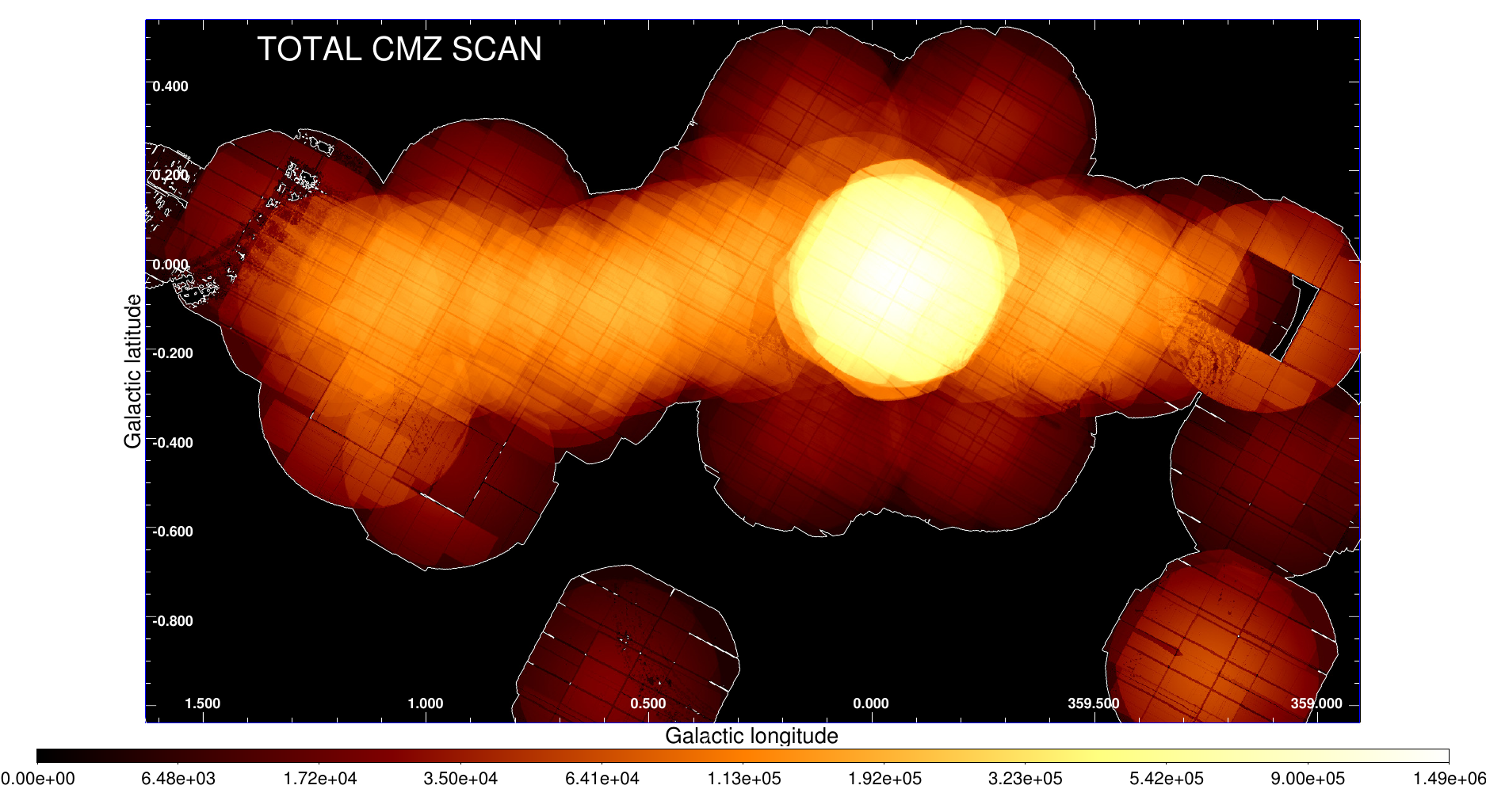}
\includegraphics[width=0.81\textwidth,angle=0]{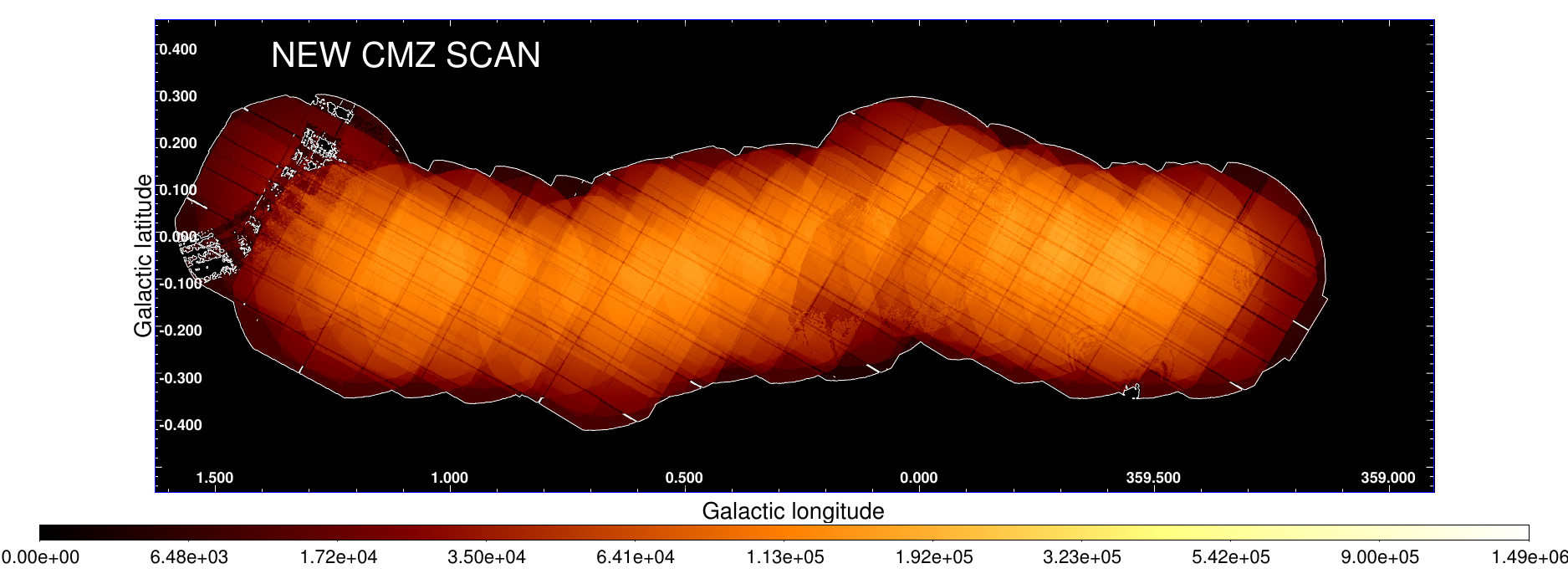}
\includegraphics[width=0.81\textwidth,angle=0]{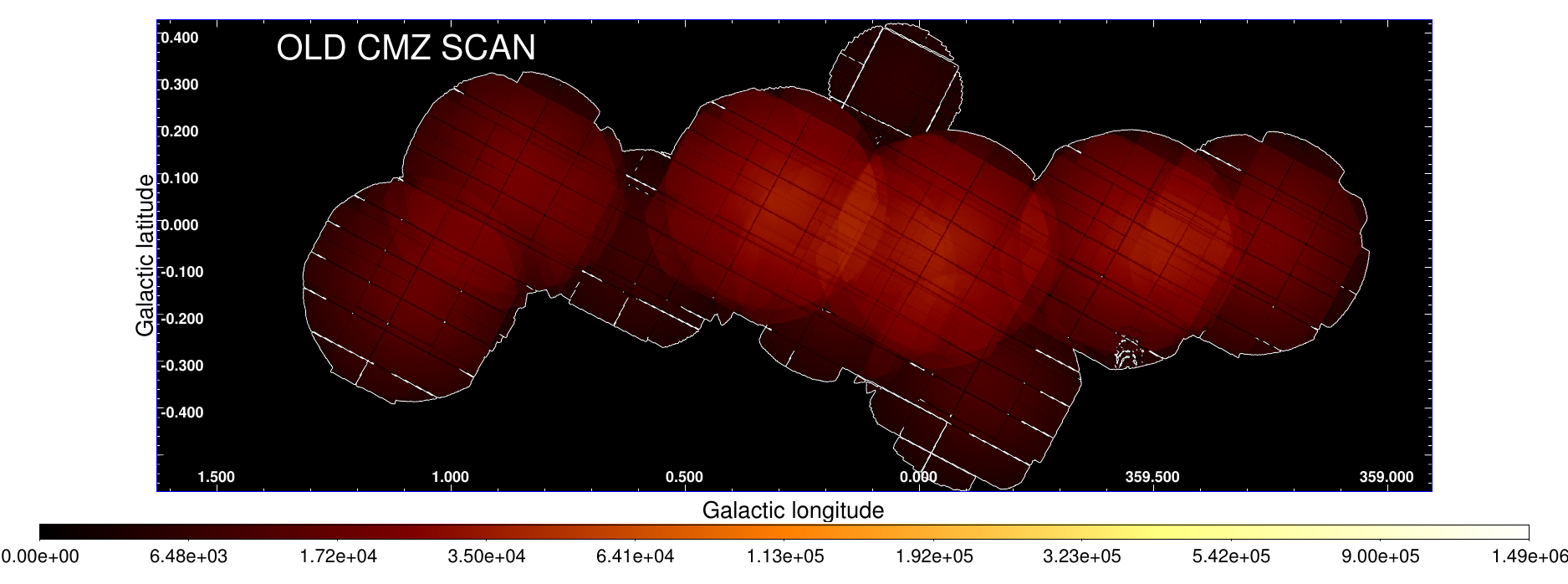}
\caption{{\it (Top panel)} Combined exposure map 
of all the \xmm\ EPIC-pn + MOS1 + MOS2 observations within one degree 
from Sgr~A$^{\star}$. Such as done in Sturm et al. (2013), EPIC-MOS1 and -MOS2 
exposure is weighted by a factor of 0.4 relative to EPIC-pn to account for 
the lower effective area. The exposure times, thus, correspond to the 
equivalent total EPIC-pn exposure time. Regions with less than 7.2 ks 
of equivalent EPIC-pn exposure have been masked out. 
The cleaned EPIC-pn equivalent exposure time is reported in seconds. 
{\it (Medium panel)} Similar exposure map for the observations of the 
new CMZ scan only. 
{\it (Bottom panel)} Similar exposure map for the old CMZ scan only 
(regions with less than 7.2 ks of equivalent EPIC-pn exposure are included).
The maximum exposure times are $\sim1.5$ Ms, $\sim190$ ks and $\sim45$ ks 
during the total, new and old scan, respectively. }
\label{exposure}
\end{figure*}

Figure \ref{exposure} shows the combined 
EPIC exposure map that covers the entire CMZ. 
Such as done in Sturm et al.\ (2013), EPIC-MOS1 and -MOS2 exposures 
are weighted by a factor of 0.4 relative to EPIC-pn, before being 
added to the latter, to account for the lower effective area. 
Therefore, the exposure times obtained correspond to the equivalent 
total EPIC-pn exposure time. This allows us to obtain a better 
combination of EPIC-pn and EPIC-MOS data for image display purposes.  
We note, however, that the fluxes can not be easily read out directly from 
these combined images. Therefore, the line profiles and the measured 
fluxes/luminosities are computed from the EPIC-pn and each EPIC-MOS 
map separately and then combined (averaged) to obtain a better signal 
to noise.

The top panel of Fig. \ref{exposure} shows that more than 1.5 Ms of 
{\it clean} (after cut of time intervals during increased particle background 
activity) exposure time (EPIC-pn equivalent) has been accumulated 
on \sgr\ and over $\sim100-200$ ks are present in each point of the 
CMZ. The few pointings above and below the plane have 
exposures between $\sim15-40$ ks. Regions with less than 7.2 ks of 
equivalent EPIC-pn exposure have been masked out. 

To check the impact of the bright transients on the images and on the physical 
quantities under investigation, two sets of maps have been created. 
The first series keeps all bright transients and point sources, while the second 
set removes their emission by excising from the data extended regions 
including the transients whenever they were in outburst (see section \ref{PS}). 
The middle and bottom panels of Fig. \ref{exposure} show the exposure 
maps (computed in the same way) for the observations of new and old 
CMZ scans, separately. The maximum exposure times are $\sim190$ ks 
and $\sim45$ ks during the new and old scan, respectively. 

\subsection{Energy bands}
\label{ParEbands}

\begin{figure} 
\includegraphics[width=0.52\textwidth,angle=0]{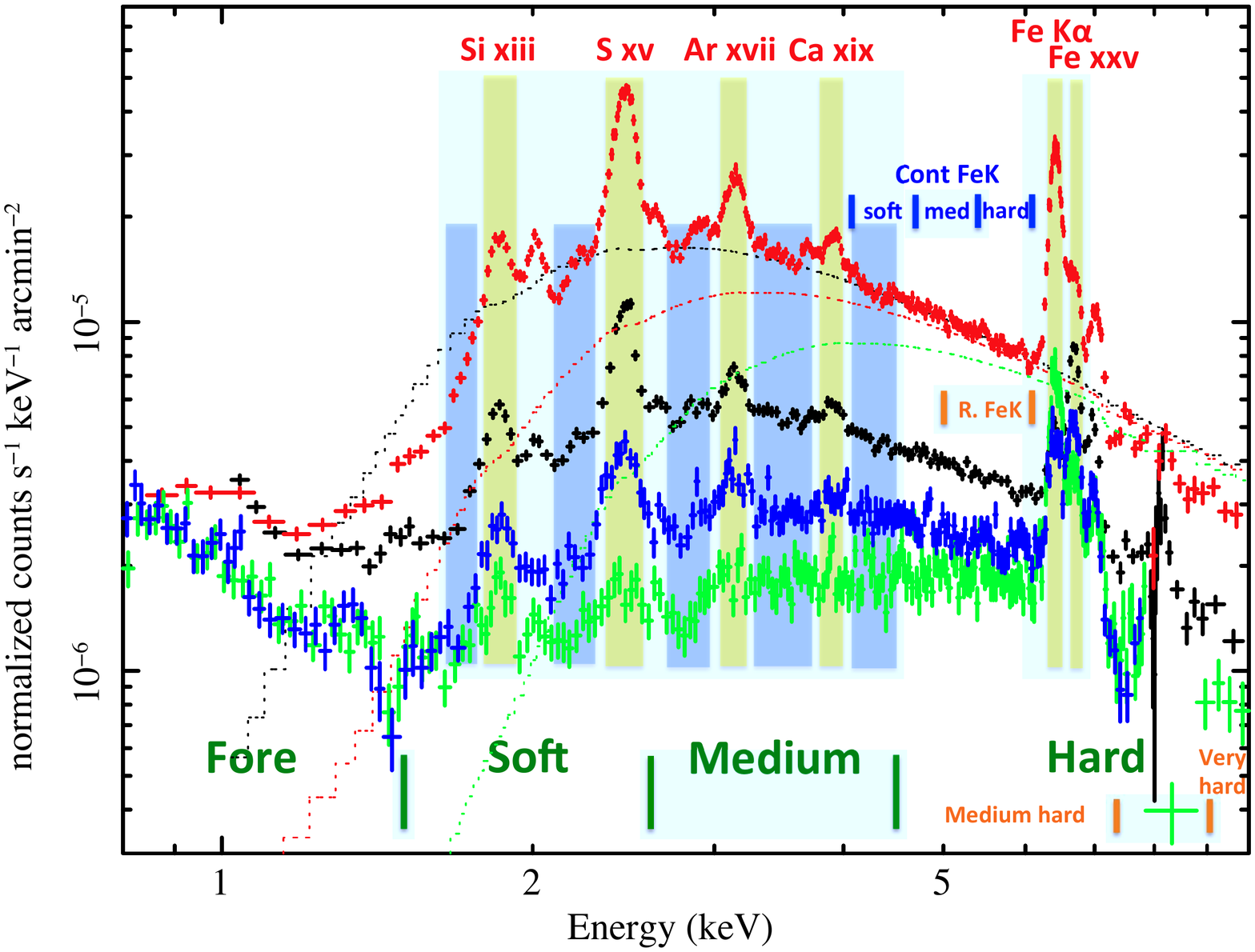}
\caption{EPIC-pn spectra of the regions: {\it G0.11-0.11} (red), 
{\it Center Superbubble} (black), {\it G0.687-0.146} (green) and 
{\it Sgr B1 soft} (blue). 
In dark green are the energy bands of the broad GC continuum. 
In orange (bottom right) is shown the part of the hard energy band 
excluded in order to avoid contamination by Ni K$\alpha$, 
Cu K$\alpha$ and ZnCu instrumental background emission lines. 
Yellow stripes show the energy bands of the soft and Fe K emission lines. 
Blue stripes indicate the regions selected for the determination 
of the amount of continuum underlying the soft lines. 
The dotted lines, from top to bottom, show the predicted emission 
of a source with a power-law spectrum (with slope $\Gamma=1.6$) 
if absorbed by a column density of $N_{\rm H}=3$, $5$ and 
$9\times10^{22}$ cm$^{-2}$, respectively. 
The blue and orange labels indicate the selected broad energy bands 
for the determination of the continuum underlying the Fe K line 
emission.}
\label{ImaEbands}
\end{figure}

We created images in several energy bands (see Tab. \ref{Ebands}). 
Figure \ref{ImaEbands} shows the EPIC-pn spectra of the extended 
emission from several regions within the CMZ. In red and black are the 
spectra from the {\it G0.11-0.11} and {\it Center Superbubble} regions, 
respectively (see Fig.~\ref{FC2}). 
Both regions are located within 15 arcmin from Sgr~A$^{\star}$, 
thus they have excellent statistics because of the large exposure. 
In green and blue are the pn spectra of {\it G0.687-0.146} and {\it Sgr~B1~soft}, 
respectively, both are located further out, thus having lower exposure. 

We first selected the standard broad energy bands for the continuum 
with the softer band being: $E=0.5-2$~keV; the medium $2-4.5$~keV 
and the hard $4.5-12$~keV (see Tab. \ref{Ebands}). 
We note that, at high energies, the EPIC-pn camera shows strong 
instrumental background emission lines due to Ni {\sc K$\alpha$} 
(at E~$\sim7.47$ keV), Cu {\sc K$\alpha$} ($\sim8.04$ keV) and 
ZnCu ($\sim8.63$ and $8.87$ keV) that strongly contribute to the observed 
X-ray emission in the hard band (see Freyberg et al. 2004). 
To avoid contamination from these strong internal background lines, 
we do not consider (for the EPIC-pn images) photons in the 7.2-9.2 keV 
range (see Fig. \ref{ImaEbands} and Tab. \ref{Ebands}). 
We chose these broad energy bands because they are typically used 
as input by the standard \xmm\ point source detection algorithm and 
for comparison to other similar surveys of nearby galaxies (e.g. 
M33: Misanovic et al. 2006; T{\"u}llmann et al. 2011; M31: Henze et al. 2014; 
Stiele et al. 2011; LMC: Haberl et al. 1999; SMC: Haberl et al. 2012; 
Sturm et al. 2013). However we note that, given the typical GC 
neutral column density of several $10^{22}$ cm$^{-2}$, the low 
energy absorption cut-off occurs at the highest energies of the standard 
soft band, making standard broad band RGB  images poorly sensitive 
to column density fluctuations. 
For this reason we define a second set of broad bands, the "GC continuum 
bands" (see Tab.~\ref{Ebands}). The first band ($E=0.5-1.5$ keV) contains 
mainly emission from foreground sources. The second band ($E=1.5-2.6$ 
keV) is selected in order to contain the low-energy GC neutral absorption 
cut-off, thus making it more sensitive to either column density or soft gas 
temperature variations. While the "GC medium" ($E=2.6-4.5$ keV) and 
the "GC hard" ($E=4.5-12$ keV) bands are similar to the standard broad bands. 

We also selected images at the energies of the soft emission lines, 
such as Si {\sc xiii}, S {\sc xv}, Ar {\sc xvii} and Ca {\sc xix}.  
To perform continuum subtracted line intensity maps as well as line 
equivalent width maps, it is essential to measure the level of the 
continuum underlying the emission line. Therefore, we created also 
several images in energy bands on each side of the soft emission 
lines\footnote{The energy band red-ward of the Si {\sc xiii} line extends 
only down to 1.65~keV, because of the presence of the strong 
Al K$\alpha$ background emission line at E~$\sim1.49$ 
keV (Freyberg et al. 2004).} (selecting, as far as possible, energy 
ranges free from line emission; see Fig. \ref{ImaEbands} and 
Tab. \ref{Ebands}). 

In the Fe K region we selected two energy bands for the Fe K$\alpha$ 
and Fe {\sc xxv} emission. At energies higher than Fe {\sc xxv} the 
presence of both Fe K$\beta$, Fe {\sc xxvi}, and of the Fe K edge can 
give a significant contribution. At even higher energies (E~$\sim7.5-8$ keV)
the contribution from internal background emission line (in the EPIC-pn 
camera) becomes very important, thus we decided to determine the 
continuum emission underlying the Fe K line emission (important to determine
the Fe~K line intensities and equivalent widths) through 
the extrapolation of the continuum red-ward of the Fe K lines (see 
Fig. \ref{ImaEbands} and Tab. \ref{Ebands}). 
The Fe~K line emission and its variations will be the prime scientific focus 
of two future publication (Ponti et al.\ in prep.; Soldi et al.\ in prep.; see 
also Ponti et al.\ 2014; Soldi et al.\ 2014) and will not be discussed 
here any further. 

All images were exposure corrected and, to remove readout streaks, 
the images from EPIC-pn were corrected for out-of-time events. 
Noisy CCDs in the MOS data (Kuntz \& Snowden 2008) have been 
searched with the SAS task {\it emtaglenoise} and removed from the 
mosaic images. 

\begin{table} 
\begin{center}
\begin{tabular}{ c c c c c c }
\hline
\multicolumn{5}{c}{{\bf Standard continuum bands:}} \\
Soft              & Medium       & Hard\dag\\
0.5-2            & 2-4.5            & 4.5-12 \\
\hline
\multicolumn{5}{c}{{\bf GC continuum bands:}} \\
Fore     & GC Soft       & GC Medium & GC Hard\dag\\
0.5-1.5 & 1.5-2.6        & 2.6-4.5         & 4.5-12 \\
\hline
\multicolumn{5}{c}{{\bf Soft emission lines:}} \\
Si {\sc xiii}& S {\sc xv}  & Ar {\sc xvii} & Ca {\sc xix} \\
1.80-1.93 & 2.35-2.56 & 3.03-3.22 & 3.78-3.99 \\
\multicolumn{5}{c}{{\bf Continuum subtraction soft emission lines:}} \\
Red-Si           & Si-S      & S-Ar      & Ar-Ca       & Blue-Ca       \\
1.65-1.77 & 2.1-2.3 & 2.7-2.97 & 3.27-3.73 & 4.07-4.5 \\ 
\hline
\multicolumn{5}{c}{{\bf Fe K lines:}} \\
 Fe K$\alpha$ & Fe {\sc xxv} \\
 6.3-6.5           & 6.62-6.8      \\
\multicolumn{5}{c}{{\bf Continuum subtraction Fe K:}} \\
CFeK & CsFeK   & CmFeK      & ChFeK \\
          & soft      & medium   & hard \\
5-6.1  & 4.0-4.7 & 4.7-5.4   & 5.4-6.1      \\ 
\hline
\end{tabular}
\caption{Energy bands used for each of the different continuum, 
and narrow line images. Also shown are the energy bands used 
to determine the continuum underlying the line emission.  
Units are in keV. Several energy bands, at lower energies compared 
to the FeK lines, have been computed to determine the best continuum 
subtraction for the FeK lines. \dag To avoid contribution from the strong 
internal detector background emission lines, present in the 
EPIC-pn camera (such as: Ni {\sc K$\alpha$}, Cu {\sc K$\alpha$} 
and ZnCu), we do not consider photons in the 7.2--9.2 keV from 
this instrument (on the other hand, we consider such photons 
detected in the EPIC-MOS cameras). }
\label{Ebands}
\end{center}
\end{table} 

\subsubsection{Stray-light rejection}
\label{stray}

\begin{figure*} 
\includegraphics[width=1.27\textwidth,angle=90]{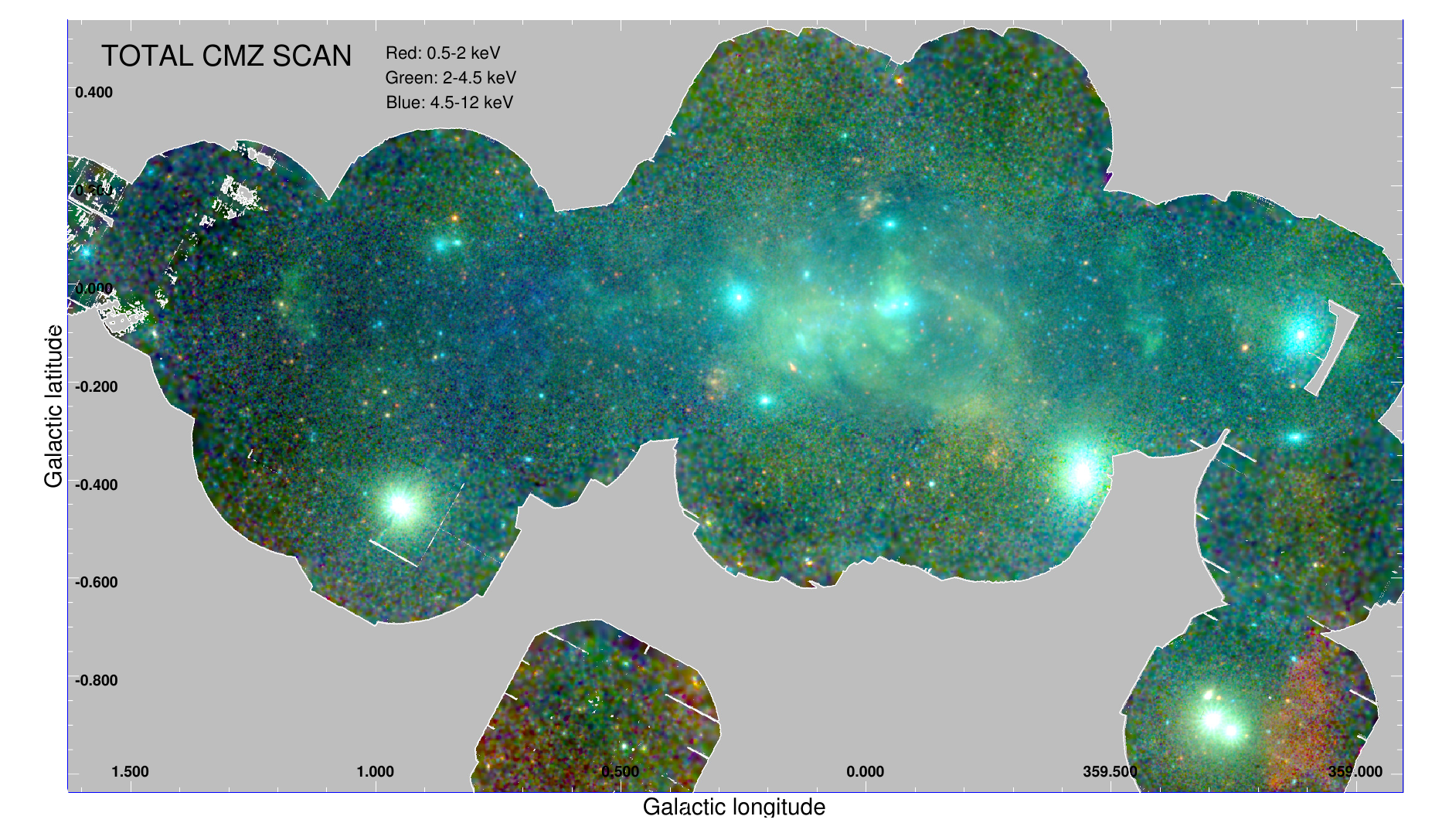}
\caption{Standard broad energy band (see Tab. \ref{Ebands} and Fig. \ref{ImaEbands}) 
RGB mosaic image of all \xmm\ observations within one degree of Sgr~A$^{\star}$ 
(see Tab. \ref{TabObs1}). This represents the deepest X-ray view of the CMZ 
region with exposure higher than $0.2$~Ms along the Galactic disc and 
$1.5$~Ms in the center (see Fig. \ref{exposure}). X-ray emission from X-ray binaries, 
star clusters, supernova remnants, bubbles and superbubbles, HII regions, PWNs, 
non-thermal filaments, nearby X-ray active stars, the supermassive BH Sgr~A$^\star$ 
and many other features are observed (see Fig.\ \ref{FC1} and \ref{FC2}). 
The detector background has been subtracted and adaptive smoothing applied. 
Residual features and holes generated by correction of the stray light from GX 3+1 
are visible (see also Fig. 1) at Galactic latitudes between $l \sim 1.2^\circ$ and 
$l \sim 1.4^\circ$ and latitudes $b \sim -0.2^\circ$ and $b \sim 0.4^\circ$. }
\label{RGBhard}
\end{figure*}

Because the GC region is crowded with many bright (transient) X-ray 
sources, several observations, including the new \xmm\ scan, are badly 
affected by stray-light (see Fig. \ref{RGBhard}). Stray-light is produced by 
photons from sources located outside of the \xmm\ EPIC instrument's 
fields of view and singly reflected by the mirror hyperbolas, thus 
creating concentric arc-like structures in the detector plane 
(see \xmm\ user handbook). The stray-light contribution is small 
(the effective collecting area for stray-light is less than $\sim0.2$ \% 
of the effective on-axis area), but a very bright source can have an 
important impact up to $\sim1.4$ deg outside the field of view. 
\begin{figure*} 
\includegraphics[width=1.27\textwidth,angle=90]{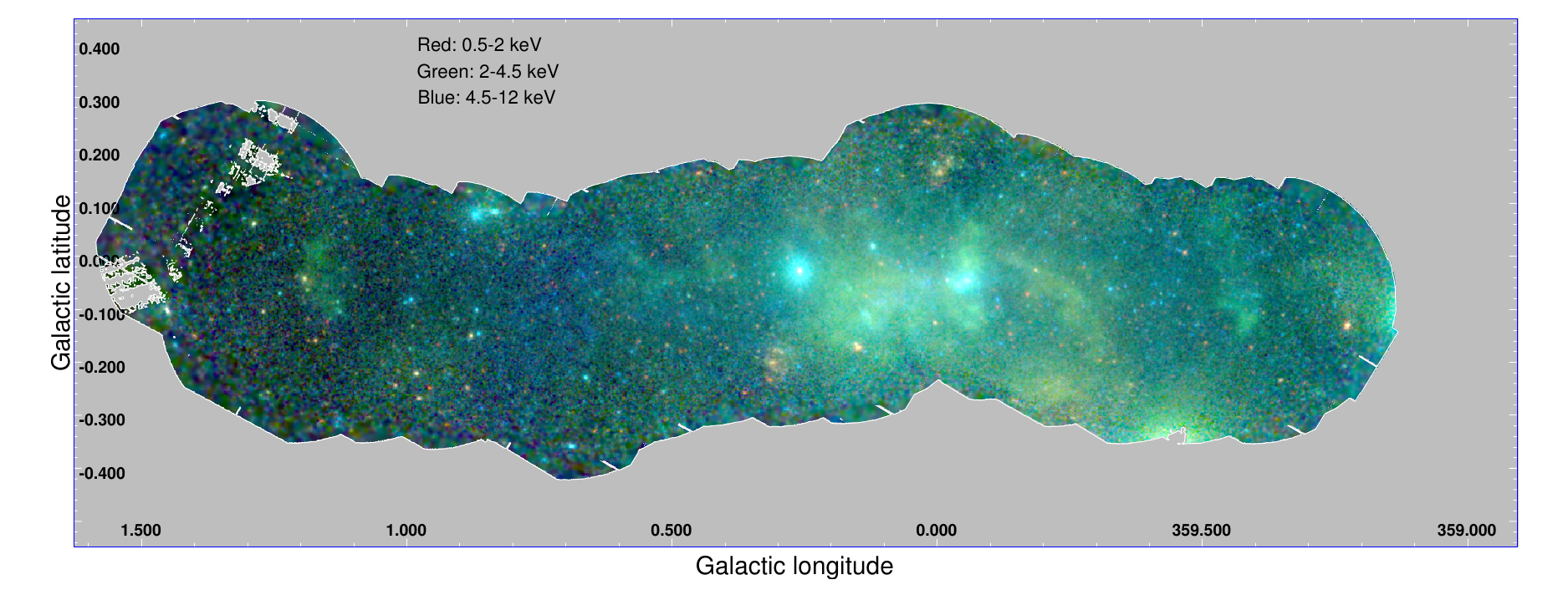}
\caption{Standard broad energy band RGB image of the \xmm\ CMZ scan performed in 2012. 
The CMZ is observed with a uniform exposure (see Fig.\ \ref{exposure}). }
\label{RGBhardNew}
\end{figure*}
\begin{figure*} 
\includegraphics[width=1.0\textwidth,angle=0]{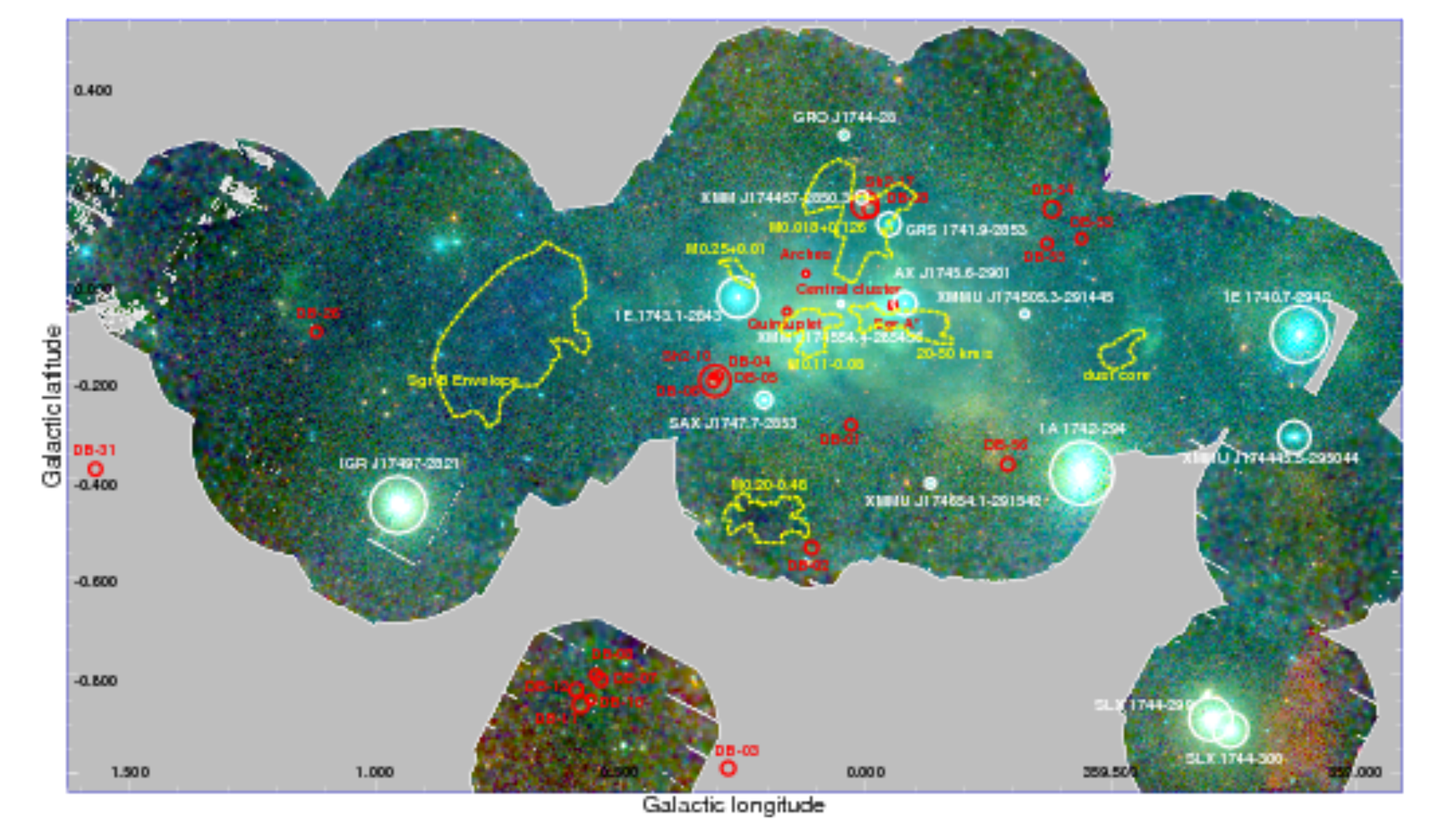}
\caption{Finding chart. The brightest X-ray point sources (all X-ray binaries)
are labelled in white (see Tab. \ref{TabPS}). In red the positions 
of some star clusters are reported, which are placed either in the GC or 
along the spiral arms of the Galaxy (see Tab. \ref{AtlasSCSNR}). 
With yellow dashed lines the location of some molecular complexes are shown. 
See {\sc www.mpe.mpg.de/heg/gc/} for a higher resolution version of this figure. }
\label{FCTSC}
\label{FC1}
\end{figure*}
\begin{figure*} 
\includegraphics[width=1.0\textwidth,angle=0]{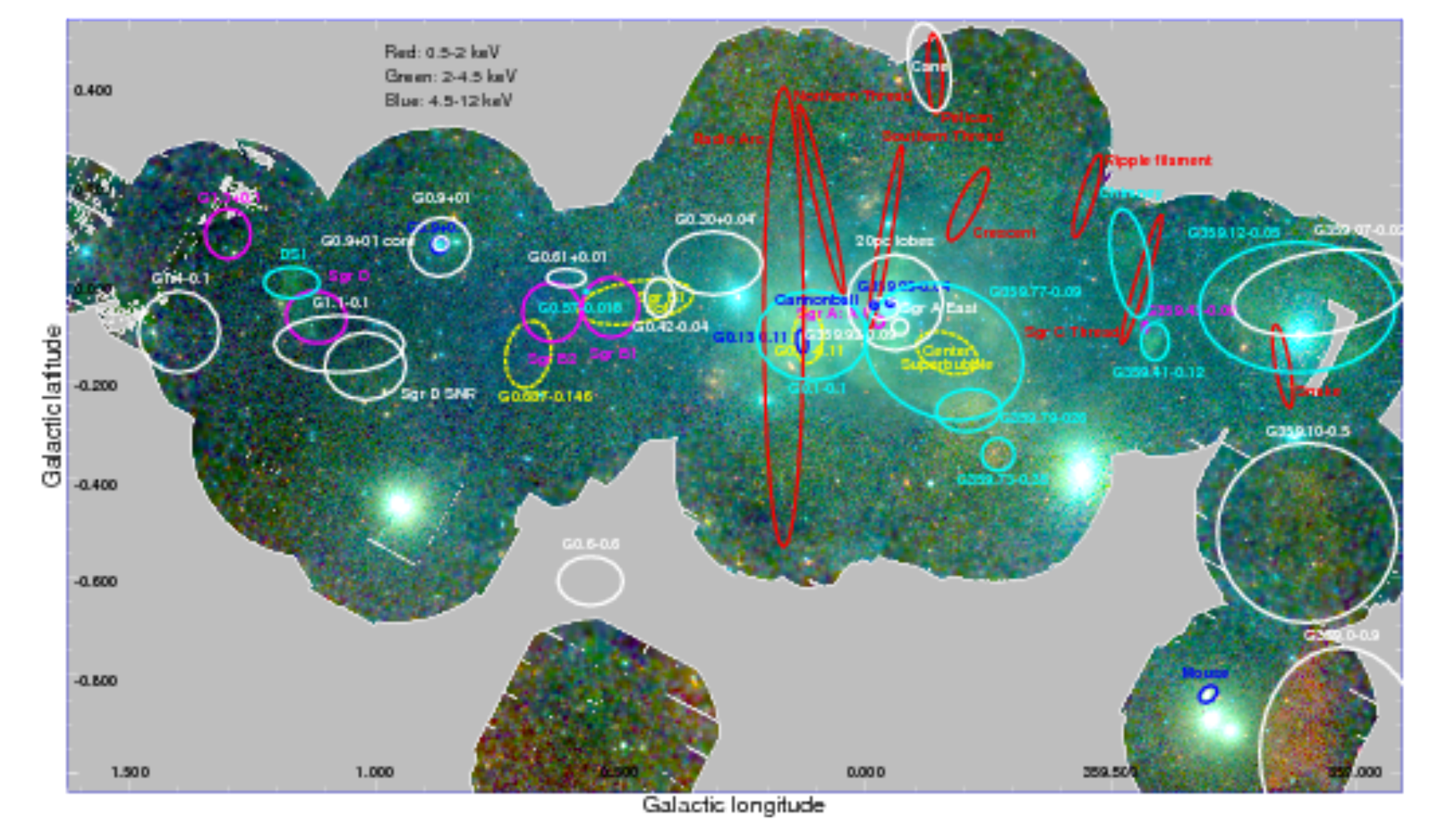}
\includegraphics[width=1.0\textwidth,angle=0]{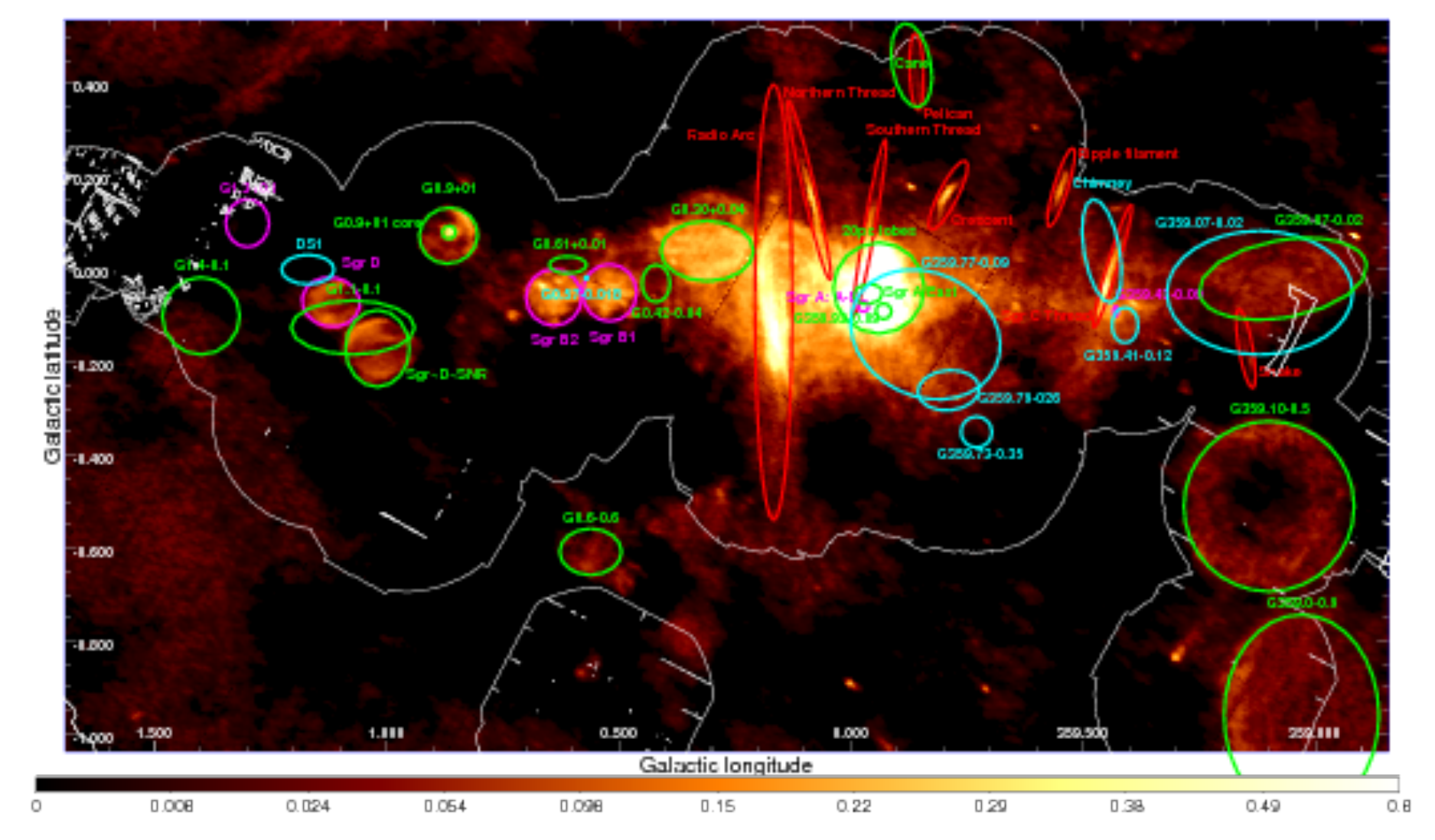}
\caption{Finding charts. {\it (Top panel)} Broadband X-ray continuum image. White ellipses 
show the position and size of known, radio-detected supernova remnants. Cyan 
ellipses indicate the position and size of bright diffuse X-ray emission possibly associated with
supernova remnants that lack a clear radio counterpart (or such in the case of G359.12-0.05 
that show X-ray emission significantly displaced from the radio emission associated to 
the radio remnant G359.07-0.02). The magenta ellipses 
show the location and dimension of some bright HII regions, while the red ellipses indicate 
some of the largest non-thermal filaments detected in radio (see Tab. \ref{AtlasNTF}). 
Blue ellipses show some PWN and the yellow dashed ellipses show the regions used 
to accumulate the spectra shown in Fig.\ \ref{ImaEbands}. 
{\it (Bottom panel)} 90-cm radio image of the CMZ region obtained with the VLA (courtesy 
of LaRosa et al. 2000). For display purposes the radio supernova remnants are shown with 
green ellipses. The other regions have the same colour code as the top panel. 
The image shows the radio flux (Jy beam$^{-1}$ unit). 
See {\sc www.mpe.mpg.de/heg/gc/} for a higher resolution version of these figures. }
\label{FCBFSNRHII}
\label{FC2}
\end{figure*}

\begin{figure*}
\includegraphics[width=1.0\textwidth,angle=0]{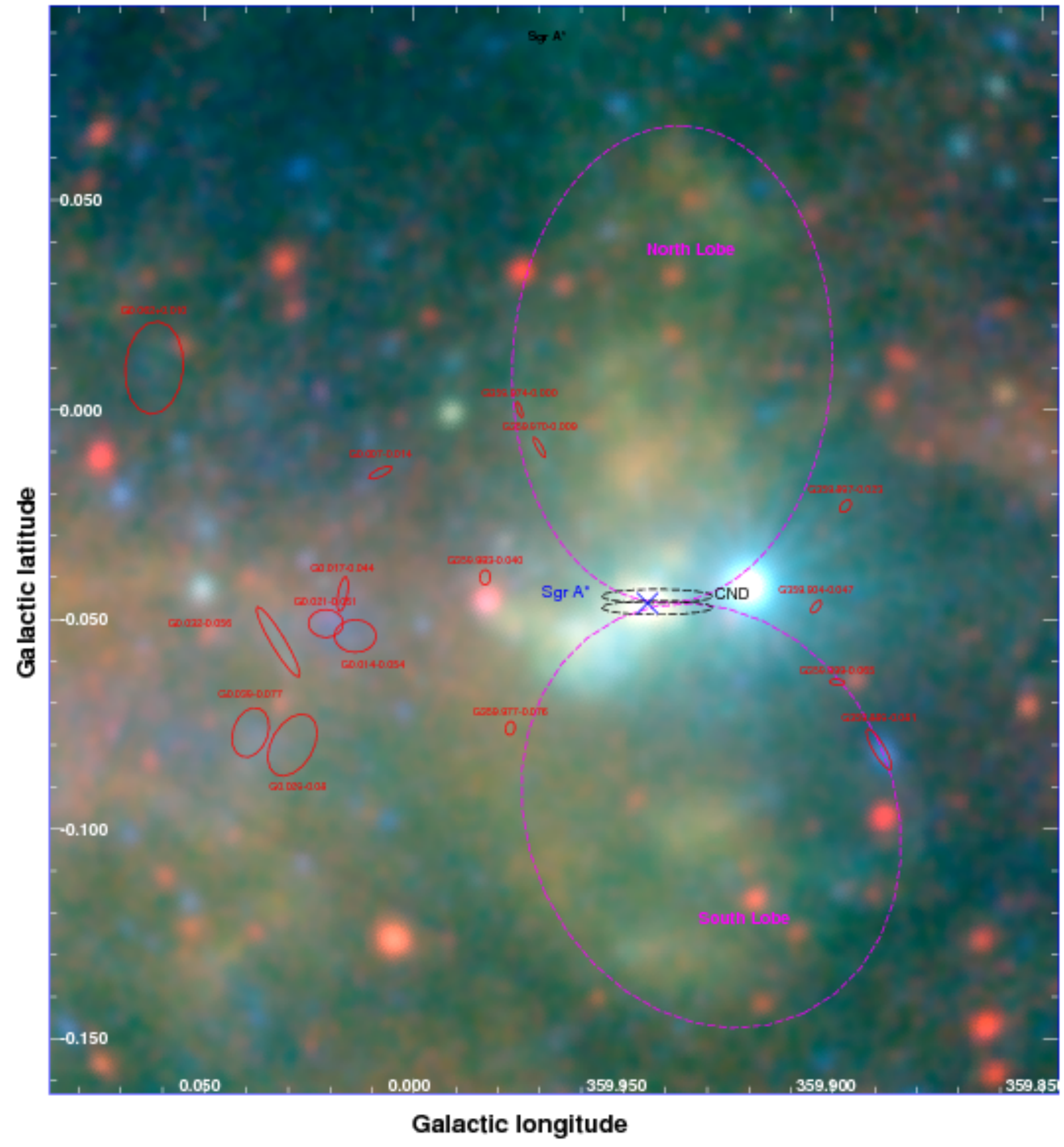}
\caption{Finding chart. Zoom of the central $\sim10$~arcmin of the Milky Way
as seen by \xmm\ (same energy bands and color scheme as in Fig.\ \ref{RGBhard}). 
The position of Sgr~A$^{\star}$ is indicated by the blue cross. 
The red ellipses show the position and extent of filamentary and diffuse X-ray emission features 
associated with, e.g., non-thermal filaments (Tab. \ref{AtlasNTF}). 
The magenta dashed ellipses show the location and extension of the 20~pc bipolar  
X-ray lobes. The black dashed ellipses indicate the position and orientation of the CND. 
See {\sc www.mpe.mpg.de/heg/gc/} for a higher resolution version of this figure. }
\label{FCZoom}
\end{figure*}

\begin{figure*}
\includegraphics[width=1.0\textwidth,angle=0]{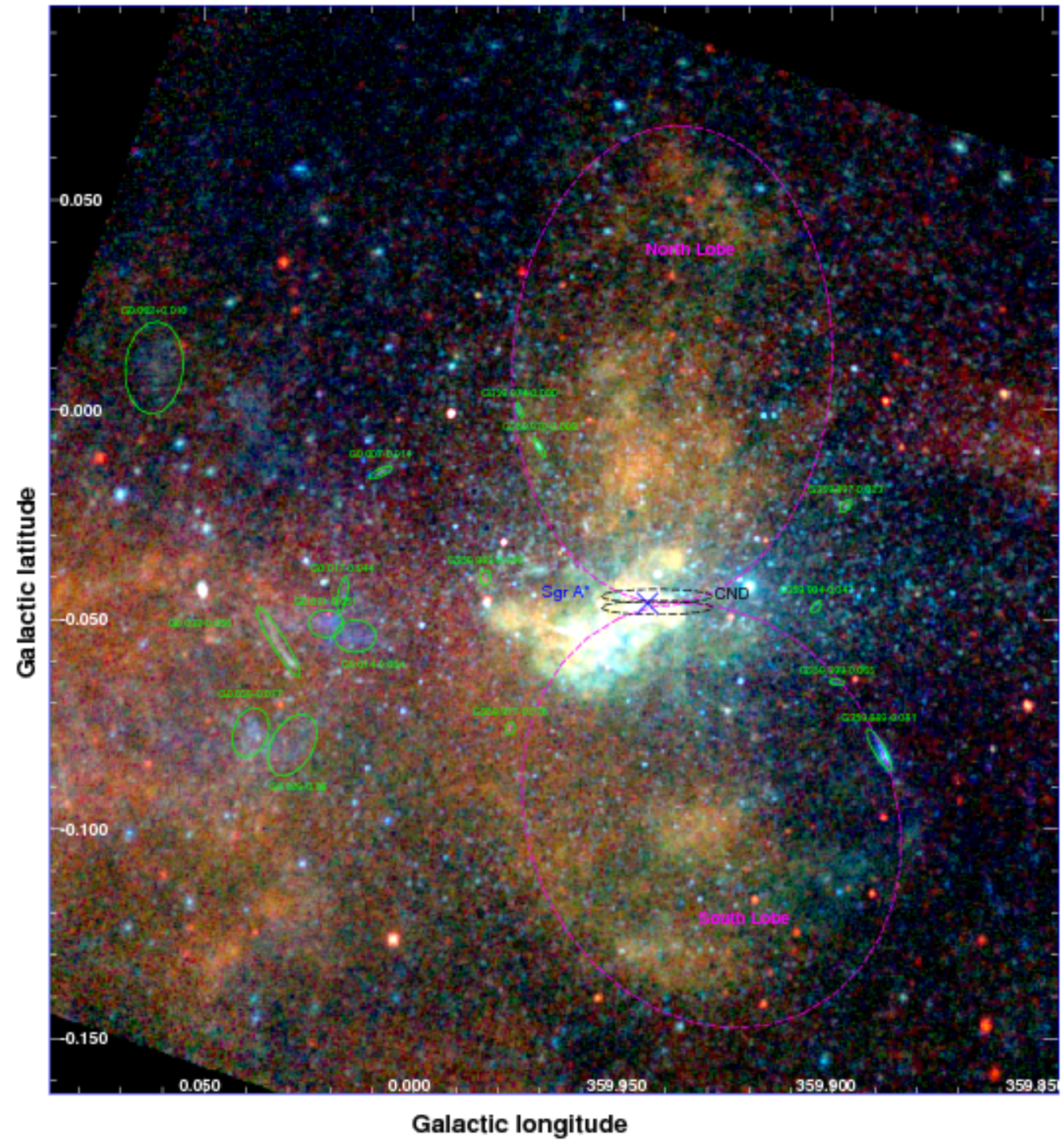}
\caption{Finding chart. \chandra\ RGB image of all the ACIS-I observations pointed at 
Sgr~A$^\star$ (see Clavel et al. 2013 for data reduction and details on the image 
production). The red, green and blue images show the GC soft (1.5-2.6~keV), 
GC medium (2.6-4.5~keV) and GC hard (4.5-8~keV) energy bands, respectively. 
The same regions displayed in Fig.\ \ref{FCZoom} are evidenced here. 
See {\sc www.mpe.mpg.de/heg/gc/} for a higher resolution version of this figure. }
\label{FCZoomChandra}
\end{figure*}

\begin{figure*}
\includegraphics[width=1.0\textwidth,angle=0]{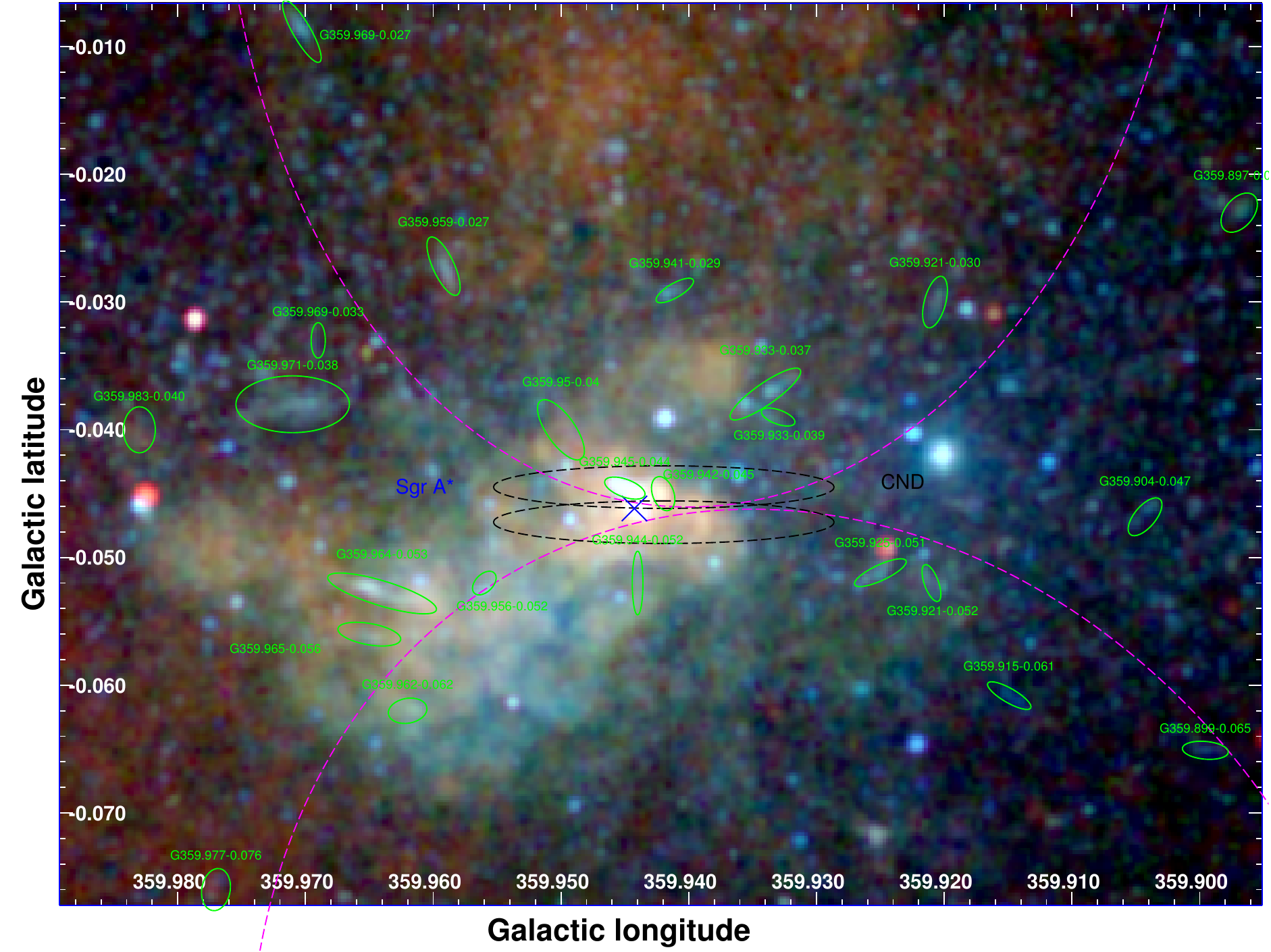}
\caption{Zoom of the finding chart displayed in Fig. \ref{FCZoomChandra}. }
\label{FCZoomChandra2}
\end{figure*}

Analogously to the removal of bright transients, we masked the strongest 
stray-light artefacts in the images of individual observations.
In most cases, affected regions are covered by other unaffected observations, 
thus leaving no features in the final mosaic map. 
To remove a stray-light artefact, we defined a rough region including
the artefact and an individual cut-off value of the surface brightness.
Using this cut-off, we created a mask from an image of this region in
the total energy band that has been smoothed with a Gaussian kernel with
a FWHM of 10\arcsec\ beforehand.
This mask was multiplied by all images and exposure maps of this
observation.

\subsubsection{Adaptive smoothing}
\label{Asmooth}

All images have been smoothed separately using the {\sc sas} tool {\sc asmooth}. 
To prevent different smoothing patterns from introducing colour artefacts in RGB 
images, we adaptively smoothed all energy bands in such images with the same 
smoothing template.  For the broadband \xmm\ continuum RGB images 
(see Fig. \ref{RGBhard}, \ref{RGBhardNew}, \ref{FC1}, \ref{FC2}, \ref{FCZoom}, 
\ref{RGBhardNH} and \ref{MorrisNH}), we required a minimum signal-to-noise ratio 
of 6 in the 0.5-12~keV energy band image (i.e., the sum of the three energy bands 
composing the RGB image), as well as the standard minimum and maximum size 
of the smoothing Gaussian kernel of 10\arcsec\ (full width half maximum) and 
200\arcsec, respectively. The signal to noise ratio at each pixel is defined as 
the value at that pixel divided by its standard deviation and the adaptive 
smoothing that we applied is making the signal to noise ratio as close at possible 
to 6, therefore fainter or less exposed areas are more smoothed than brighter or 
better exposed regions. 
For the narrower-band soft-line images (see Fig. \ref{RGBSoftLines1}, 
\ref{RGBSoftLines2}, \ref{Face} and \ref{MIPS}), we use the S~{\rm xv} map as 
a template, requiring a minimum signal-to-noise ratio of 4 and the same standard minimum 
and maximum of the smoothing kernel. The same smoothing kernel 
is then applied to all the other bands of the RGB images. 

\subsubsection{Internal particle background subtraction}
\label{PB}

Unless otherwise specified, internal particle background has been 
removed from each broad-band image. Following Haberl 
et al. (2012) we first create, for each selected energy band, both 
the {\it total emission} and the {\it filter wheel closed} images. 
We then re-normalise and subtract the {\it filter wheel closed} images from 
the {\it total emission} images. The {\it filter wheel closed} image re-normalisation 
factor is computed by equating, for each instrument, the number of 
photons in the unexposed corners of the detector to that in the {\it filter wheel closed} 
images (see Haberl et al.\ 2012 for more details).  
This procedure is reliable and accurate for reasonably long exposures 
($t\simeq5-10$~ks). For this reason datasets with total clean EPIC-pn 
exposure shorter than $5$ ks have not been considered in this analysis. 

\subsubsection{Continuum subtraction}
\label{ContSub}

To subtract the continuum emission from an emission-line image, 
we define a narrow band (B) containing the line, typically sandwiched 
by two nearby but generally wider energy bands (A and C) that are dominated 
by continuum emission. Under the assumptions that the emission 
in the A and C bands is dominated by the continuum and that 
the continuum emission can be described by a simple power-law, 
we could in principle determine the intensity of the continuum for each pixel 
of the band B image. 
Indeed, using the fluxes in A and C bands, it is possible to 
derive the continuum parameters (spectral index $\Gamma$ and 
intensity). However, this requires the solution of non-linear equations. 
Therefore, we prefer to implement a different technique based on interpolation. 
Using {\it Xspec} we simulate, for power-law spectral indices going from 
$\Gamma=0.3$ to $3.6$, the ratio between the observed flux  
(e.g., number of photons measured) in the continuum in bands A and C, 
compared to the simulated continuum flux in band B (e.g. $N_B/(2\times N_A)$ 
and $N_B/(2\times N_C)$). 
We record these ratios and then plot them as a function 
of the hardness ratios $(N_C-N_A)/(N_C+N_A)$), which is 
a proxy for the spectral index $\Gamma$. 
We then find the best-fitting linear relationship between these values, 
thus determining $Con_{AB}$ and $Lin_{AB}$ 
that are then allowing us 
to measure the continuum emission underlying the line emission in 
band B ($N_B$) from the intensity in band A ($N_A$) and the hardness 
ratio ($N_B=2\times N_A \times [Con_{AB}+Lin_{AB}\times 
(N_C-N_A)/(N_C+N_A)]$.  To reduce the uncertainties, we perform the 
same procedure for band C, determining $Con_{CB}$ and $Lin_{CB}$. 
We then average these values obtaining, for each pixel: 
$N_B= N_A \times [Con_{AB} + Lin_{AB} \times (N_C-N_A)/(N_C+N_A)] 
        + N_C \times [Con_{CB} + Lin_{CB} \times (N_C-N_A)/(N_C+N_A)]$. 
We finally subtract this continuum emission image from the total 
emission image B to determine the line intensity map.

\section{The \xmm\ broadband view of the Galactic Centre}

Figure \ref{RGBhard} shows the broad energy band mosaic image 
of all existing \xmm\ observations within 1 degree from Sgr~A$^{\star}$.
Figure~\ref{RGBhardNew} shows the Galactic centre image obtained only 
with data from the 2012 \xmm\ campaign. At the GC distance of 7.8 kpc 
(Boehle et al.\ 2015), 1~arcmin corresponds to 2.3~pc, 10~pc subtends 
$\sim4.3$' and $\sim0.2^\circ$ corresponds to 28~pc. 
In red, green and blue, the soft (0.5-2~keV), medium (2-4.5~keV) 
and hard (4.5-12~keV) continuum bands are shown, respectively. 
Hundreds of point sources and strong diffuse emission are clearly 
observed in the map. These point sources are characterised by a 
wide variety of colours, ranging from distinctive red to dark blue. 

\subsection{Bright and transient GC point sources during the 
new (2012) \xmm\ CMZ scan}
\label{PS} 

Many X-ray point sources are clearly visible in Fig. \ref{RGBhard} and 
\ref{RGBhardNew}. 
A detailed catalogue of the properties of all the detected point sources 
is beyond the scope of this paper. Here we briefly describe the 
brightest GC sources detected by \xmm\ and the X-ray transients in the 
field of the 2012 scan (see Degenaar et al. 2012 for a compendium of
previously noted transients). 

The brightest GC point source of the 2012 \xmm\ CMZ scan 
is 1E 1743.1-2843 (Porquet et al. 2003; 
Del Santo et al. 2006) a persistently accreting neutron star binary 
detected at an observed flux level of F$_{\rm 2-10 keV}\sim1.1\times
10^{-10}$ erg cm$^{-2}$ s$^{-1}$ (implying an unabsorbed flux of 
F$_{\rm 2-10 keV, unab}\sim2.6\times10^{-10}$ erg cm$^{-2}$ s$^{-1}$; 
N$_{\rm H}\sim2\times10^{23}$ cm$^{-2}$; obsid: 0694641201). 
During the 2012 \xmm\ campaign we also detected an outburst from a 
new, very faint X-ray transient that we name XMMU J174505.3-291445. 
The source has a typical quiescent luminosity at or below $L_X \sim 
10^{33}$ erg s$^{-1}$, but on 2012 August 31$^{st}$ (during obsid 0694640201) 
it was observed to go into outburst for about $\sim2$ hr and to reach a peak X-ray 
luminosity of $L_X \sim 10^{35}$~erg~s$^{-1}$. The detailed spectral and 
multiwavelength analysis of this new transient will be presented in a 
separate paper (Soldi et al.\ in prep.; but see also Soldi et al.\ 2014). 

Another faint X-ray transient, XMM J174457-2850.3, is detected in
two 2012 observations; obsid: 0694641101 - 0694640301. 
The observed 2-10~keV flux is $1.1\pm0.3$ and 
$2.9\pm0.6\times 10^{-13}$~erg~cm$^{-2}$~s$^{-1}$, 
respectively. A power-law fit to the spectrum with the photon 
index fixed to the value reported by Sakano et al. (2005) yields 
a column density of ${\rm N_H}=(1.4\pm0.4)\times10^{23}$ cm$^{-2}$.
This source was discovered in 2001 by Sakano et al. 
(2005) who reported a tentative detection of an X-ray pulsation 
of $\sim5$~s, during the $\sim25$ ks \xmm\ observation. 
Both a visual inspection and timing analysis of the X-ray light curve 
show no evidence for bursts and/or dips. However, even considering 
the 4 times longer exposure of the new data, we cannot exclude 
or confirm the $\sim5$~s periodic modulation because of the lower flux observed.  
In fact, during the 2001 \xmm\ observation (obsid: 0112972101) XMM J174457-2850.3 
had a flux about $10-40$ times higher ($\sim45\times 10^{-13}$ erg 
cm$^{-2}$ s$^{-1}$) than in 2012 (in quiescence XMM J174457-2850.3 
has a typical 2-10 keV flux lower than $0.2\times 10^{-13}$ erg 
cm$^{-2}$ s$^{-1}$). 

Only upper limits are measured for the other well-known X-ray transients 
within the field of view. The two bursters GRS 1741.9-2853 and AX J1745.6-2901 
(Sakano et al. 2002; Trap et al. 2009; Ponti et al. 2014; 2015) have flux limits of 
F$_{\rm 2-10 keV} < 2\times10^{-14}$ erg cm$^{-2}$ s$^{-1}$
and F$_{\rm 2-10 keV} < 10^{-13}$ erg cm$^{-2}$ s$^{-1}$
(obsid: 0694641101, 0694640301), respectively. 
Closer to Sgr~A$^{\star}$, we find an upper limit on the 2-10 keV flux of 
F$_{\rm 2-10 keV} < 5\times10^{-12}$~erg~cm$^{-2}$~s$^{-1}$ toward three other sources:
CXOGC J174540.0-290031, the low-mass X-ray 
binary showing X-ray eclipses (Porquet et al. 2005; Muno et al. 2005), 
CXOGC~J174540.0-290005 (Koch et al. 2014), and the magnetar discovered 
on April 25, 2013 (Degenaar et al.\ 2013; Dwelly \& Ponti 2013; Mori et al.\ 2013; 
Rea et al.\ 2013; Kaspi et al.\ 2014; Coti-Zelati et al.\ 2015), located 
at distances from Sgr~A$^{\star}$ of only $\sim2.9$, $\sim23$ 
and $\sim2.4$~arcsec, respectively. 

Finally we observe that both XMMU J174554.4-285456, the faint transient 
with a possible pulsation period of about $172$ s (Porquet et al. 2005), 
and SAX J1747.7-2853, the bursting (showing also superbursts) X-ray 
transient (Wijnands et al. 2002; Natalucci et al. 2004; Werner et al. 2004), 
have flux limits of F$_{\rm 2-10 keV} < 2\times10^{-13}$~erg~cm$^{-2}$~s$^{-1}$. 

\subsubsection*{Bright sources outside the 2012 scan}
Three very bright sources are outside the field of view during the 2012 CMZ 
scan, however they imprint their presence through bright stray-light arcs. 
The arc features between and south of the Sgr~A and C complexes 
($l\sim359.6-359.9^{\circ}$, $b\sim-0.15-0.4^{\circ}$) testify 
that the bright X-ray burster 1A~1742-294 (Belanger et al. 2006; 
Kuulkers et al. 2007) was active during the 2012 \xmm\ campaign. 
The very bright arcs east of the Sgr D complex (obsid: 0694641601) are most 
probably produced by the very bright neutron star low-mass X-ray binary 
GX 3+1 (Piraino et al. 2012) located about $1.18^{\circ}$ northeast of the 
arcs\footnote{This region is covered only by the observations of the 2012 \xmm\ 
scan, therefore the removal of the stray-light arcs generates regions with 
null exposures in the final mosaic maps (e.g., Fig.~\ref{RGBhard} and \ref{RGBhardNew}).}. 
On the far west edge of the 2012 scan a brightening is observed. 
This is due to 1E 1740.7-2942 (Castro et al. 2013; Reynolds \& Miller 2010;
Natalucci et al. 2014), a bright and persistent accreting microquasar, 
at only $\sim1.5$~arcmin from the edge of the 2012 field 
of view (see Fig.~\ref{RGBhard}).
The lack of straylight south of the Sgr B region suggests that the BH 
candidate IGR J17497-2821 (Soldi et al. 2006; Paizis et al. 2009) 
was in quiescence during these observations. 
Two bright X-ray bursters have been active during the 2003 \xmm\ observation pointed 
to the pulsar wind nebula called The Mouse, i.e., SLX 1744-299 and SLX 1744-300 
(Mori et al. 2005). 

\subsection{Very soft emission: Foreground emission}

Despite the presence of distinctively soft (red) point sources, 
Fig. \ref{RGBhard} shows no strong, diffuse, very soft X-ray emission. 
This is mainly due to the very high column density of neutral hydrogen 
toward the GC (with typical values in the range 
N$_H\sim3-9\times10^{22}$ cm$^{-2}$; see also \S \ref{NH}). 
Almost no Galactic centre 
X-ray radiation reaches us below $E\lsimeq1.3$, $1.7$ or $2.3$~keV 
for column density values of N$_H\simeq3$, $5$, or 
$9\times10^{22}$ cm$^{-2}$, respectively (see Fig. \ref{ImaEbands}). 
The majority of the "red" sources present in the 0.5-1.5 keV 
band are point-like and are associated with foreground active 
stars characterised by an unabsorbed soft X-ray spectrum. 

Two clearly extended and soft X-ray emitting sources are present in 
Fig. \ref{RGBhard}. These correspond to Sh2-10 and Sh2-17 (Wang et 
al. 2002; Dutra et al. 2003; Law et al. 2004; Fukuoka et al. 2009), 
two stellar clusters located in one of the Milky Way spiral arms and 
thus characterised by a lower column density of absorbing material, 
consequently appearing stronger in the 1-2.5 keV range 
(visible in Fig. \ref{RGBhard} with orange 
colours). 

\subsection{Soft and hard GC emission}

Galactic centre radiation with energies above $\sim2-3$ keV can 
typically reach us and be detected (in green and blue in Fig.~\ref{RGBhard}). 
GC sources with a significant 
continuum component (e.g. power-law or Bremstrahlung), 
such as observed from most GC point sources, the GC stellar clusters (e.g. the Arches, 
the Quintuplet and the Central cluster) as well as some supernova 
remnants (such as SNR G0.9+0.1 and Sgr~A~East) appear with a bright 
light blue colour. 

Colour gradients confirm the presence of at least two components 
of the diffuse emission, each having a different spatial 
distribution (see Figs. \ref{RGBhard} and \ref{RGBhardNew}). 
One component dominates the emission in the soft and medium energy 
bands, thus appearing with a distinctively green colour. 
Its distribution appears to be very patchy, peaking typically at the position of
known supernovae remnants. 
Another, harder component appears with a dark blue colour (Fig. \ref{RGBhard}).
This harder emission is known to consist of at least two separate 
contributions. 
One, which is associated with intense high-ionisation Fe K lines, 
is smoothly distributed and peaks right at the GC; it is likely produced 
by faint point sources (Muno et al.\ 2004; Revnivtsev et al.\ 2009; Heard \& 
Warwick 2013a). 
The other, which is associated with neutral Fe K emission lines, has a patchy 
distribution peaking at the position of molecular cloud complexes; 
it is likely due to an ensemble of X-ray reflection nebulae (see Ponti et al. 2013 
for a review). 
We also note that the Galactic plane emission is dominated 
by dark blue colours (Fig. \ref{RGBhard}), while regions located at 
$b\gsimeq0.2^{\circ}$ and $b\lsimeq-0.35^{\circ}$ have a significantly greener colour. 
We address this in more detail in \S~\ref{SpecDec2}, \ref{NH} and \ref{Fermi}.

\section{Soft line emission}

Figure \ref{ImaEbands} shows the spectra of the diffuse emission from the 
regions marked in magenta in Fig. \ref{FC2}. 
The $\sim1.5$ to $\sim5$ keV band shows strong, narrow emission lines, 
the strongest of which are Si {\sc xiii}, S {\sc xv}, Ar {\sc xvii} and Ca {\sc xix}. 
This line emission, as well as the underlying continuum and the intra-line 
emission, are typically well fitted by a thermal model (e.g. {\sc apec} in 
{\sc Xspec}) with temperatures in the range $0.6-1.5$~keV (Kaneda et al. 1997; 
Tanaka et al. 2000; Muno et al. 2004; Nobukawa et al. 2010; Heard \& Warwick 2013b). 
At higher energies a power-law component with intense Fe~{\sc xxv} and 
Fe~{\sc xxvi} lines is also observed over the entire GC region. 
Additionally, neutral Fe~K$\alpha$ and K$\beta$ lines are also observed. 
The neutral Fe~K emission lines are associated with different processes, therefore 
they will be the focus of separate publications. 

\subsection{RGB images of soft emission lines}
\label{RGBsoft}

\begin{figure*}
\includegraphics[width=1.05\textwidth,angle=0]{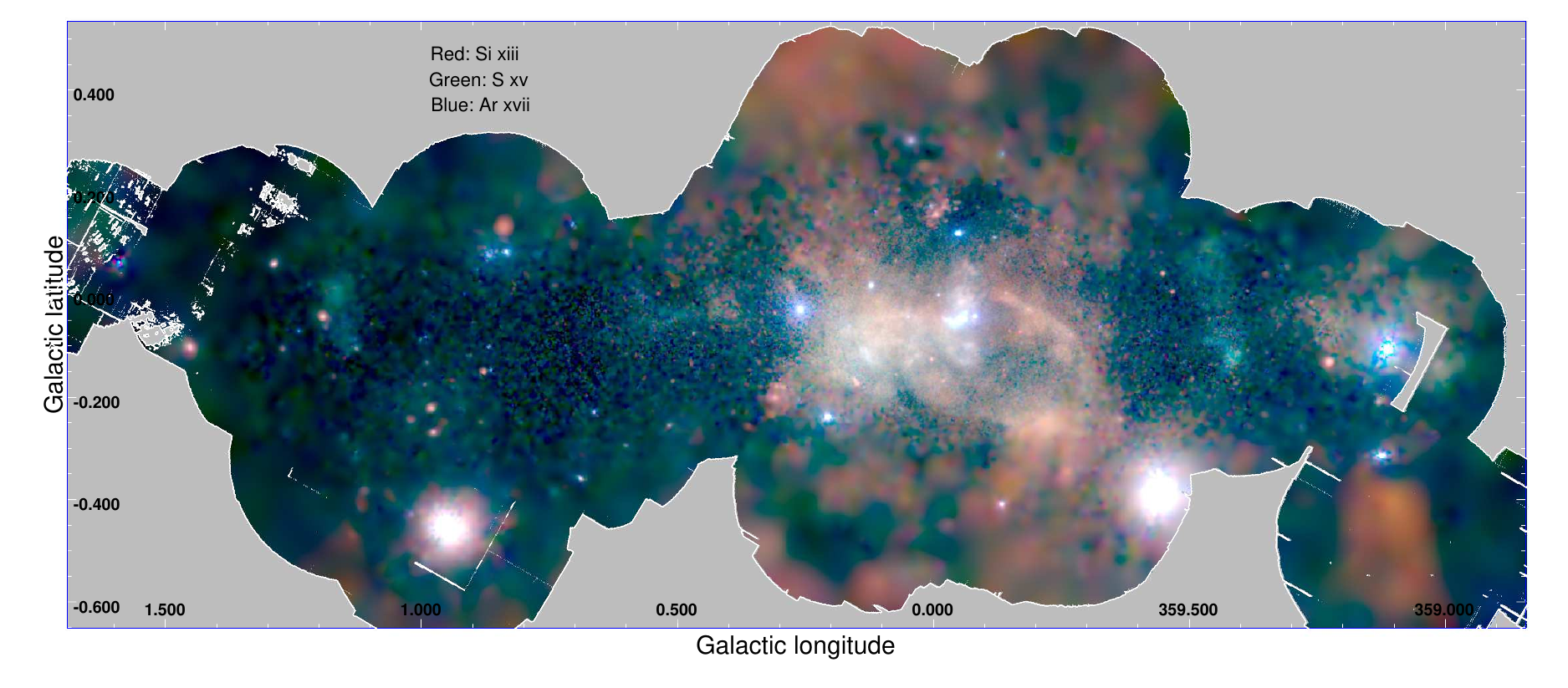}
\includegraphics[width=1.05\textwidth,angle=0]{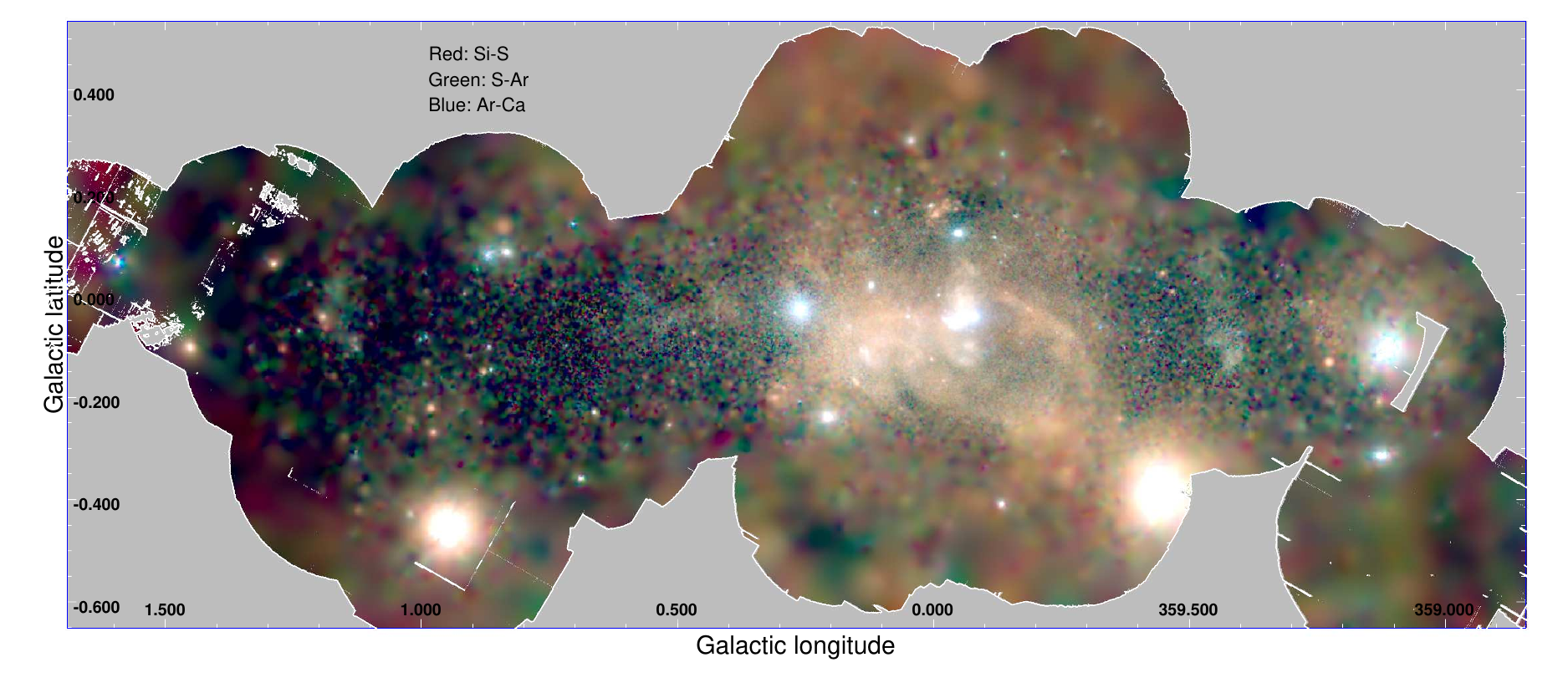}
\caption{{\it (Top panel)} Soft lines (continuum unsubtracted) RGB 
image of the CMZ. The Si {\sc xiii} line emission is shown in red, S {\sc xv} 
in green and Ar {\sc xvii}$+$ Ca {\sc xix} in blue. 
{\it (Bottom panel)} RGB image of the energy bands between soft 
emission lines. The Si-S band emission is shown in red (see Tab. 
\ref{Ebands} for a definition of the energy bands), S-Ar in green and 
Ar-Ca $+$ Blue-Ca in blue. The diffuse emission in this inter-line 
continuum is very similar to the soft emission line one, 
suggesting that the same process is producing both the lines 
and the majority of the soft X-ray continuum. 
The colour variations within the map are modulated primarily by abundance 
variations (top), temperature of the emitting plasma, continuum shape and 
absorption. }
\label{RGBSoftLines1}
\end{figure*}

\begin{figure*}
\includegraphics[width=1.05\textwidth,angle=0]{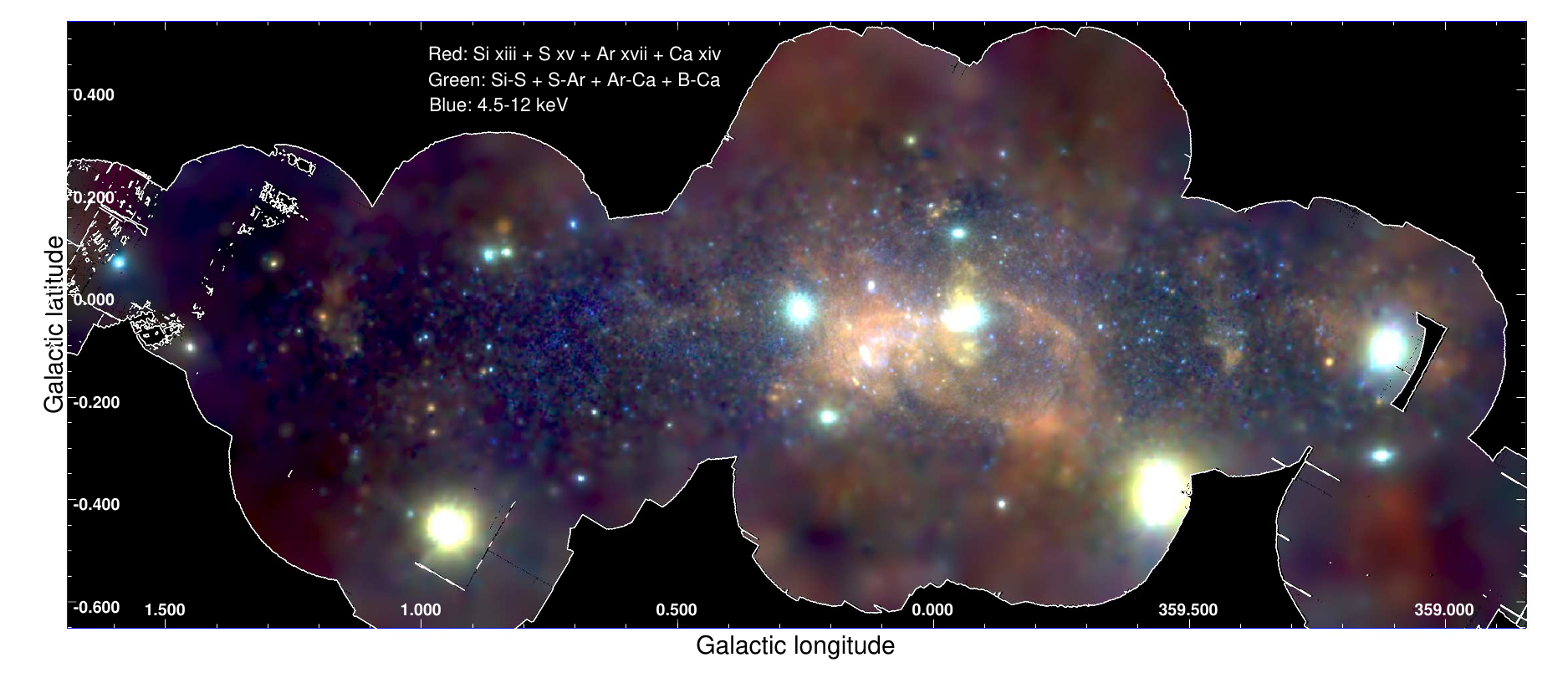}
\includegraphics[width=1.05\textwidth,angle=0]{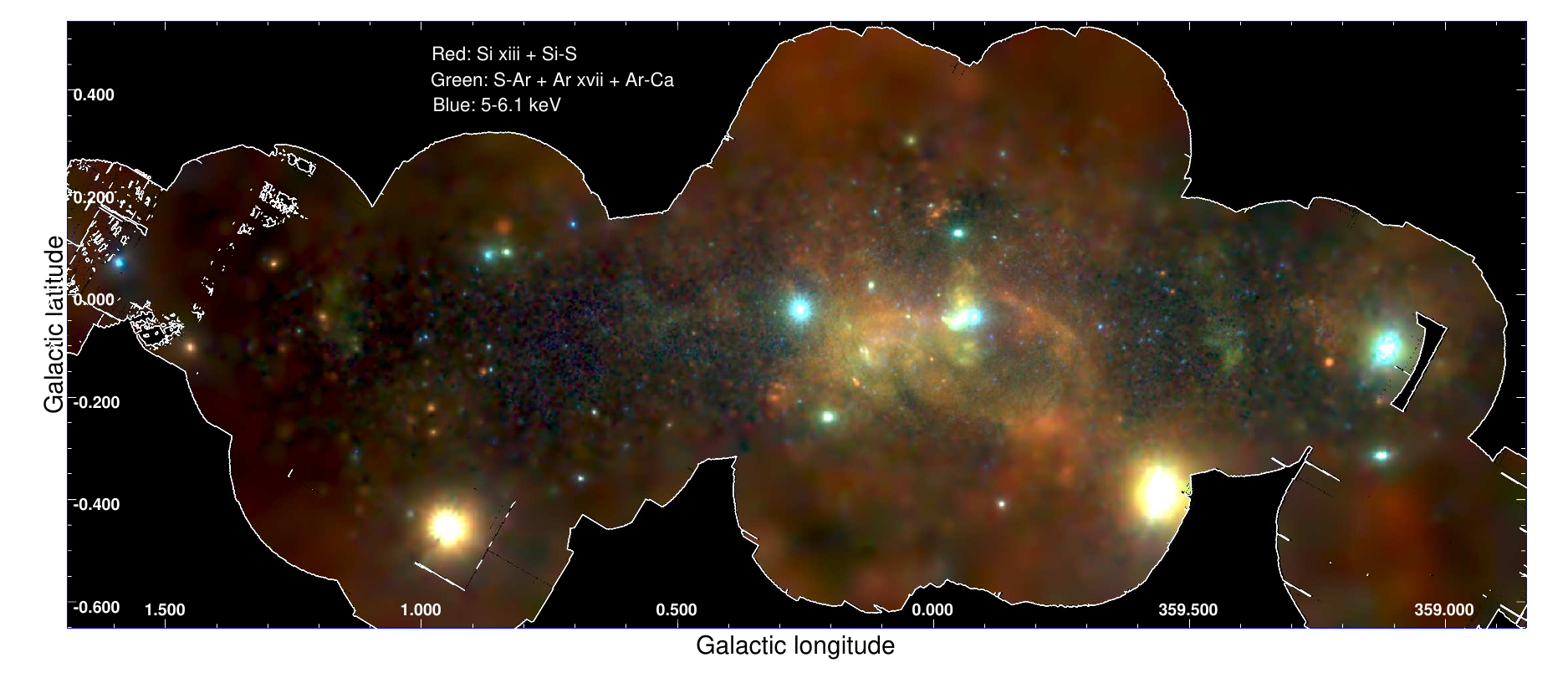}
\caption{{\it (Top panel)} RGB image composed of summed, 
continuum-subtracted line emission (Si {\sc xiii} $+$ S {\sc xv} 
$+$ Ar {\sc xvii} $+$ Ca {\sc xix}) in red, the sum 
of the interline continua (Si-S $+$ S-Ar $+$ Ar-Ca $+$ Blue-Ca) 
in green and the CFeK emission in blue (see Tab. \ref{Ebands}). 
{\it (Bottom panel)} RGB image, in red the Si {\sc xiii} $+$ Si-S emission, 
in green the S-Ar $+$ S {\sc xv} $+$ Ar-Ca and in blue the CFeK emission. }
\label{RGBSoftLines2}
\end{figure*}

The top and bottom panels of Fig. \ref{RGBSoftLines1}
show the line (continuum non-subtracted) RGB image and the inter-line 
continuum RGB image (see caption of Fig.~\ref{RGBSoftLines1} and 
Tab.~\ref{Ebands} for more details).
We note that the soft X-ray line image shows very strong colour 
gradients (less dramatic colour variations are observed in the continuum 
image). In particular, the sources DS1 (the core of Sgr~D), the western part of Sgr~B1 
(i.e., G0.52-0.046, G0.570-0.001), Sgr~C, as well as the Chimney above it, all have 
a distinctively green-blue colour, while G359.12-0.05, G359.10-0.5, 
G359.79-0.26, G359.73-0.35 and the entire G359.77-0.09 superbubble 
are characterised by orange-brown colours.  G0.1-0.1, the Radio Arc, 
the arched filaments (see Fig. of Lang et al. 2002), G0.224-0.032, 
and G0.40-0.02 are also characterised 
by red-brown colours, however here a gradation of white and green is also 
present (please refer to Tab.\ \ref{AtlasSCSNR} and \ref{AtlasNTF} and Figs.\ 
\ref{FC1} and \ref{FC2} for the positions of the regions listed here). 
The lobes of Sgr A appear with a whiter colour than the surroundings. 
In addition, we observe bright red-brown emission along two broad, linear ridges 
having relatively sharp edges to the northwest and northeast of \sgras. 
This latter feature is discussed in detail in section \ref{Lobes}.  

In spite of the fact that these images show different components (one being 
dominated by emission lines, the other by continuum emission), 
they are remarkably similar. No clear diffuse emission component 
is present in one and absent from the other image. This indicates 
that most of the diffuse soft X-ray continuum and line emission are, indeed, 
produced by the same process. 
In addition, the differences in the ratio between photons emitted in the 
lines and in the continuum can add valuable information for understanding 
the radiative mechanism.  In fact, such differences could be due, for example, 
to different cosmic abundances and/or variations in the relative contributions 
of various thermal and nonthermal radiation mechanisms. 
In order to better highlight these differences, we map the sum of the 
interline continua in the same image (see caption of Fig.~\ref{RGBSoftLines2} 
and Tab.~\ref{Ebands}). 
As expected, none of the point sources is a strong soft line emitter 
(they in fact appear brighter in the interline image). 
We also note that the intense soft X-ray emitting regions in the Galactic 
plane, such as Sgr~D, Sgr~B1, Sgr~C, the Chimney and G359.9-0.125 
are characterised by distinctively orange-red colours, 
indicating they are strong line emitters. 

For an alternative perspective, the bottom panel of Fig. \ref{RGBSoftLines2} 
shows the Si {\sc xiii} $+$ Si-S bands in red, S-Ar $+$ S {\sc xv} $+$ Ar-Ca in green, 
and CFeK in blue. These energy bands are chosen to 
highlight any energy dependence in the soft X-ray emission that 
could be due to column density variations of the obscuring matter 
or temperature fluctuations of the emitting gas. 
In fact, the softer energy bands (Si {\sc xiii} $+$ Si-S) will be more 
affected by absorption or low temperature plasma emission compared 
to the medium (S-Ar $+$ S {\sc xv} $+$ Ar-Ca) or high energy bands.
We defer the detailed 
discussion of the features present in these images to the 
discussion of the various physical components presented in \S~\ref{Disc} 
and subsections. 

\subsection{Continuum subtracted soft emission line maps and profiles} 
\label{ParSoftLines}

\begin{figure*}
\hbox{\hspace{-1.3cm}\includegraphics[width=1.15\textwidth,angle=0]{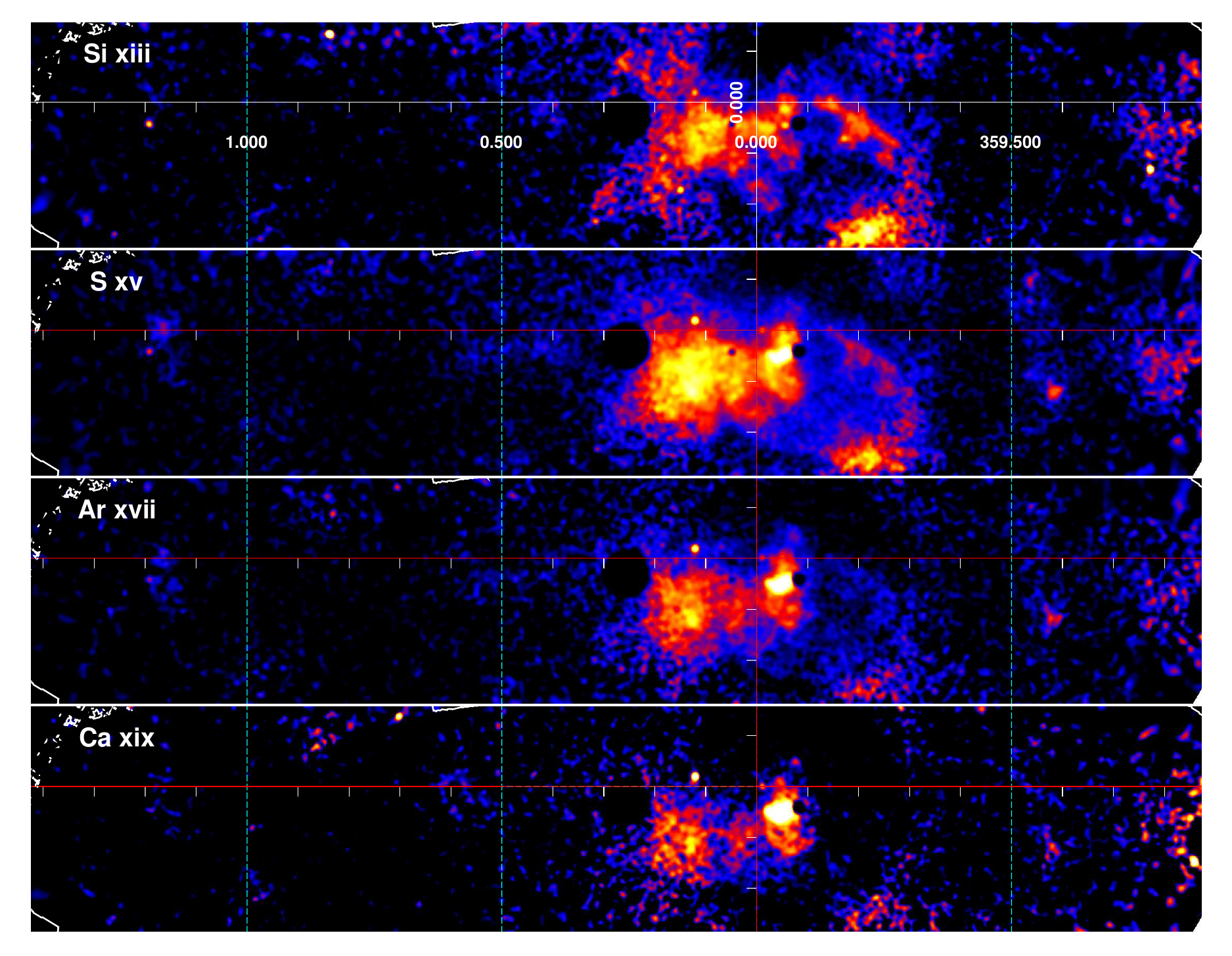}}
\caption{From top to bottom, continuum subtracted Si {\sc xiii}, 
S {\sc xv}, Ar {\sc xvii}, Ca {\sc xix} intensity maps of all the stacked 
\xmm\ observations of the CMZ. }
\label{SoftLines}
\end{figure*}
\begin{figure}
\includegraphics[width=0.5\textwidth,angle=0]{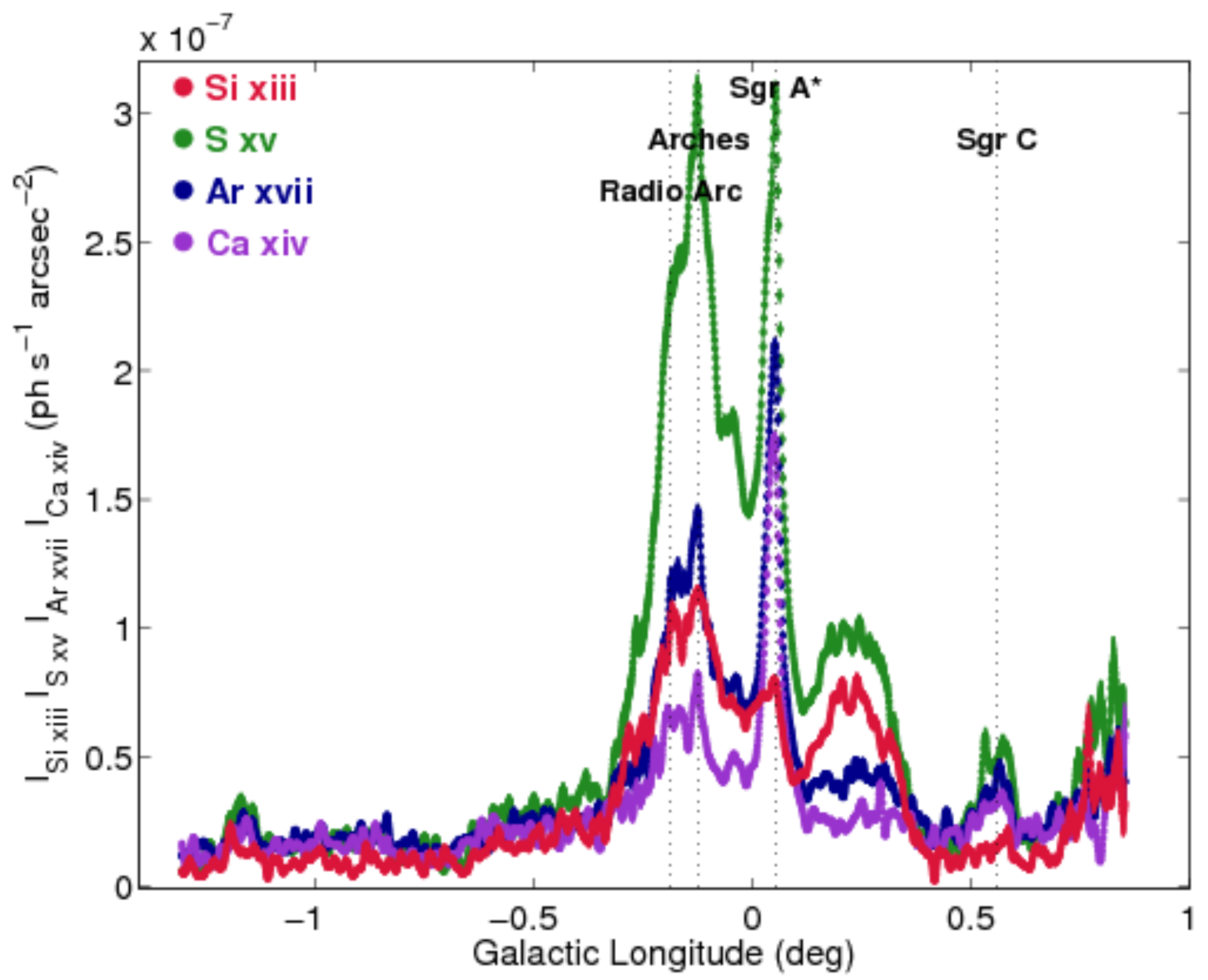}
\caption{Longitudinal intensity profiles of the Si {\sc xiii} (red), S {\sc xv} (green), 
Ar {\sc xvii} (blue) and Ca {\sc xix} (violet) emission lines, integrated over Galactic
latitude within the magenta rectangular region shown in Fig. \ref{SoftLines}. }
\label{ProfSoftLines}
\end{figure}

From top to bottom, the panels of Fig. \ref{SoftLines} 
show the continuum-subtracted Si {\sc xiii}, S {\sc xv}, Ar {\sc xvii}, Ca {\sc xix} 
intensity maps. Although the continuum subtraction procedure should naturally 
remove the emission from the line-free point sources, small fluctuations in the 
continuum subtraction, in the case of the brightest sources, sometimes leave
significant residuals. For this reason, we have masked out the brightest point 
sources in our computation of these maps. 

The different curves of Fig. \ref{ProfSoftLines} show the continuum-subtracted 
emission profiles (integrated over latitude from the magenta rectangle 
in Fig. \ref{SoftLines}) for the individual soft emission lines. 
The same four line profiles are compared in Fig.~\ref{ProfSoftLines} with 
similar profiles in which the contribution of specific bright structures has been removed.

\section{Spectral decomposition} 
\label{SpecDec2}

In order to better trace the relative contributions of the diffuse thermal 
(soft and hot) and non-thermal components, we have performed a 
simple component separation using a list of images depicting various 
energy bands. 
We use a total of 17 energy bands: 11 for the continuum\footnote{The 
eleven continuum energy bands used are: 1.0--1.5 keV; 1.5--1.8 keV; 
2.0--2.35 keV; 2.55--3.05 keV; 3.25--3.75 keV; 3.95--4.70 keV; 
4.70--5.40 keV; 5.40--6.30 keV;  6.50--6.60 keV;  6.80--7.80 keV and 
8.20--9.50 keV.} and 6 for the lines (tracing Si~{\sc xiii}, S~{\sc xv}, 
Ar~{\sc xvii}, Ca~{\sc xix}, Fe~K$\alpha$ and Fe~{\sc xxv}, see Tab. 
\ref{Ebands}).
This treatment of the data allows us to be more confident 
about the spectral decomposition, e.g. compared to single RGB maps, 
retaining most of the morphological information on sufficiently large 
scales (i.e. beyond few arcmin scales). 

The general assumption is that the emission at any position can be 
represented by the linear sum of three main components, namely
i) a soft plasma with a temperature of 1~keV (Kaneda et al. 1997; 
Bamba et al. 2002); ii) a hot plasma of temperature 6.5 keV;
and iii) a non-thermal component modeled by an absorbed power-law 
plus a neutral, narrow iron line (with 1~keV equivalent width), that are 
subject to an additional absorbing column of N$_H = 10^{23}$ cm$^{-2}$.
All three components are also absorbed by gas in front of the GC region 
and both thermal plasmas are modelled using the {\sc ape} model in 
{\sc XSpec}. The resulting model is therefore {\sc phabs (apec + apec + 
phabs (powerlaw + Gauss))}) and has only three free parameters:
the relative normalizations of the three components. 
The hot plasma component represents the emission associated with 
faint unresolved point sources, whose cumulative spectrum is well 
described by a thermal spectrum (Revnivtsev et al. 2009) plus a possibly
truly diffuse hot plasma component (Koyama et al. 2007). 
The spectral index of the non-thermal component is assumed to be
$\Gamma = 2$, consistent with the values measured through the combined 
spectral fits of \xmm\ spectra with higher energy data (e.g., \integral\ and/or 
\nustar; Terrier et al. 2010; Mori et al. 2015; Zhang et al. 2015).

The strongest assumptions in this approach are that the emission can be 
represented everywhere with these three components. This obviously fails on 
bright point sources or on regions where the emission is much hotter (e.g. Sgr East). 
For the soft components the even stronger assumption is that absorption to 
the GC is assumed to be uniform over the CMZ at a value of 
N$_H = 6 \times 10^{22}$ cm$^{?2}$ (Sakano et al. 2002; Ryu et al. 2009), 
putting aside absorption in the GC region itself. Clear column density modulations 
are observed towards different lines of sight (e.g. Ryu et al. 2009; Ryu et al. 2013). 
We tested significantly different column densities (up to N$_H = 1.5 \times 
10^{23}$ cm$^{-2}$ characteristic of several GC sources; see e.g. Baganoff et al. 2003; 
Rea et al. 2013; Ponti et al. 2015). We found that if the soft plasma normalization 
is significantly modified, the overall morphology is consistent.  
We tested various values of the other parameters (spectral index or temperatures) and 
did not find strong effects on the soft plasma morphology or normalization. 

We first produced counts, exposure and background maps for each 
observation and each instrument. Background was obtained from 
cal-closed datasets distributed in the 
{\sc esas}\footnote{http://xmm2.esac.esa.int/external/xmm\_sw\_cal/background/epic\_esas.shtml} 
calibration database. 
For each observation and instrument, an average {\sc rmf} is computed as 
well as the un-vignetted {\sc arf}. For each instrument, individual observation 
images were reprojected using the final image astrometry and then 
combined to compose a mosaic. Average {\sc arf} and {\sc rmf} for each instrument 
were obtained with the 
{\sc ftools}\footnote{https://heasarc.gsfc.nasa.gov/ftools/ftools\_menu.html} 
routines {\sc addrmf} and {\sc addarf}. 

For each pixel of the final maps, we fit the measured numbers of counts 
in all the energy bands and instruments with a model consisting of the 
three aforementioned components as well as the background events 
number and the Out-of-Time (OoT) events for the EPIC-pn camera. 
The free parameters are the normalization of each individual component. 
We apply Cash statistics (Cash 1979) to take into account the low statistics 
in each pixel.
This analysis allows us to perform a rough spectral decomposition, 
better separating the spectral emission components, although retaining 
the maximum spatial resolution.

\begin{figure*}
\hbox{\includegraphics[width=1\textwidth,angle=0]{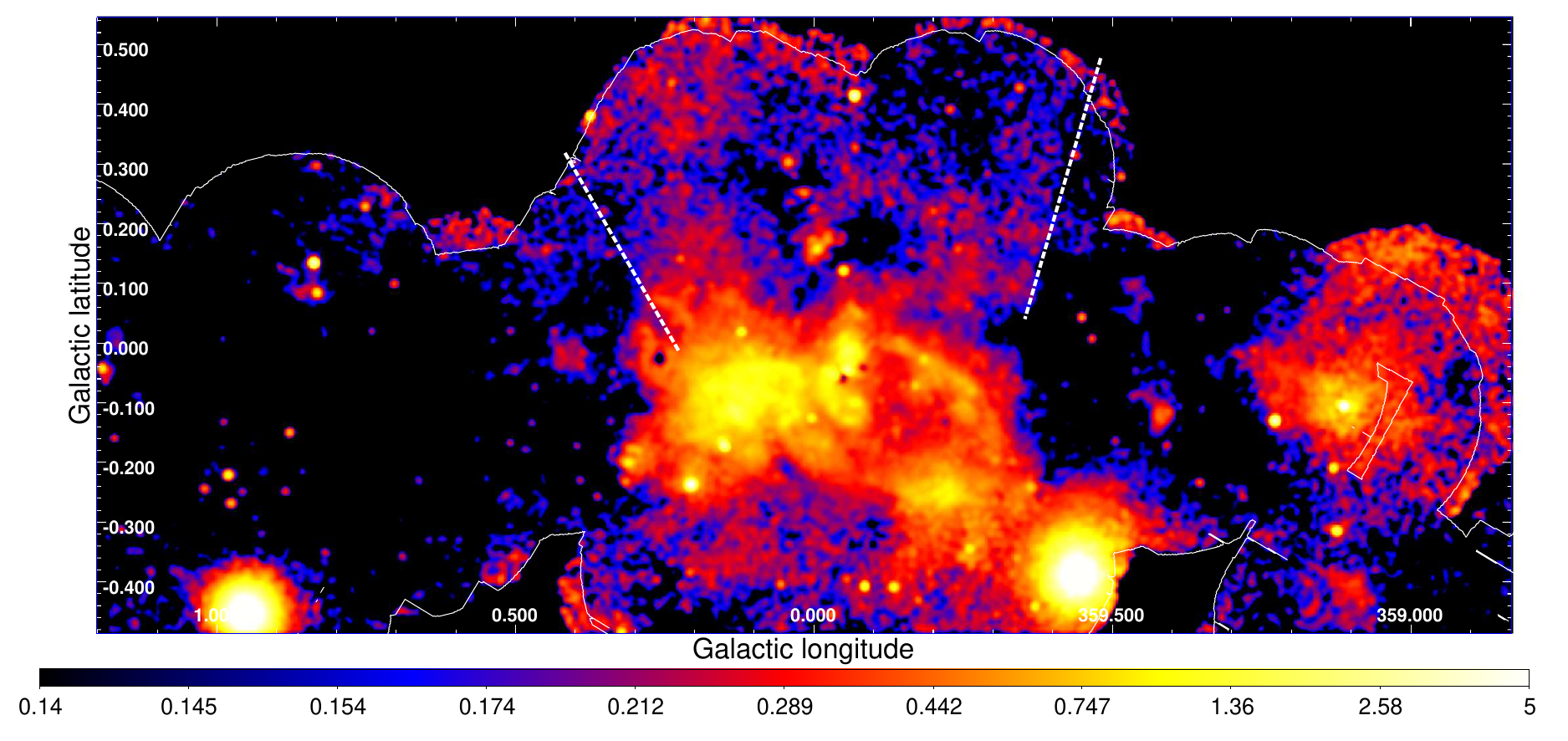}}
\caption{Map of the normalisation of the soft thermal gas component 
(in units of $10^{-4}$ times the {\sc apec} normalisation). The white lines 
indicate the extent of the survey having more than 7.2~ks exposure. 
The white dashed lines show the position of the two sharp edges in the 
distribution of the high latitude plasma. Some bright point sources (i.e., 1E~1743.1-2843, 
AX~J1745.6-2901, 1E~1740.7-2942, GRS~1741.9-2853) have been removed, 
thereby producing artificial holes in the maps at their respective locations. }
\label{Terrier}
\end{figure*}
Figure \ref{Terrier} presents the map of the normalisation (in units of $10^{-4}$ 
times the {\sc apec} normalisation) of the soft thermal emission component. 
The normalisation of the soft thermal component has a distribution similar 
to the one traced by the soft lines and the continuum (Fig.~\ref{RGBhard}, 
\ref{RGBSoftLines1}, \ref{RGBSoftLines2} and \ref{SoftLines}). 
Enhanced high-latitude soft plasma emission is observed. 
The white dashed lines show the position of two sharp edges in the distribution
of this high latitude emission (see also Fig. \ref{RGBhard}, \ref{RGBSoftLines1}
and \ref{RGBSoftLines2}). The white solid line shows the edge of the region having 
more than 7.2~ks of exposure (see Fig.\ \ref{exposure}). 

\section{An atlas of diffuse X-ray emitting features}

The patchy and non-uniform distribution of the diffuse emission makes 
the recognition of the shape, the border and connection of the different 
structures and components difficult. Occasionally, different works report 
the same X-ray feature with different names and shapes and, in extreme 
cases, the same X-ray emitting feature is associated with different larger 
scale complexes. 

In Tab.~\ref{AtlasSCSNR} and \ref{AtlasNTF} we report all the 
new X-ray features discussed in this paper, plus many GC features 
presented in previous works. 
The main purpose of these tables is to provide a first step towards the building of an atlas 
of  diffuse X-ray emitting GC features. The table is available online at: 
{\sc www.mpe.mpg.de/HEG/GC/AtlasGCdiffuseX-ray} and will be updated, 
should the authors be notified of missing extended features. 
This exercise is clearly prone to incompleteness and deficiencies, however 
we believe this might help in providing a clearer and more systematic picture 
of the diffuse X-ray emission from the GC region. 
The spatial location and size of all these features is shown in the finding 
charts in Fig. \ref{FCTSC} and \ref{FCBFSNRHII}. 

\begin{table*} 
\begin{center}
\footnotesize
\begin{tabular}{ l l r c } 
\hline
\multicolumn{4}{c}{{\bf BRIGHT AND TRANSIENT POINT SOURCES}} \\
\hline
Source name & Coordinates\ddag & Flux\dag & References \\
\hline
\multicolumn{4}{l}{{\bf within the 2012 CMZ scan}} \\
1E~1743.1-2843                   & 0.2608,-0.0287     & $110$    & 90,92,93 \\ 
XMMU~J174505.3-291445   & 359.6756,-0.0634 & $14$      & 94 \\
XMMU~J174457-2850.3       & 0.0076,-0.1743     & $0.3$     & 90,95 \\
GRS~1741.9-2853                & 359.9528,+0.1202 & $<0.02$ & 90,59,97 \\
AX~J1745.6-2901                 & 359.9203,-0.0420  & $<0.1$   & 59,90,91,124 \\
CXOGC~J174540.0-290031 & 359.9435,-0.0465  & $<5$      & 98,99 \\
SGR~J1745-2900                  & 359.9441,-0.0468 & $<5$      & 101,102,103,104 \\
XMMU~J174554.4-285456    & 0.0506,-0.0429     & $<0.2$   & 98 \\
SAX~J1747.7-2853                & 0.2073,-0.2385     & $<0.2$   & 90,105,106,107 \\
CXOCG~J174540.0-290005  &359.9497,-0.04269& $<5$      & 100 \\ 
\hline
\multicolumn{4}{l}{{\bf within the total GC scan}} \\
1E~1740.7-2942                    & 359.1160,-0.1057 & & 111,112,113 \\
1A~1742-294                         & 359.5590,-0.3882 & & 90,108,109 \\
IGR~J17497-2821                  & 0.9532,-0.4528    & & 90,114,115 \\
GRO~J1744-28                      & 0.0445,+0.3015    & & 90 \\
XMMU~J174654.1-291542    & 359.8675,-0.4086 & & 90 \\
XMMU~J174554.4-285456    & 359.1268,-0.3143 & & 84,85,86 \\
SLX~1744-299                       & 359.2961,-0.8892 & & 37,59,87,88,89 \\
SLX~1744-300                       & 359.2565,-0.9111 & & 37,59,87,88,89 \\
\hline
\end{tabular}
\caption{List of bright and transient point sources during 
the 2012 \xmm\ scan as well as bright point sources observed 
in all scans of the region (see Fig.\ \ref{FC1}). 
To avoid exessive crowding around \sgras, CXOCG~J174540.0-290005 
and SGR~J1745-2900 are not shown. 
\dag Fluxes are given in units of $10^{-12}$ erg cm$^{-2}$ s$^{-1}$ 
and correspond to the mean flux observed during the 2012 \xmm\ 
scan of the CMZ. \ddag Coordinates are in Galactic format. 
References:  
(1) Wang et al. 2006a;
(2) Yusef-Zadeh et al. 2002;
(3) Capelli et al. 2011;
(4) Tatischeff et al. 2012;
(5) Sakano et al. 2003;
(6) Habibi et al. 2013;
(7) Habibi et al. 2014;
(8) Krivonos et al. 2014;
(9) Clavel et al. 2014;
(10) Dutra et al. 2003;
(11) Law et al. 2004;
(12) Fukuoka et al. 2009;
(13) Wang et al. 2002a;
(14) Tsuru et al. 2009;
(15) Mori et al. 2008;
(16) Mori et al. 2009;
(17) Heard \& Warwick 2013a;
(18) Maeda et al. 2002;
(19) Park et al. 2005;
(20) Koyama et al. 2007a;
(21) Kassim \& Frail 1996;
(22) Nobukawa et al. 2008;
(23) Senda et al. 2002;
(24) Renaud et al. 2006;
(25) Mereghetti et al. 1998;
(26) Gaensler et al. 2001;
(27) Porquet et al. 2003a;
(28) Aharonian et al. 2005;
(29) Dubner et al. 2008;
(30) Nobukawa et al. 2009;
(31) Sawada et al. 2009;
(32) Morris et al. 2003;
(33) Morris et al. 2004;
(34) Markoff et al. 2010;
(35) Zhang et al. 2014;
(36) Nynka et al. 2013;
(37) Gaensler et al. 2004;
(38) Pedlar et al. 1989;
(39) Cotera et al. 1996;
(40) Figer et al. 1999;
(41) Johnson et al. 2009;
(42) Lu et al. 2008;
(43) Lu et al. 2003;
(44) Yusef-Zadeh et al. 2005;
(45) Baganoff et al. 2003;
(46) Ho et al. 1985;
(47) Bamba et al. 2002;
(48) LaRosa et al. 2000;
(49) Morris \& Yusef-Zadeh 1985;
(50) Lang et al. 1999;
(51) Anantharamaiah et al. 1991;
(52) Yusef-Zadeh \& Morris 1987a;
(53) Yusef-Zadeh \& Morris 1987b;
(54) Yusef-Zadeh \& Morris 1987c;
(55) Muno et al. 2008;
(56) Uchida et al. 1992;
(57) Predehl \& Kulkarni 1995;
(58) Senda et al. 2003;
(59) Sakano et al. 2002;
(60) Coil et al. 2000;
(61) Murakami 2002;
(62) Yusef-Zadeh et al. 2007;
(63) Dutra \& Bica 2000;
(64) Zoglauer et al.\ 2014;
(65) Koyama et al. 2007b;
(66) Nakashima et al. 2010;
(67) Downes \& Maxwell 1966;
(68) Tanaka et al. 2009;
(69) Tanaka et al. 2007;
(70) Wang et al. 2006b;
(71) Wang et al. 2002b;
(72) Phillips \& Marquez-Lugo 2010;
(73) Hewitt et al. 2008;
(74) Reich \& Fuerst 1984;
(75) Gray 1994;
(76) Roy \& Bhatnagar 2006;
(77) Marquez-Lugo \& Phillips 2010;
(78) Borkowski et al. 2013;
(79) Yamauchi et al. 2014;
(80) Inui et al. 2009;
(81) Green 2014;
(82) Yusef-Zadeh et al. 2004;
(83) Nord et al. 2004;
(84) Uchiyama et al. 2011;
(85) Heinke et al. 2009;
(86) Muno et al. 2006;
(87) Mori et al. 2005;
(88) Skinner et al. 1990;
(89) Pavlinski et al. 1994;
(90) Degenaar et al. 2012;
(91) Ponti et al. 2014;
(92) Porquet et al. 2003b;
(93) Del Santo et al. 2006;
(94) Soldi et al. 2014;
(95) Sakano et al. 2005;
(97) Trap et al. 2009;
(98) Porquet et al. 2005a;
(99) Muno et al. 2005b;
(100) Kock et al. 2014;
(101) Degenaar et al. 2013;
(102) Dwelly \& Ponti 2013;
(103) Rea et al. 2013;
(104) Kaspi et al. 2014;
(105) Wijnands et al. 2002;
(106) Natalucci et al. 2004;
(107) Werner et al. 2004;
(108) Belanger et al. 2006;
(109) Kuulkers et al. 2007;
(110) Piraino et al. 2012;
(111) Castro et al. 2013;
(112) Reynolds \& Miller 2010;
(113) Natalucci et al. 2014;
(114) Soldi et al. 2006;
(115) Paizis et al. 2009 
(116) Lu et al. 2013;
(117) Do et al. 2013; 
(118) Yelda et al. 2014;
(119) Bamba et al. 2000;
(120) Bamba et al. 2009;
(121) Ohnishi et al. 2011;
(122) Zhao et al. 2013;
(123) Hales et al. 2009;
(124) Ponti et al. 2015. }
\label{TabPS}
\end{center}
\end{table*} 

\begin{table*} 
\begin{center}
\scriptsize
\begin{tabular}{ l l l c c }
\hline
\multicolumn{5}{c}{{\bf ATLAS OF DIFFUSE X-RAY EMITTING FEATURES}} \\
\hline
Name & Other name or & Coordinates & Size & References \\
   & associated features & (l, b)              & arcmin & \\
\hline
\multicolumn{5}{l}{{\bf STAR CLUSTERS:}} \\
Central cluster & & 359.9442, -0.046 & $0.33$ & 45,116,117,118 \\ 
Quintuplet & & 0.1604, -0.0591 & $0.5$ & 1,63,11 \\
Arches & G0.12+0.02 & 0.1217, 0.0188 & $0.7$ & 1,2,3,4,5,6,7,8,9,39,40,11 \\
Sh2-10$\flat\dag$ & DB-6   & 0.3072,-0.2000 & $1.92$ & 10,11,12,63,11 \\
Sh2-17$\flat\dag$ & DB-58 & 0.0013, 0.1588 & $1.65$ & 13,63,11 \\
DB-05$\flat\dag$ & G0.33-0.18 & 0.31 -0.19 & $0.4$ & 22,63,11 \\ 
\hline
\multicolumn{5}{l}{{\bf SNR and Super-bubbles candidates:}} \\
G359.0-0.9$\dag\dag\dag$& G358.5-0.9 - G359.1-0.9 & 359.03,-0.96 & $26\times20$ & X-R 48,51,75,76,81,119,120 \\
G359.07-0.02& G359.0-0.0             & 359.07,-0.02 & $22\times10$ & R 14,48,51,66 \\ 
                                           & G359.12-0.05         & 359.12,-0.05 & $24\times16$ & X 66 \\ 
G359.10-0.5$\dag\dag\dag$&                                & 359.10,-0.51 & $22\times22$ & X-R 37,48,51,56,74,75,81,120,121 \\        
G359.41-0.12                     &                                & 359.41,-0.12 & $3.5\times5.0$ & X 14 \\
Chimney$\flat\dag$           &                                 & 359.46,+0.04 & $6.8\times2.3$ & X 14 \\
G359.73-0.35$\dag\ddag$   &                                 & 359.73,-0.35 & $4$ & X 58 \\
G359.77-0.09   & Superbubble & 359.77,-0.09 & $20\times16$   & X 15,16,17,58 \\
                            & G359.9-0.125 & 359.84,-0.14 & $15\times3$    & X 15,16,17,58 \\
                       & G359.79-0.26$\natural$ & 359.79,-0.26 & $8\times5.2$ & X 15,16,17,58 \\
                       & G0.0-0.16$\dag\natural$      &     0.00,-0.16 & & X This work \\  
20~pc lobes & & 359.94, -0.04 & $5.88$ & X-R 32,33,34,17 \\
G359.92-0.09\ddag & Parachute - G359.93-0.07 & 359.93,-0.09 & $1$ & R 35,38,43,47,58,60,61 \\
Sgr A East & G0.0+0.0 & 359.963, -0.053 & $3.2\times2.5$ & X-R 5,18,19,20,48,75,81 \\
G0.1-0.1 & Arc Bubble & 0.109,-0.108 & $13.6\times11$ & X This work \\
               & G0.13,-0.11$\flat$ & 0.13,-0.12 & $3\times3$ & X 17 \\
G0.224-0.032&        & 0.224,-0.032 & $2.3\times4.6$ & X This work \\
G0.30+0.04 & G0.3+0.0    & 0.34,+0.045 & $14\times8.8$ & R 21,48,51,81,82 \\
                    & G0.34+0.05 \\
                    & G0.33+0.04 \\
G0.42-0.04 & Suzaku J1746.4-2835.4 & 0.40,-0.02 & $4.7\times7.4$ & X 22 \\
                   & G0.40-0.02 \\
G0.52-0.046&          & 0.519,-0.046$\diamond$ & $2.4\times5.1$ & This work \\
G0.57-0.001&          & 0.57,-0.001   & $1.5\times2.9$ & This work \\
G0.57-0.018\dag &CXO~J174702.6-282733 & 0.570,-0.018 & $0.2$ & X 23,24,58,59,68,80 \\
G0.61+0.01\dag & Suzaku~J1747.0-2824.5 & 0.61,+0.01 & $2.2\times4.8$ & X 22,65,79 \\
G0.9+01$\heartsuit$ & SNR~0.9+0.1 & 0.867,+0.073 & $7.6\times7.2$ & R 25,26,27,28,29,48,75,81,82 \\
DS1                                       & G1.2-0.0     & 1.17,+0.00 & $3.4\times6.9$ & X 31 \\ 
Sgr~D~SNR                           & G1.02-0.18 & 1.02,-0.17 & $10\times8.0$ & R 30,31,48,51,75,77,81,82 \\
                                              & G1.05-0.15 \\
                                              & G1.05-0.1 \\
                                              & G1.0-0.1 \\
G1.4-0.1   & & 1.4,-0.10   & $10\times10$ & R 73,81,82 \\
\hline
\end{tabular}
\caption{Atlas of diffuse X-ray emitting features. 
The first two columns in the table indicate the name primarily used in this work 
to refer to the feature as well as the other names used in the previous literature. 
The third and fourth columns show the coordinates of each feature as well as 
its approximate projected size. Finally, the fifth column provides 
references to selected works discussing the feature. 
For convenience, we report in Tab.~\ref{TabPS} all the references 
ordered according to the numbering used in this table. 
The other names column shows the different designations used in previous 
literature. In the case of bubbles, these features are not necessarily referring 
to the same structure but to features forming the bubble candidate. 
\dag Possibly due to a thermal filament. \ddag The interpretation as a SNR 
is probably obsolete. $\dag\ddag$ Most probably a foreground feature. 
$\natural$ This feature appears to be part of the superbubble G359.77-0.09. 
$\dag\natural$ New extended X-ray feature, possibly part of the superbubble 
G359.77-0.09.
$\flat$ This feature appears to be part of the Arc bubble. 
$\diamond$ Possibly connected to G0.61+0.01. 
$\heartsuit$ X-ray emission primarily non-thermal, therefore it appears also in the next table. 
$\dag\dag$ New extended X-ray feature, possibly part of the superbubble G359.77-0.09. 
$\flat\dag$ The low X-ray absorption towards these star clusters indicate that they are 
located in front of the GC region.
$\dag\dag\dag$ Because of the low X-ray absorption column density 
($N_{H}\sim2\times10^{22}$~cm$^{-2}$) this is most probably a foreground source 
(Bamba et al. 2000; 2009). $\flat\dag$ The Chimney is most probably either part of 
a large scale structure (see \S 8.7) or an outflow from G359.41-0.12 (Tsuru et al. 2009), 
therefore most probably it is not a separate SNR. }
\label{AtlasSCSNR}
\end{center}
\end{table*}

\begin{table*} 
\begin{center}
\scriptsize
\begin{tabular}{ l l l c c }
\hline
\multicolumn{5}{c}{{\bf ATLAS OF DIFFUSE X-RAY EMITTING FEATURES}} \\
\hline
Name & Other name or & Coordinates & Size & References \\
     & associate features & (l, b)              & arcsec & \\
\hline
\multicolumn{5}{l}{{\bf Radio and X-ray filaments and PWN candidates:}} \\
Snake & G359.15-0.2   & 359.15,-0.17 & $312\times54$ & R 48 \\
G539.40-0.08 &            & 359.40,-0.08 & $27.5\times5.1$ & X 41 \\
G359.43-0.14 &            & 359.43,-0.14 & $21.4\times3.9$ & X 41 \\
Sgr~C Thread &           & 359.45,-0.01 & $500\times42$ & R 48,51 \\
Ripple filament & G359.54+0.18 & 359.548,+0.177 & $320\times55$ & R 43,44,48,51 \\
G359.55+0.16 & X-ray thread     & 359.55,+0.16     & $56.1\times8.0$ & X 13,41,42,43,79 \\
                        & Suzaku J174400-2913 \\
Crescent & G359.79+0.17 & 359.791,+0.16 & $300\times74$ & R 63,50,51 \\
                        & Curved filament \\
Pelican    & G359.85+0.47 & 359.859,+0.426 & $300\times54$ & R 48,50,51 \\
Cane               & G359.87+0.44 & 359.87,+0.44 & $420\times50$ & R 48 \\
                        & G359.85+0.39 \\
Sgr~A-E & G359.889-0.081- wisp & 359.889,-0.081 & $20\times5$ & R X 5,35,41,42,43,44,50,55 \\
                       & XMM J174540-2904.5 \\
                       & G359.89-0.08          \\
G359.897-0.023 & & 359.897,-0.023 & $6.4\times4$ & X 55 \\
G359.899-0.065 & Sgr~A-F & 359.899,-0.065 & $6.5\times2.5$ & X 42,44,55 \\
                       & G359.90-0.06        \\
G359.904-0.047 & & 359.904,-0.047 & $6.5\times3$ & X 55 \\
G359.915-0.061 & & 359.915,-0.061 & $7\times2$ & X 55 \\
G359.91-1.03 & & 359.919,-1.033 & $138\times36$ & R 48 \\
G359.921-0.030 &F7& 359.921,-0.030 & $7.5\times3$ & X 42,55 \\
G359.921-0.052 & & 359.921,-0.052 & $5.5\times2$ & X 55 \\
The Mouse & G359.23-0.82 & 359.30-0.82 & $156\times108$ & PWN F 37,48,57,123  \\
G359.925-0.051 & & 359.925,-0.051 & $8\times2.2$ & X 55 \\
G359.933-0.037 &F2& 359.934,-0.0372 & $12\times3$ & X 41,42,55 \\
G359.933-0.039 &F1& 359.933,-0.039 & $5\times2$ & X 42,55 \\
G359.941-0.029 & & 359.941,-0.029 & $6\times2$ & X 41,55 Stellar wind \\
G359.942-0.045 & & 359.942,-0.045 & $5\times3$ & X 55 \\
G359.944-0.052 & & 359.944,-0.052 & $9\times1.5$ & X 41,55 \\
G359.945-0.044 & & 359.945,-0.044 & $6\times2.5$ & X 41,1,42,55 PWN \\
G359.95-0.04   & & 359.950,-0.043 & $10\times4$ & X 55,70 PWN \\
G359.956-0.052 & & 359.956,-0.052 & $4\times2.5$ & X 55 \\
G359.959-0.027 &F5& 359.959,-0.027 & $9\times3$ & X 41,42,55 \\
Southern thread & G359.96+0.09 & 359.96,+0.11 & $500\times40$ & R 48,50,51 \\
                        & 359.96+0.09 \\
G359.962-0.062 & & 359.962,-0.062 & $5.5\times3.5$ & X 55 \\
G359.964-0.053 &F3& 359.964,-0.053 & $16\times3.5$ & X 41,42,45,55 PWN \\
G359.965-0.056 &F4& 359.965,-0.056 & $9\times3$ & X 42,55 \\
G359.969-0.033 & & 359.969,-0.033 & $5\times2$ & X 55 \\
G359.970-0.009 &F8& 359.970,-0.009 & $10\times2.5$ & X 41,42,55 PWN \\
G359.971-0.038 &F6& 359.971,-0.038 & $16\times8$ & X 41,42,55 PWN \\
G359.974-0.000 &F9& 359.974,-0.000 & $7\times2$ & X 42 \\
G359.977-0.076 & & 359.977,-0.076 & $6\times4$ & X 55 \\
Cannonball & J174545.5-285829 & 359.983,-0.0459 & $30\times15$ & X-R PWN 36,122 \\
G359.983-0.040 & & 359.983,-0.040 & $6.5\times4.5$ & X 42,55 \\
G359.98-0.11     & & 359.979,-0.110 & Streak & R 50  \\
G0.007-0.014 & G0.008-0.015 & 0.008,-0.015 & $11\times3.5$ & X 41,55 \\
G0.014-0.054 & & 0.014,-0.054 & $18\times14$ & X 55 \\
G0.017-0.044 & MC2 & 0.017,-0.044 & $15\times4$ & X 41,42 FeKa \\
G0.02+0.04    & & 0.0219,+0.044 & Streak & R 50 \\
G0.021-0.051 & & 0.021,-0.051 & $15\times12$ & X 55 \\
G0.029-0.08   & & 0.029,-0.08 & $29\times18$ & X 55 \\
G0.032-0.056 & G0.029-0.06 - F10 & 0.0324,-0.0554 & $35\times6$ & FeKa 41,42,55 PWN \\
                   & G0.03-0.06  \\
G0.039-0.077 & & 0.039,-0.077 & $22\times15$ & X 55 \\ 
G0.062+0.010 & G0.06+0.06 & 0.062,+0.010 & $40\times25$ & R 50,55 \\
Northern thread & G0.09+0.17 & 0.09,+0.17 & $714\times48$ & R 48,49,50,51 \\ 
                    & G0.08+0.15 \\
G0.097-0.131 & & 0.097,-0.131 & $70\times50$ & X 55 \\ 
Radio Arc & GCRA & 0.167,-0.07 & $1690\times145$ & R 38,48,49,50,51\\
           & G0.16-0.15 \\
G0.116-0.111 & & 0.116,-0.111 & $50\times40$ & X 55 \\
G0.13-0.11 & & 0.13,-0.11 & $55\times12$ & 71,17,41,42 PWN \\
G0.15-0.07 & Steep spectrum of Radio Arc & 0.138,-0.077 & & R 50 \\ 
XMM~J0.173-0.413 & G0.17-0.42 & 0.173,-0.413 & $180\times18$ & X This work\\
                               &  S5               & 0.17,-0.42     & $912\time100$ R 82 \\
G0.223-0.012 & & 0.223,-0.012 & $50\times20$ & X 41 \\
G0.57-0.018\dag &CXO~J174702.6-282733 & 0.57,-0.0180 & 0.33 & X 23,24,58,59,68,79,80 \\
G0.61+0.01\dag  & & 0.61,+0.01 & $132\times288$ & X 22,65,79 \\
G0.9+0.1             & & 0.9,+0.1     &  & PWN \\
\hline
\end{tabular}
\caption{Atlas of diffuse X-ray emitting features. This table 
has the same structure as Tab.~\ref{AtlasSCSNR}.  
Because of possible misplacements between the peak emission of the 
radio and X-ray counterparts of filaments, SNR, PWN and other diffuse 
structures (generally related to the different ages of the population of 
electrons traced at radio and X-ray bands), when available we give the 
best X-ray position (following the preference: \chandra, \xmm, \suzaku), 
otherwise we state the radio position. We cite the literature results 
separating thes between the X-ray (X) from the radio (R) detections.
\dag Possibly part of a young supernova remnant.  
For convenience, we report in Tab.~\ref{TabPS} all the references 
ordered according to the numbering used in this table. }
\label{AtlasNTF}
\end{center}
\end{table*}

\section{The foreground column density}
\label{NH}

\begin{table} 
\begin{center}
\begin{tabular}{ c c c c c}
\hline
$N_H$&$F_{2-4.5}/F0$&$F_{4.5-10}/F0$\\
(10$^{22}$)&\\
\hline
0.01 & 1           &           1            \\
     1 & 0.776    &     0.968	     \\
     3 & 0.488    &     0.912	     \\
     5 & 0.322    &     0.857	     \\
     7 & 0.223    &     0.809	    \\
   10 & 0.137    &     0.737	     \\
   15 & 0.0698  &     0.636	      \\
   30 & 0.0156  &     0.414	      \\
   50 & 0.0033  &     0.244	       \\
   70 & 0.0008  &     0.149	      \\
100  &$1.2\times10^{-4}$&0.0754	\\
150  &$6.2\times10^{-5}$&0.0273 \\
\hline
\end{tabular}
\caption{Expected ratio of the obscured flux to the un-obscured flux for different values of the 
column density of obscuring material. A thermally emitting gas with temperature 
of $kT=1$~keV ({\sc phabs*apec} model) is assumed in the computations. 
The predicted observed flux in both the $2-4.5$ and $4.5-10$~keV bands ($F_{2-4.5}$ and 
$F_{4.5-10}$) is computed and compared to the respective un-absorbed 
($F0$) flux. }
\label{TabNH}
\end{center}
\end{table} 

\begin{figure*} 
\begin{center}
\includegraphics[width=1.03\textwidth,angle=0]{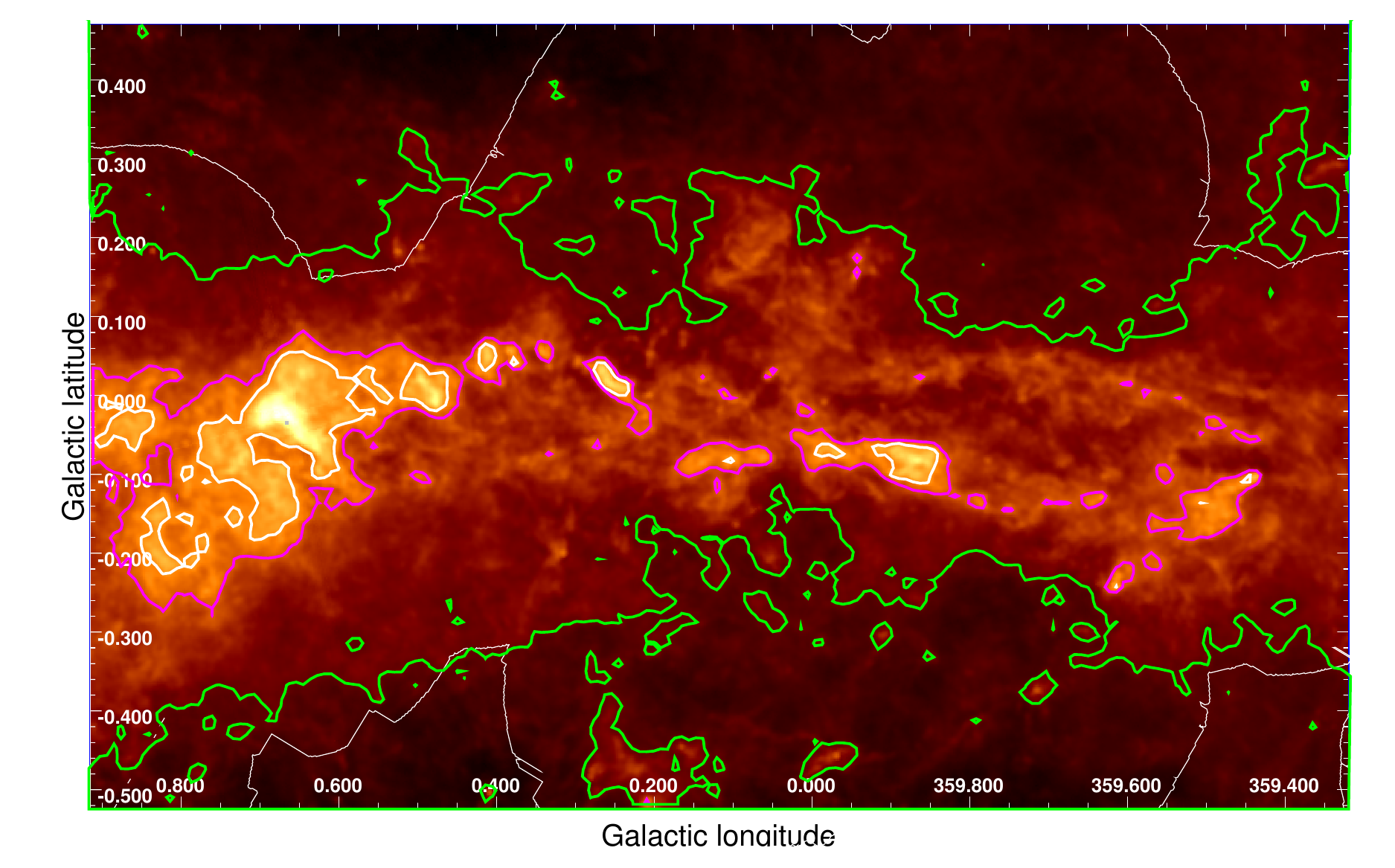}
\includegraphics[width=1.03\textwidth,angle=0]{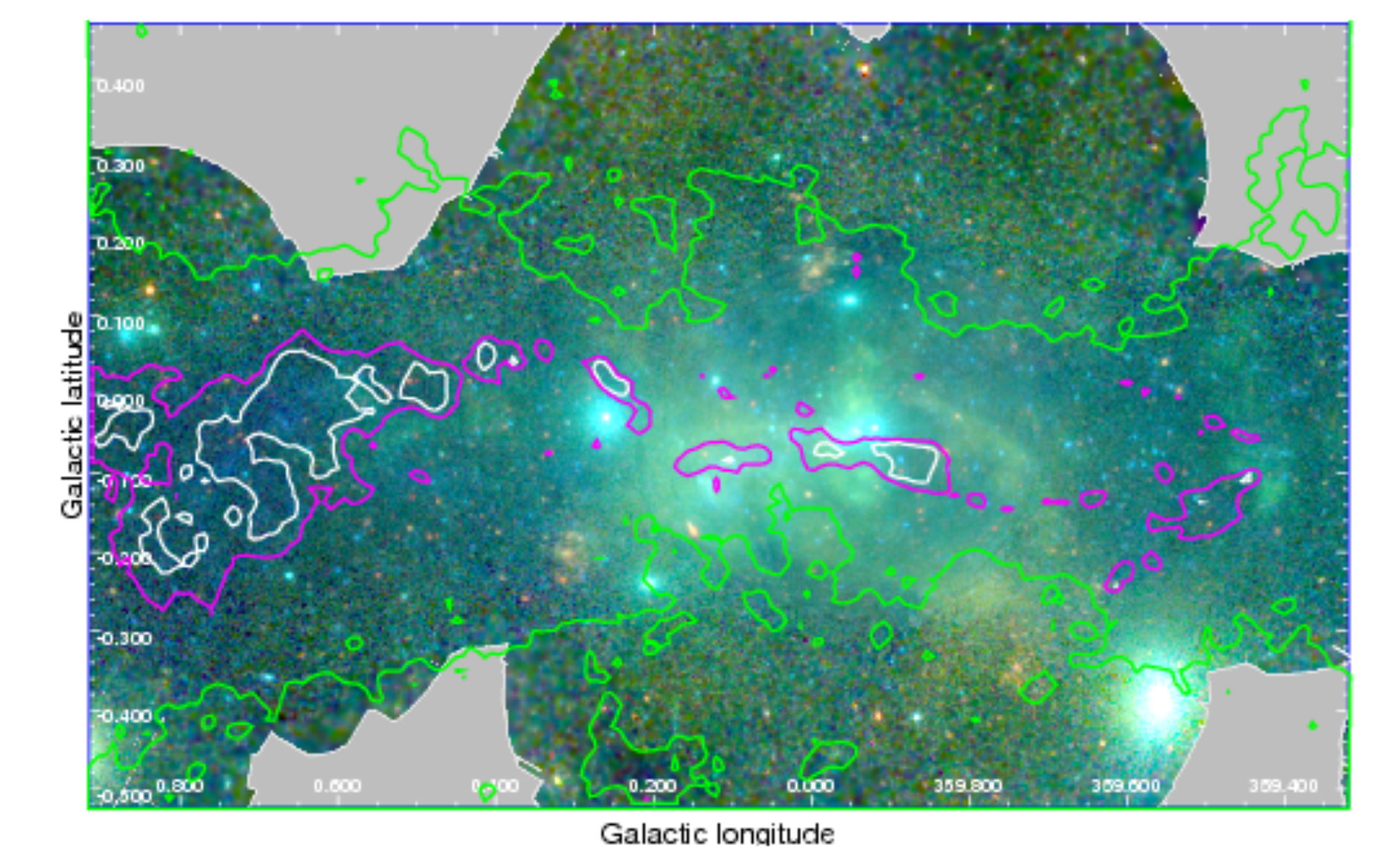}
\caption{{\it Top panel:} Neutral Hydrogen column density distribution as derived 
from dust emission (Molinari et al.\ 2011).
The image shows the $N_H$ distribution in logarithmic scale from 
N$_{\rm H}= 4.5\times10^{22}$ up to $3.8\times10^{25}$ cm$^{-2}$. 
The green, magenta and white contour levels correspond to $N_H=1.5\times10^{23}$, 
$7\times10^{23}$ and $1.5\times10^{24}$~cm$^{-2}$, respectively. 
{\it Bottom panel:} X-ray continuum RGB map (Fig.~\ref{RGBhard}) with the 
column-density contours overlaid. }
\label{RGBhardNH}
\end{center}
\end{figure*} 
Given the high column densities of neutral or weakly ionized
material absorbing the soft X-ray radiation, it is important to estimate the effects of 
X-ray obscuration.  For example, a molecular complex 
having a column density of N$_H\sim10^{25}$ cm$^{-2}$, such as 
the Sgr~B2 core, would completely obscure the radiation below 
about $4$~keV, if placed in front of the GC; see Fig. \ref{ImaEbands}.

To calculate the effects of absorption of the X-ray emission, we computed the 
flux generated by a thermally emitting plasma with temperature of $kT=1$~keV 
(using a {\sc phabs*apec} model), in both the $2-4.5$ and $4.5-10$~keV bands, after 
being absorbed by a given column density of neutral material (see also 
Fig.\ \ref{ImaEbands}). 
For each column density explored, we report in Tab. \ref{TabNH} the ratio of the 
observed flux ($F_{2-4.5}$ and $F_{4.5-10}$) over the respective un-absorbed 
($F0$) flux. We note that the hard X-ray band starts to be affected 
(corresponding to flux reductions up to a factor of 2) 
for column densities up to $N_H\sim3\times10^{23}$~cm$^{-2}$ while it 
is heavily affected (flux reduction of a factor of 10 or more) for 
$N_H\sim10^{24}$~cm$^{-2}$ or higher (see Tab.~\ref{TabNH}). 
At lower energies, the obscuration effect is even more pronounced. Already, 
for $N_H\sim3\times10^{22}$~cm$^{-2}$, the observed flux is less than half and 
for $N_H~\gsimeq~5\times10^{23}$~cm$^{-2}$ it is less than 0.1~\% of its 
un-obscured flux. This indicates that the softer band is expected to be heavily 
affected by absorption. 

\subsection{Column density distribution}

The top panel of Fig. \ref{RGBhardNH} shows the neutral Hydrogen column 
density distribution as derived from dust emission (Molinari et al. 
2011)\footnote{We do not show the entire CMZ, because of the limited coverage 
of the {\it Herschel} dust emission map (Molinari et al.\ 2011).}.
The image shows the $N_H$ distribution in logarithmic scale in the range 
N$_{\rm H}= 4.5\times10^{22} - 3.8\times10^{25}$ cm$^{-2}$. 
This total column density estimated from the dust has large uncertainties that 
can be mainly ascribed to the uncertainty associated with
the dust-to-$N_H$ ratio. In particular, the column densities shown in this map 
appear to be systematically larger than what is measured with other methods.
For example, the column density of the G0.11-0.11 massive cloud 
is estimated to be $N_H\sim5-6\times10^{23}$~cm$^{-2}$ in this map, while 
Amo-Baladron et al.\ (2009) measure $N_H\sim2\times10^{22}$~cm$^{-2}$, 
through a detailed modelling of the molecular line emission. 
The core of Sgr~B2 is estimated by Molinari et al. (2011) to 
have $N_H\sim3\times10^{25}$~cm$^{-2}$, 
while modelling of the X-ray emission (Terrier et al.\ 2010) suggests 
$N_H\sim7\times10^{23}$~cm$^{-2}$, more than an order of magnitude lower. 
Moreover, the average column densities of G0.40-0.02, G0.52-0.046, 
G0.57-0.018 and a fourth region (the magenta ellipse in Fig.\ \ref{MorrisNH}) are 
estimated to be $N_H\sim4\times10^{23}$, $4\times10^{23}$, 
$1.2\times10^{24}$ and $1.5\times10^{24}$~cm$^{-2}$, respectively, 
from the dust map, while they are measured to be in the range 
$N_H\sim7-10\times10^{22}$~cm$^{-2}$, from modelling of the X-ray 
emission. Therefore, the total normalisation of the $N_H$ map built 
from the dust emission appears to be overestimated. However, the method 
employed to produce it does not suffer from self-absorption, so it is 
presumably giving unbiased relative $N_H$ ratios.

\subsection{X-ray emission modulated by absorption}

The bottom panel of Fig. \ref{RGBhardNH} shows the X-ray map with the 
column-density contours overlaid for comparison. We observe that, 
as expected, no soft X-ray emission is observed toward the central part of the 
most massive molecular cores. In particular we observe depressed X-ray emission 
from: i) the Sgr~B2 nucleus and its envelope (with $N_H>7\times10^{23}$~cm$^{-2}$); 
ii) the almost perfect coincidence between the hole in soft X-ray emission 
east of G0.224-0.032 (see Fig.\ \ref{FC1} and \ref{MorrisNH}) and the shape of the 
so-called "Brick" molecular cloud, M0.25+0.01 
(see Fig. \ref{FC1}; Clark et al. 2013); iii) the core of the Sgr~C complex\footnote{ 
At this location a sharp transition in the soft X-ray emission, with an arc-like shape, 
is observed. This is spatially coincident to the edge of a very dense core 
of dust, suggesting that the modulation in the soft X-ray emission is induced 
by obscuration by the molecular cloud. }; iv) the regions around DB-58 and at 
Galactic position $l\sim0.2$, $b\sim-0.48^\circ$ also appear to have darker colours 
and, once again, it is possible to find molecular complexes (M0.018+0.126 and 
M0.20-0.48) covering roughly the same region (see Fig. \ref{FC1} 
and \ref{RGBhardNH}). 
All these clouds are characterised by very high column densities 
N$_H\gsimeq3-7\times10^{23}$~cm$^{-2}$ and they most probably lie in 
front of Sgr~A$^{\star}$ and of most of the GC (e.g., according to the twisted ring 
model of Molinari et al. 2011). Therefore, they are absorbing the GC's extended,
soft X-ray emission. 

All this evidence suggests that at least the most massive clouds located 
in front of the GC do actually modulate (obscure) the soft X-ray emission. 
However, fluctuations in column densities cannot be the only cause for the 
observed distribution of soft X-ray emission for two reasons. 
First, we do observe only weak soft X-ray emission along some lines of 
sight having a low column density of molecular material (such as around the 
Sgr~C and Sgr~D complexes and south of the Sgr~B1 region). 
Second, we do detect intense (among the brightest) 
soft X-ray emission from several regions such as the Sgr~A complex 
and the cores of the Sgr~C and Sgr~D complexes, where some of the 
highest column-density clouds are found. 
In particular, in Sgr~A very intense soft X-ray 
emission is observed along the line of sight toward the 50 km s$^{-1}$, 
the Bridge (Ponti et al.\ 2010) and the G0.11-0.11 clouds, some of the highest 
column density clouds in the CMZ. Although this might be explained by 
placing these clouds on the far side of the CMZ, it appears that these regions 
are characterised by truly enhanced soft X-ray emission (see e.g. 
\S 8.4, 8.5 and 8.6). We defer the detailed disentangling of these effects 
to an elaborate spectral study of these regions.

\section{Discussion} 
\label{Disc}

In the process of systematically analysing all \xmm\ observations of the central degrees 
of the Galaxy, we have discovered several new extended features 
and have produced an atlas of known, extended soft X-ray features. 
Here we discuss their general properties and investigate the origin/existence 
of several specific features. 

\subsection{General properties}
\label{general}

To compute the total observed (absorbed) flux from the CMZ 
we first mask out the emission from the brightest binaries, by 
excluding a circle with a radius of: $1'$ around SAX~J1747.7-2853; 
$1.5'$ for AX~J1745.6-2901 and GRS1741.9-2853; $2'$ for 
XMMU~J174445.5-295044; $2.5'$ for 1E~1743.1-2843; 
$3.5'$ for IGR~J17497-2821 and 1E~1740.7-2942 (see white circles in Fig.\ 
\ref{FC1}). 
We then compute the total observed count rate from two boxes, 
one with a size of $1.5^\circ\times0.35^\circ$ ($l~\times~b$) and centered on \sgras\ 
and one with a bigger size of $2.08^\circ\times0.413^\circ$ centered on 
$l=0.232^\circ$, $b=0.080^\circ$. We then measure the total count rate within 
these regions and convert it into a flux\footnote{To perform this task 
we use {\sc webpimms}: 
https:\/\/heasarc.gsfc.nasa.gov\/cgi-bin\/Tools\/w3pimms\/w3pimms.pl.
As explained in \S~2.1, the combined count rate is the sum of the EPIC-pn 
plus the EPIC-MOS count rates after scaling the latter exposure maps by 0.4.
We then use the EPIC-pn, medium-filter, rate-to-flux conversion 
computed with {\sc webpimms}. } assuming that the soft X-ray diffuse 
emission is dominated by a thermally emitting plasma with a temperature of 
$kT=1$~keV (Kaneda et al. 1997; Bamba et al. 2002). 
In particular, we assumed an {\sc apec} emission component with temperature 
$kT=1.08$~keV, absorbed by a column density of neutral material of 
N$_H=6\times10^{22}$ and with Solar abundance. 

The fluxes of the integrated continuum and line intensities (line plus continuum) 
within the big and small boxes are reported in Tab.\ \ref{LumBande}. 
This corresponds to an observed $2-12$~keV luminosity of 
$L_{2-12}=3.4\times10^{36}$~erg~s$^{-1}$ and 
$L_{2-12}=2.6\times10^{36}$~erg~s$^{-1}$ for the big and small boxes, respectively, 
assuming a distance of 8~kpc to the Galactic center (Reid 1993; Reid et al. 2009). 

\begin{table*}
\begin{center}
\begin{tabular}{ c c c c }
\hline
\multicolumn{2}{c}{{\bf Big box}} & \multicolumn{2}{c}{{\bf Small box}} \\
\hline
Flux & Surf. Bright. & Flux & Surf. Bright. \\
\hline
$F_{1-2~keV}=19.0$ & $f_{1-2~keV}=6.2$ & $F_{1-2~keV}=12.9$ & $f_{1-2~keV}=6.9$ \\
$F_{2-4.5~keV}=155.0$ & $f_{2-4.5~keV}=50.7$ & $F_{2-4.5~keV}=119.3$ & $f_{2-4.5~keV}=64.3$ \\
$F_{4.5-12~keV}=290.7$ & $f_{4.5-12~keV}=95.1$ & $F_{4.5-12~keV}=219.0$ & $f_{4.5-12~keV}=118.0$ \\
$F_{Si~{\sc xiii}}=4.4$ & $f_{Si~{\sc xiii}}=1.5$ & $F_{Si~{\sc xiii}}=3.4$ & $f_{Si~{\sc xiii}}=1.8$ \\
$F_{S~{\sc xv}}=15.1$ & $f_{S~{\sc xv}}=4.9$ & $F_{S~{\sc xv}}=12.0$ & $f_{S~{\sc xv}}=6.4$ \\
$F_{Ar~{\sc xvii}}=15.1$ & $f_{Ar~{\sc xvii}}=4.9$ & $F_{Ar~{\sc xvii}}=11.9$ & $f_{Ar~{\sc xvii}}=6.4$ \\
$F_{Ca~{\sc xix}}=18.4$ & $f_{Ca~{\sc xix}}=9.9$ & $F_{Ca~{\sc xix}}=14.3$ & $f_{Ca~{\sc xix}}=7.7$ \\
\hline
\end{tabular}
\caption{Fluxes (F) and surface brightnesses (f) of the continuum and the line intensities 
(line plus continuum) integrated over the small and big boxes described in 
\S~\ref{general}. The fluxes are given in units of $10^{-12}$~erg~cm$^{-2}$~s$^{-1}$,
while the surface brightnesses are given in $10^{-15}$~erg~cm$^{-2}$~s$^{-1}$~arcmin$^{-2}$.}
\label{LumBande}
\end{center}
\end{table*} 

\subsection{A new X-ray filament, XMM~0.173-0.413}

\begin{figure}
\hbox{\includegraphics[width=0.5\textwidth,angle=0]{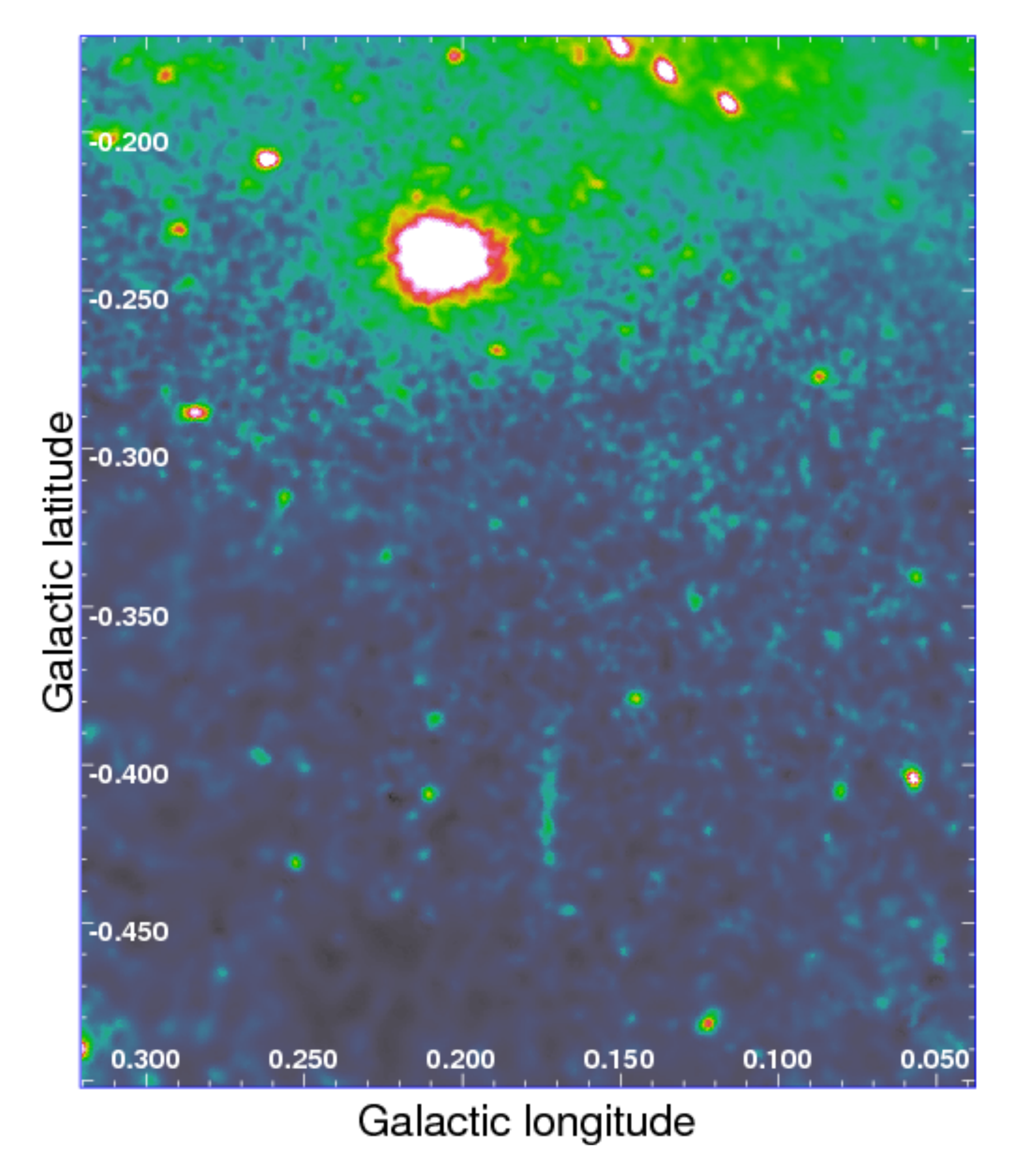}}
\caption{\xmm\ image in the 2-12~keV band showing the new X-ray filament located 
south of the Radio Arc. The filament is located at $l\simeq0.173^\circ$ and 
$b\simeq-0.413^\circ$ and appears as a thin ($<0.15$ arcmin) and long 
($\sim2.7$ arcmin) filament running along the north-south direction, as the Radio Arc. 
The brightest source in the image is SAX~J1747.7-2853. }
\label{NewFil}
\end{figure}
We observe a new X-ray filament extending $\sim2.2$~arcmin perpendicular 
to the Galactic plane and situated at $l = 0.173^\circ$, $b = -0.413^\circ$, which is almost
directly toward negative longitudes from the GC Radio Arc. It coincides
with the brightest segment of a much longer radio filament that extends
toward the southernmost extensions of the filaments of the Radio Arc (Yusef-Zadeh
et al. 1989; 2004; see figures 7a, 16b,c, and 17a,b,c of the latter reference, where
the filament is labelled "S5"), but because it is not continuous or exactly parallel with the 
filaments of the Arc, it is not completely evident that it is an extension of the Arc in 
three dimensions.  The X-ray filament has a hard X-ray colour and does not appear 
in the soft line images, indicating a non-thermal emission spectrum.  

Three other nonthermal radio filaments have been found to have X-ray emission 
along some portion of their lengths: G359.54+0.18, G359.89-0.08, and G359.90-0.06 
(Lu, Wang \& Lang 2003; Sakano et al.\ 2003; Lu et al.\ 2008; 
Johnson et al.~2009; Morris, Zhao \& Goss 2014; Zhang et al.\ 2014). 
XMM~0.173-0.413 is the only one of the four known cases where the X-ray 
emission is not at or near a location where the radio filament shows unusually strong 
curvature.  

\subsection{SNR excavated bubbles within the CMZ?}
\label{SgrB1SNR}

\begin{figure*}
\begin{center}
\includegraphics[width=1\textwidth,angle=0]{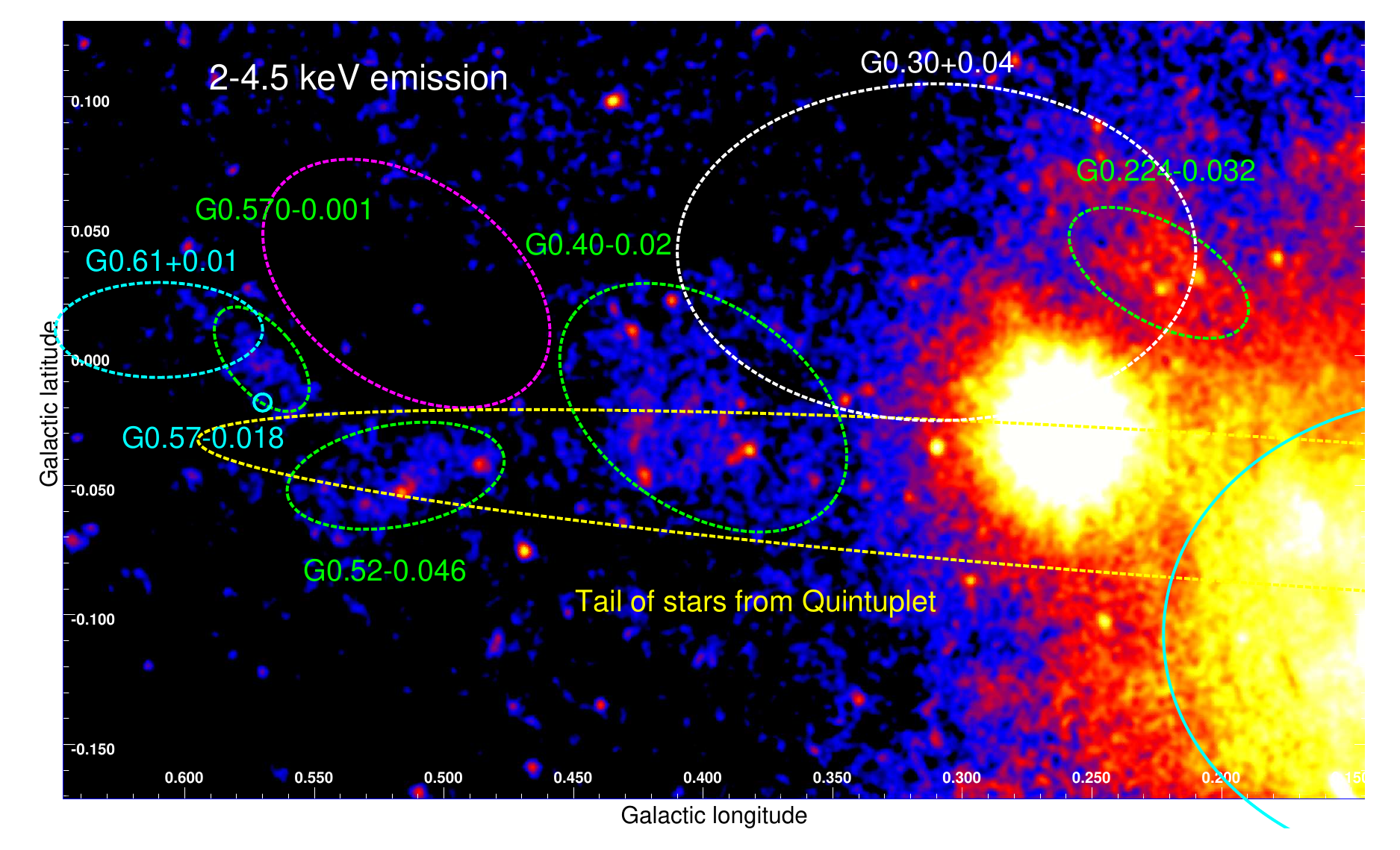}
\includegraphics[width=1\textwidth,angle=0]{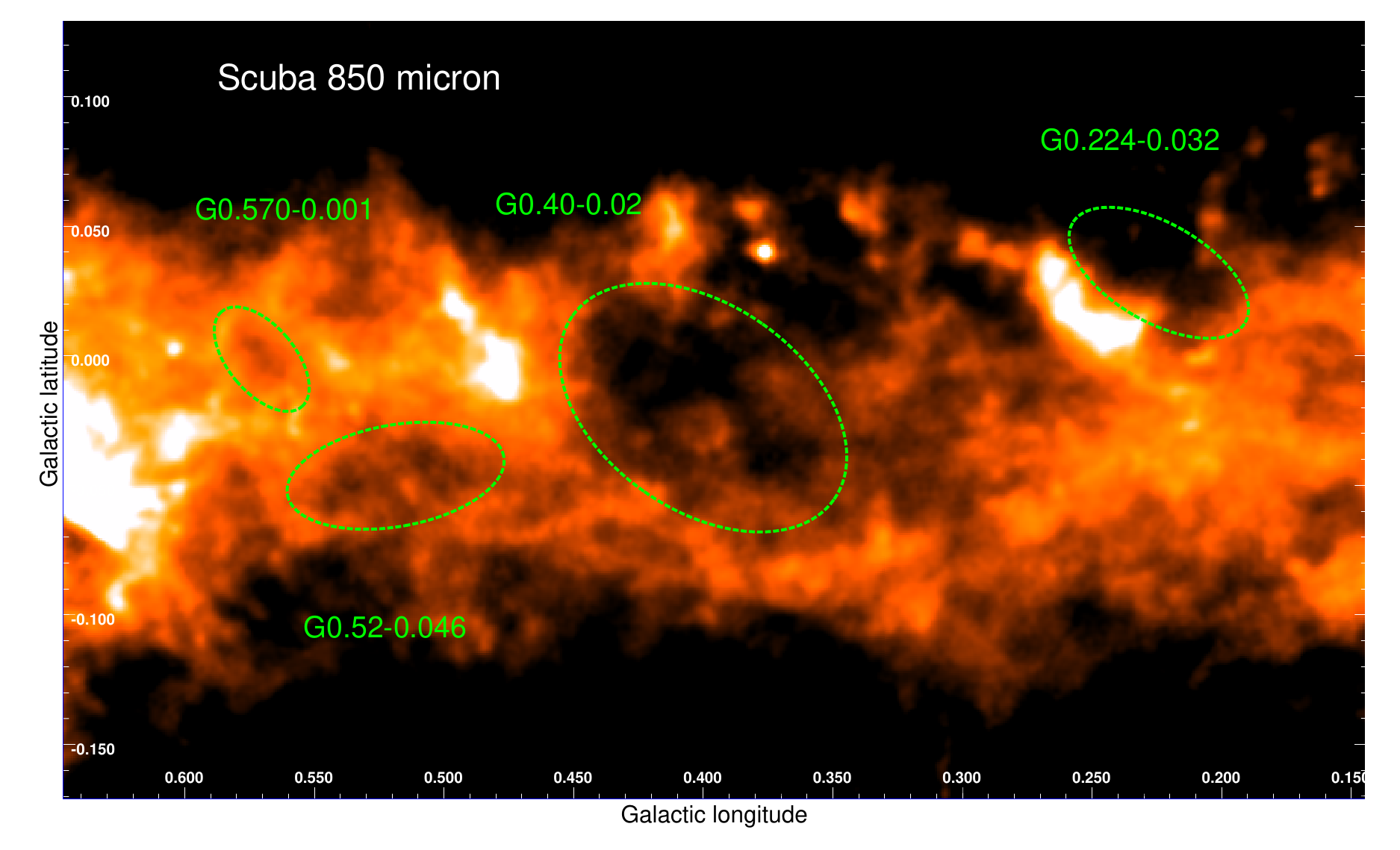}
\caption{{\it (Top panel)} 2-4.5 keV map. The dashed white ellipse shows the location 
of the radio SNR G0.30+0.04. The dashed and solid cyan ellipses show the 
positions of known X-ray SNRs and superbubbles, respectively. 
The green dashed ellipses indicate enhancements of soft X-ray emitting gas
(G0.40-0.02 was already observed in X-rays by Nobukawa et al.\ 2008), 
the magenta ellipse is used in the spectral analysis as a background region (Back 
in Tab. 7). The dashed yellow ellipse shows the 
location of the massive stars that could be in the tidal tail of the Quintuplet cluster 
(Habibi et al.\ 2013; 2014). 
{\it (Bottom panel)} $850~\mu$m map of the GC obtained with the SCUBA bolometer 
(Pierce-Price et al. 2000). The four dashed ellipses indicated by the green ellipses
and showing enhanced X-ray emission correspond to holes in the 2mm gas distribution. }
\label{MorrisNH}
\end{center}
\end{figure*}
The top and bottom panels of Fig. \ref{MorrisNH} show the X-ray (2.5--4.5~keV) 
and $850~\mu$m (Pierce-Price et al. 2000) 
maps of the Sgr~B1 region. Four enhancements of X-ray emission 
are clearly present (shown by the green dashed ellipses in Fig. \ref{MorrisNH}). 
In particular, G0.570-0.001, G0.52-0.046 and G0.40-0.02 correspond to holes 
in the dust distribution derived from the $850~\mu$m radiation (see bottom panel 
of Fig. \ref{MorrisNH}). To reinforce this evidence, we observe that their X-ray edges 
can be traced in the dust distribution all around the X-ray enhancements, suggesting 
a tight connection between the two. 
Such phenomenology is typical of SNe exploding within or near molecular clouds
and interacting with them, creating bubbles in the matter distribution
(Ferreira \& de Jager 2008; Lakicevic et al.\ 2014). 
Indeed, in this case, the SN ejecta might have cleared the entire region that is not 
filled with hot, X-ray emitting plasma, pushing away the ambient molecular material. 
However, this is not the only possibility. The apparent X-ray enhancements
could have resulted instead from a higher obscuration surrounding the submm holes, 
leading to higher X-ray extinction at the edges. 
We note that none of these regions has a known radio SNR counterpart (see 
Fig.\ \ref{MorrisNH}). However, the radio emission might be confused 
within the very high radio background of diffuse emission created by G0.30+0.04, 
the several HII regions present in this region (see Fig.\ \ref{FC2}), and the bright, 
extended synchrotron background of the GC.
Enhanced X-ray emission has already been reported towards G0.40-0.02 
(Nobukawa et al.\ 2008) and close to G0.570-0.001 (see cyan ellipses in 
Fig.\ \ref{MorrisNH}). 

\subsubsection{Spectral analysis}

To further investigate the origin of these structures, we extracted a spectrum 
from each of these features (in either obsid 0694641301 or 0694641201). 
We fitted each spectrum with a model composed of a SNR emission component 
(fitted with a {\sc pshock} model; a constant temperature, plane-parallel plasma 
shock model, meant to reproduce the X-ray emission from a supernova remnant 
in the Sedov phase) plus the emission components typical of the GC environment 
such as a hot thermal plasma (with temperature in the range: $kT=6.5-10$~keV), 
and an Fe~K$\alpha$ emission line, all absorbed by foreground neutral material 
({\sc phabs*(pshock+apec+gaus)} in {\sc XSpec}). 
We assume that all these components have Solar abundances. 
Possible confusion effects produce uncertainties associated with 
the determination of the correct sizes of these candidate SNRs, so
some of the results presented here, such as the dynamical time-scales, 
could thereby be affected. 

G0.40-0.02 shows a best-fit temperature and normalisation of the warm plasma 
associated with the SNR component of $kT=0.55\pm0.1$~keV and 
$A_{\rm psho}=2.4^{+5}_{-1}\times10^{-2}$, while the column density is observed 
to be $N_H=7.7\pm0.8\times10^{22}$~cm$^{-2}$ (see also Nobukawa et al.\ 2008). 
To test whether the observed X-ray enhancement is due to a real variation 
of the intensity of the soft X-ray emission or whether it is the product of lower 
extinction, we also extracted a spectrum from a nearby background comparison 
region (magenta in Fig.\ \ref{MorrisNH} and "Back" in Tab.\ \ref{TabSNR}). 
This second region has the same size as G0.40-0.02 and 
it is located in a fainter region in X-rays, characterised by higher $N_H$,
as suggested by the $850~\mu$m map (see bottom panel of Fig.\ \ref{MorrisNH}). 
This background region shows a slightly higher absorption column density, 
$N_H\sim8.8\pm1\times10^{22}$~cm$^{-2}$, and no significant warm plasma component. 
If we impose the presence of a warm plasma component having the same temperature 
and $\tau$ (the ionisation timescale of the shock plasma model) as observed 
in G0.40-0.02 we obtain an upper limit to its normalisation of $A_{\rm phsho}<6\times10^{-3}$. 
This suggests that the enhanced X-ray emission towards G0.40-0.02 is due to 
a real excess of X-ray emission and is not a simple byproduct of lower extinction. 
The thermal energy, the dynamical timescale and the size of G0.40-0.02 are 
$E_{th}\sim1.9\times10^{50}$~erg, $t_{dy}\sim3700$~yr and $8.6\times5.5$~pc$^2$, 
respectively, as expected for a young SNR in the Sedov-Taylor phase (derived from 
the equations shown in Maggi et al. 2012). 

Similar parameters characterise G0.52-0.046 ($kT=0.77\pm0.3$~keV, 
$A_{\rm psho}=5^{+7}_{-3}\times10^{-3}$, $N_H=7.9\pm1.1\times10^{22}$~cm$^{-2}$). 
Therefore, this feature also appears to be consistent with an SNR origin. 
However, although we derive a dynamical age of the same order ($t_{dy}\sim1700$~yr), 
the energy inferred for the SN explosion is substantially lower, 
$E_{th}\sim5\times10^{49}$~erg. 

The spectrum of G0.57-0.001 is characterised by significantly lower statistics
and a higher column density of absorbing material. The best fit prefers a low 
temperature plasma, at the limit of detection. We fix its temperature to a relatively 
low value of $kT=0.6$~keV and find $A_{\rm psho}=4.3^{+11}_{-4}\times10^{-3}$ 
and $N_H=9.5\pm2\times10^{22}$~cm$^{-2}$. 
The derived thermal energy and dynamical times are 
$E_{th}\sim2.6\times10^{49}$~erg and $t_{dy}\sim1600$~yr, respectively. 
The soft X-ray excess of this feature lies spatially very close to a region of X-ray excess that 
has been traced by Fe~{\sc xxv} line emission (Nobukawa et al.\ 2008). 
We also note that the region defining G0.57-0.001 almost completely contains 
a diffuse X-ray source detected both by the \chandra\ and \asca\ satellites 
(Senda et al.\ 2002). The \chandra\ image shows a very compact ($\sim10"$
radius) and hot shell, G0.57-0.018, possibly the youngest SNR in the Galaxy, 
less than about 100~yr old. However, Renaud et al.\ (2006), because they found neither radio 
nor nucleosynthetic decay products (such as $^{44}$Ti), questioned 
such an interpretation. Further investigations are required to understand the link, 
if any, between these features. 

The morphology of G0.224-0.032 appears more complex, compared to the other SNR candidates. 
The edge of the X-ray emission 
is well defined only towards the Brick molecular cloud (designated M0.25+0.01 in Fig.\ 5) 
that, with its very high column density, can obscure the soft X-ray emission there. In any 
case, the fit of its X-ray spectrum shows 
parameters typical of a SNR. In particular, we derive a thermal energy and a 
dynamical time of $E_{th}\sim2.6\times10^{50}$~erg and $t_{dy}\sim1800$~yr, 
respectively. Therefore, G0.224-0.032 might be a new SNR 
partly obscured by the brick molecular cloud. In such a case, the true size and 
the energy estimate are likely larger. 

Overall, we remark that, if these SNR candidates are real, their dynamical 
timescales are extremely short, which would imply an extremely high supernova rate. 
The SN rate in the CMZ has been estimated to he as high as 0.4 SN per millenium 
(Crocker et al. 2011). 
However, we caution that our dynamical time-scales could be off either because 
of a higher ambient density than we have assumed, or because absorption or confusion 
effects do not allow us to distinguish the proper border of the SNRs, or to detect any 
colder and more extended portions that might be present. 
\begin{table*}
\begin{center}
\footnotesize
\begin{tabular}{ c c c c c c c }
\hline
\hline
parameter  &  unit   &G0.40-0.02                           & G0.52-0.046                      & G0.570-0.001                            & G0.224-0.032                       & Back \\
\hline                   
$kT$            & keV          & $0.55\pm0.1$                      & $0.77_{-0.2}^{+0.7}$         & $0.6$\ddag                             & $0.54\pm0.1$                       & $0.55$\dag \\
$A_{psho}$ &                  & $2.4_{-1}^{+5}\times10^{-2}$ & $5_{-3}^{+9}\times10^{-3}$ &$4.3_{-4}^{+11}\times10^{-3}$& $3_{-2}^{+6}\times10^{-2}$ & $<6\times10^{-3}$ \\
$\tau$\dag\ddag&s cm$^{-3}$&$>1.7\times10^{11}$ & $>7.5\times10^{10}$         &$>1.4\times10^{10}$               & $>3.5\times10^{11}$             & $7.5\times10^{11}$\dag \\
Size             & pc            & $8.6\times5.5$                     & $5.9\times2.7$                  & $4.2\times2.1$                       & $5.4\times2.6$                      & $8.6\times5.5$ \\
$N_H$ & $10^{22}$ cm$^{-2}$ & $7.7\pm0.8$             & $7.9\pm1.1$                      & $9.5\pm2$                             & $7.4\pm1$                             & $8.8\pm1$ \\
$\chi^2$/dof &                 & 807/777                            & 342/364                          & 141/134                              & 369/338                              & 834/741 \\
\hline
$n_e$         & cm$^{-3}$ & $1.4$                                   & $1.4$                                 & $2.1$                                      & $11$                        \\
$t_{dy}$      & yr              & $3.7\times10^3$                  & $1.7\times10^{3}$             & $1.6\times10^3$                    & $1.8\times10^3$      \\
$E_{th}$     & erg            & $1.9\times10^{50}$              & $5.0\times10^{49}$           & $2.6\times10^{49}$               & $2.6\times10^{50}$ \\
\hline
\end{tabular}
\caption{Best-fit and derived parameters of the SNR candidates described 
in \S~\ref{SgrB1SNR}. \dag Parameter unconstrained. \ddag Value weakly 
constrained by the high column density of neutral material, therefore 
fixed for the corresponding fit. \dag\ddag~Ionisation time-scale of the 
shock plasma model. }
\label{TabSNR}
\end{center}
\end{table*} 

\subsubsection{Expanding molecular shells}

The observed temperatures, ages and sizes of these SNR candidates are 
consistent with a Sedov-Taylor framework expanding into an average ambient 
density between $1-10$~cm$^{-3}$ (higher density  environments would 
result in older and cooler SNRs; Ostriker \& McKee 1988). 

If the X-ray enhancements described above truly arise from SNRs interacting with 
and carving bubbles inside or near the surfaces of molecular clouds (preceded, 
perhaps, by the wind of the massive progenitor), we should observe clear traces 
of such events also in the kinematics of the surrounding molecular matter. 
Such a complex and delicate investigation is beyond the scope of the present paper. 
Nevertheless, we note that Tanaka et al.\ (2009) discovered an expanding 
SiO shell (SiO0.56-0.01) centered at $l\sim0.56^\circ$, $b\sim-0.01^\circ$ 
and having a size of $\sim3.0\times3.4$~pc$^2$. 
The center and size of the expanding SiO shell closely match the peak and size 
of the X-ray emission of G0.57-0.001 and suggest an association between the two.
In particular, high-velocity clumps have been found consistent with 
the idea that the SiO shell consists of swept-up material. Tanaka et al.\ (2009)
calculated a kinetic energy of $E_{kin}\sim10^{50.4}$~erg for SiO0.56-0.01. 
This strongly suggests that G0.57-0.001 is indeed a SNR caught in the process of 
carving its bubble. Further studies of the gas kinematics around the other X-ray 
enhancements are required to establish their real nature. 

We note that the thermal energy estimated for these SNRs is observed to be 
systematically lower than the theoretical value for the remnant of a standard 
type II SN expanding into the interstellar medium. 
This might result from a relatively higher ambient density in the GC, leading to 
greater energy dissipation, or from a significant fraction of the energy budget 
going into the inflation of the bubbles and the production of cosmic rays.

We also note that G0.570-0.001, G0.52-0.046 and G0.40-0.02 are located 
within the trail of massive stars that have been hypothesised to have tidally 
escaped from the Quintuplet cluster (see Fig.\ \ref{MorrisNH} and 
Habibi et al.\ 2013; 2014). This raises the possibility that some of these 
SNRs might be associated with SN explosions from stars originating in 
this massive, young stellar cluster. 

\subsection{Origin of the Sgr~A X-ray lobes} 
\label{Lobes}
All the soft X-ray maps (see Figs. \ref{RGBhard}, \ref{RGBSoftLines1}, 
\ref{RGBSoftLines2}, \ref{FCZoom}, \ref{SoftLines} and \ref{Terrier})
show the presence of two extended features, with a size of roughly $5-10$~pc, 
located to the Galactic north and south of \sgras, the so-called "bipolar Sgr~A lobes" 
(Morris et al. 2003; 2004; Markoff et al. 2010; Heard \& Warwick 2013). 

\subsubsection{Lobe morphology}

The lobes appear to have roughly oval shapes with co-aligned major axes
oriented perpendicular to the Galactic plane.  They appear joined at the position 
of \sgras\, suggesting the latter is their point of origin (see Fig. \ref{FCZoomChandra}).  
The top panel of Fig.\ \ref{RGBSoftLines2} and Fig.\ 
\ref{FCZoom} show that the lobes' emission is characterised by a 
smaller ratio of soft X-ray lines to continuum (therefore characterised by a 
greener colour) compared to the surrounding regions (appearing with 
a redder colour) such as the superbubble, G0.1-0.1 and the Radio Arc 
(Fig. \ref{RGBSoftLines2}, \ref{FCZoom} and \ref{FCZoomChandra}). 
This suggests that the lobes, although they show thermal emission 
lines (see \S \ref{ParSoftLines}), have either a stronger non-thermal 
component or significantly hotter thermal emission than the surrounding 
regions\footnote{A hot plasma, with temperatures of $\sim2-4$~keV, produces intense X-ray 
emission but weaker soft X-ray lines, compared to a plasma having a temperature 
around $1$~keV.}. 
We also note that the eastern portion of what appears to be part of the 
southern lobe has a colour as red as G0.1-0.1 and the superbubble 
regions (see Fig.\ 10). Therefore, this emission might not be associated 
with the lobes, but rather with G0.1-0.1 or the edge of the superbubble 
(however, a gap such as might be produced by a foreground dust lane, 
appears to separate the lobes' emission from G0.1-0.1).
Figures \ref{FCZoom} and \ref{FCZoomChandra} also show that the 
surface brightness of the northern  
lobe decreases with distance from \sgras\ (Heard \& Warwick 2013). 

\begin{figure*}
\begin{center}
\includegraphics[width=0.495\textwidth,angle=0]{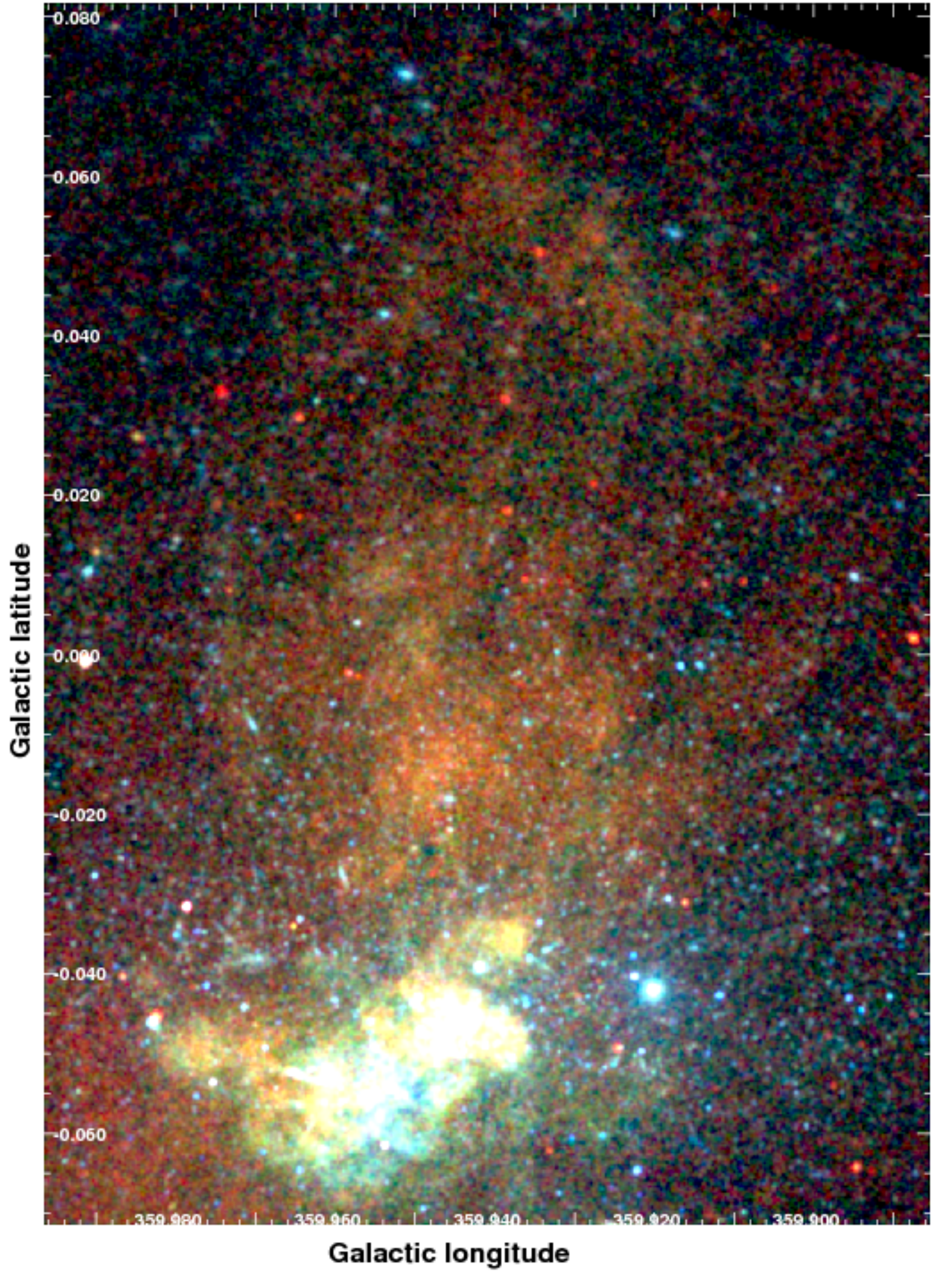}
\includegraphics[width=0.495\textwidth,angle=0]{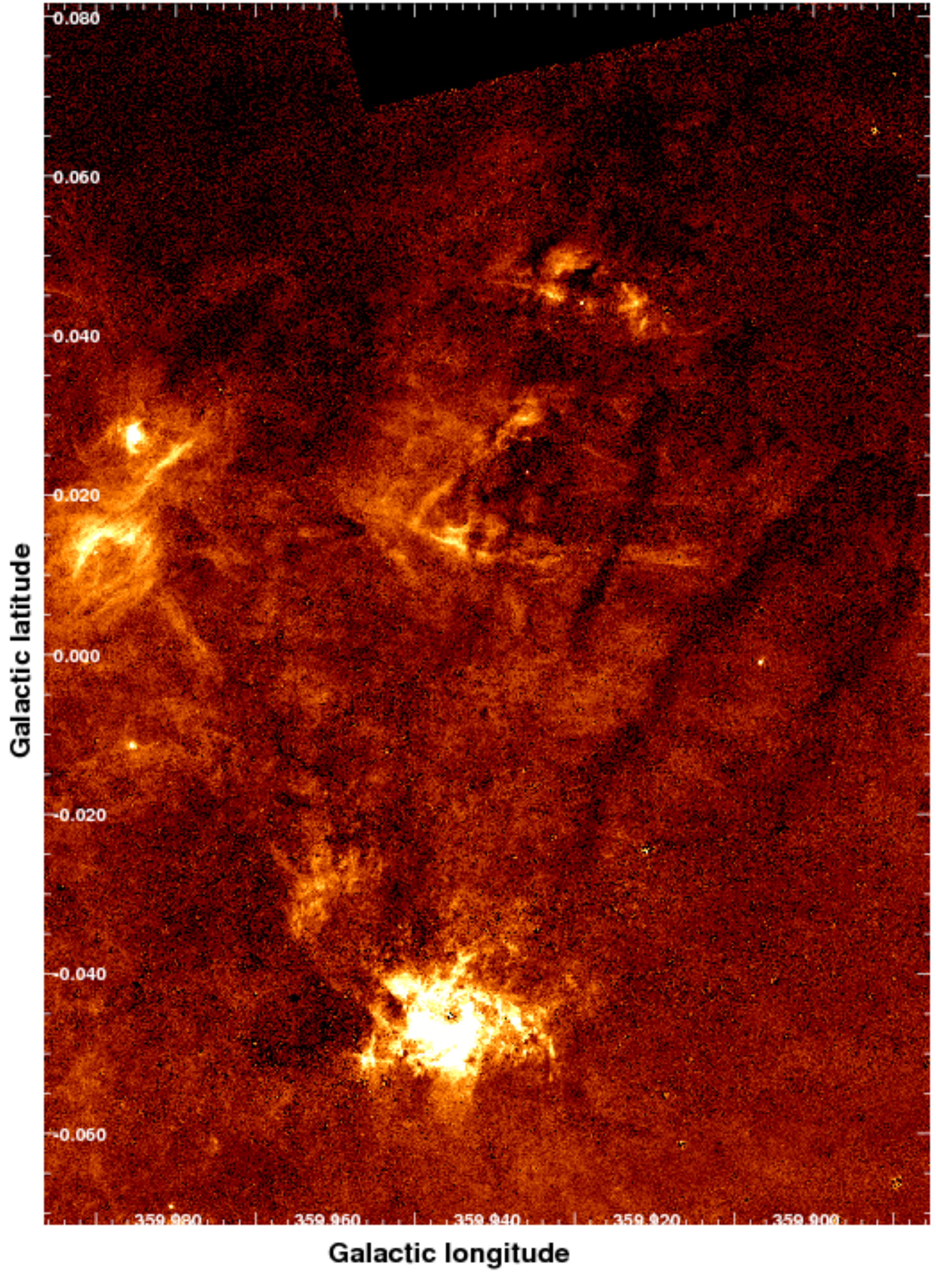}
\caption{{\it (Left panel)} \chandra\ RGB image of the northern lobe. See Figure 8 
for more details. 
{\it (Right panel)} Pa-$\alpha$ image of the northern lobe, tracing thermal, ionized gas
(from Wang et al. 2010). Continuum emission from stars has been removed (Dong et al.
2011), so the only stars that appear are those that have strong Pa-$\alpha$ emission lines.
Two Pa$\alpha$-emitting luminous stars located at 
$l\sim359.935^\circ$, $b\sim0.21^\circ$ and $l\sim359.925^\circ$, $b\sim0.45^\circ$ 
(Mauerhan et al. 2010; Dong et al. 2012) are probably responsible for at least part of 
the ionisation indicated by the Pa-$\alpha$. 
See {\sc www.mpe.mpg.de/heg/gc/} for a higher resolution version of these figures. }
\label{PaAlpha}
\end{center}
\end{figure*}

The bottom panel of Fig. \ref{RGBSoftLines2} shows that the lobes 
have an orange colour, indicating harder soft X-ray emission compared 
to the surrounding regions. In particular, a brighter and harder (yellow-green) 
linear structure outlines the northern lobe, shaping it to have 
a well-defined and symmetric spade structure, with a sharp transition 
at the border. The sharpness of the transition suggests the presence of a 
limb-brightened shock, indicating that the lobe is a bubble 
enclosed by a thin shell of hot, compressed 
material. This claim is strengthened by our analysis of the \chandra\ data (see Fig.\ 
\ref{FCZoomChandra} and \ref{FCZoomChandra2}, Baganoff et al.\ 2003; 
Lu et al.\ 2008; Muno et al.\ 2008). 
The superior \chandra\ spatial resolution, in fact, allows us to note that these 
projected linear features are running right along the lobes' edges, indeed confirming 
the presence of a shock (Fig.\ \ref{FCZoomChandra} and \ref{FCZoomChandra2}). 
We also note that the emission from the northern half of the lobe seems to be 
mainly due to three harder filaments converging at the top in a cusp, having 
radio continuum counterparts (Zhao et al.\ 2015) and associated Paschen-$\alpha$
emission, indicating that these are thermal features (see Fig.\ \ref{PaAlpha} 
and \ref{Face}). 

In the southern lobe, two bright knots are observed in the center and at the tip. 
Interestingly these appear to be located approximately at the same distance 
and in the opposite direction, compared to \sgras, as two 
enhancements present in the northern lobe. Moreover, the two bright 
knots have a green-yellow colour (upper panel of Fig. 11) similar 
to their apparent counterparts in the northern lobe. 
This suggests both: i) a similar physical origin for these features in 
both the north and south lobes and; ii) that the process that created the 
lobes is symmetric about the Galactic plane and its engine 
is (or is located close to) \sgras. The obvious interpretation of this
morphology is that energetic events simultaneously ejected diametrically 
opposed blobs of hot gas.  

However, upon closer inspection, the northern and southern lobes do 
not appear completely symmetric. For example, compared to the northern lobe, 
the western side of the southern lobe and the region close to Sgr~A$^\star$ 
appear suppressed (see Fig. \ref{FCZoom}, \ref{FCZoomChandra} and 
discussion following).
On the other hand, the eastern side of the southern lobe appears to extend 
further (e.g. further east compared to G359.977-0.076) than the corresponding 
boundary of the northern lobe (located close to e.g. G359.974-0.000). 
As described before, the emission around and east of G359.977-0.076 has 
a different colour and might therefore be associated with either the 
superbubble, with G0.1-0.1, or it could be a feature that is independent of
either of these and of the lobes. 

The bottom panel of Fig. \ref{RGBhardNH} shows that the region with 
depressed soft X-ray emission south of Sgr~A$^\star$ and on the western side 
of the southern lobe spatially coincides with the presence of the 20 km s$^{-1}$ 
molecular cloud, which is thought to be located in front of \sgras\ (Coil et al. 
2000; Ferri\`ere 2009) and to have a large column density. 
Soft X-ray emission could be produced there but be completely obscured 
to us by this intervening cloud (see Fig. \ref{ImaEbands}). 
To reinforce this idea, we note that, in fact, at this position, hard X-ray radiation 
(between $4.5$ and $12$ keV) is observed by \chandra\ to 
have a non-thermal spectrum and to be extended (Morris et al. 2003). 
Therefore, this hard radiation, which is able to penetrate the cloud, 
could be produced by strong shocks at the bubble's border.
Similar hard non-thermal filaments are observed in several places 
at the border of the northern lobe (Morris et al. 2003). 
This strongly suggests that the actual border of the southern lobe 
is located further west than the images reveal, and that the lobe's 
soft X-ray emission is obscured there. 
Therefore, once the effect of absorption by molecular clouds 
(e.g. the 20 km s$^{-1}$ cloud) is considered, the symmetry between the northern 
and southern lobes appears more clearly. The linear or filamentary structures (such as 
G359.974-0.000, G359.970-0.009, G359.959-0.027, G359.945-0.044, 
G359.942-0.045, 359.933-0.039, see Fig.\ \ref{FCZoomChandra}) 
observed in the northern lobe might be present in the southern one as well, 
but be suppressed by the intervening absorption.

Figure \ref{PaAlpha} shows the comparison between the \chandra\ and Pa$\alpha$
emission. The Pa$\alpha$ map clearly show the presence of tendrils of 
foreground absorbing material, running north-south along the extension of the lobe. 
Interestingly, the same tendrils are also evident as absorption lanes in the X-ray image. 
We also note that two Pa$\alpha$ emitting luminous stars 
are contributing to, if not dominating, the ionisation of the gas surrounding them.  However,
there is a close correspondence between the Pa$\alpha$ emission and the soft X-ray 
emission at the northern tip of the lobe, so strong shocks might be contributing 
both to the ionization and to the production of very hot post-shock gas that 
can emit X-rays.  
Indeed, we hypothesize that some portion of the hot wind that we see in the softest 
X-ray bands, presumably emanating from near \sgras, is undergoing a shock where 
it encounters ambient interstellar material
at b $\sim$ 0.02 degrees, and that it is thereby blocked from continuing to higher latitudes. 
That shock is manifested both as a horizontal feature in the Pa$\alpha$ image, and 
as a diminution in the brightness of the X-ray emission proceeding north from that latitude.
There is still some weaker, extended X-ray emission at higher latitudes, 
indicating that not all of the outflowing wind is completely blocked. In addition, the portion 
of the X-ray emitting plasma lying behind the shock front appears to 
be absorbed along the shock front, creating a horizontal shadow in the extended X-ray
emission, and suggesting that the shock at b $\sim$ 0.02 degrees has created a thick, 
compressed layer that absorbs the X-rays coming from behind it. 

\subsubsection{Lobes collimated by the Circumnuclear Disk (CND)}

Morris et al. (2003) observed that the CND has a size and orientation that are 
consistent with it being the agent that collimates an isotropic outflow from 
the Sgr~A$^\star$ region, thereby creating the bipolar lobes of hot plasma.  
Those authors further suggested that the sequence of enhancements along the 
axis of the lobes might have resulted from a series of energetic mass ejections 
from the immediate environment of Sgr~A$^\star$.  A similar scenario has recently 
been discussed by Heard \& Warwick (2013b). At an outflow velocity of 
$10^3$~km~s$^{-1}$, it would have taken $2-6\times10^3$~yr to inflate the lobes. 
Assuming a thermal emission model, an electron density $n_e$ 
in the range $1-10$~cm$^{-3}$ can be inferred (Morris et al. 2003;
Heard \& Warwick 2013b) and, assuming that the line-of-sight depth is equal 
to the projected width, the total hot plasma mass involved in the X-ray 
lobes is only about $1-3$ Solar masses. 

\subsubsection{Energetics of the Lobes}

Except for the broad intensity enhancements along the axis of the lobes, the 
surface brightness is relatively smoothly distributed.  Assuming a continuous 
and constant outflow, we examine the energy budget of the lobes.
In particular, integrating the measured energy density over a cylinder of 5~pc 
radius and $12$~pc height (the approximate sizes of the lobes), 
a thermal energy of $E_{th}\sim9\times10^{49}$~erg is estimated. 

It has been estimated that the massive stars in the central parsec collectively
lose $\sim5\times10^{-3}$~M$_{\odot}$~yr$^{-1}$ in stellar winds, with velocities 
ranging from $\sim300$ to $1000$~km~s$^{-1}$ (Geballe et al.\ 1987; Najarro 
et al.\ 1997; Paumard et al.\ 2001). 
The total kinetic energy thermalised by shocks is then $E\sim5\times10^{38}$~erg~s$^{-1}$ 
for such a mass outflow rate and an outflow velocity of $1000$~km~s$^{-1}$ 
(Quataert \& Loeb 2005). Therefore the energy released within the time needed 
to inflate the lobes ($\sim4\times10^3$~yr) is equivalent to $E\sim5\times10^{49}$~erg,
therefore giving an important contribution to the generation of the lobes. 

The lobes might also be traceable to the accretion flow onto \sgras.
We note that, as estimated by Wang et al.\ (2013), only $\sim1$~\% of the 
matter initially captured at the Bondi radius presently reaches the innermost regions around \sgras. 
The rest of the accretion power, estimated to be $\sim10^{39}$~erg~s$^{-1}$ 
(Wang et al.\ 2013), is probably converted to kinetic energy and used to drive an 
outflow that carries away the bulk of the inflow, sculpting the environment with its ram
pressure. If so, within the lobe inflation time, a total energy of $\sim10^{50}$~erg would 
have been deposited. 

Within the CMZ, X-ray reflection nebulae indicate that a few hundred years 
ago Sgr~A$^\star$ was more active, being $\sim10^6$ times brighter than at present
for approximately $5-10$~\% of the time in the past millennium (experiencing 
$L_X\sim10^{39}$~erg~s$^{-1}$; see Ponti et al. 2013 for a review). 
Could the lobes have been created by similar events that occurred over the 
past $10^4$~yr? 
The light crossing time of the CMZ limits our capability to trace 
Sgr~A$^\star$'s past activity beyond about $10^3$~yr ago, so 
it is difficult to directly trace the echoes of possible energetic events on
such time scales. However, if the process has been 
active over the past $(5-10)\times10^3$ yr at roughly the same rate,
(therefore active at $L\sim2\times10^{39}$~erg~s$^{-1}$ for 
$\sim10^3$~yr in the past $10^4$~yr), a total integrated energy equal 
to $\sim5\times10^{49}$~erg should have been generated. 
If \sgras's past activity was characterised by outbursts with associated outflows 
having kinetic luminosity comparable to the radiated power (therefore much higher
than in soft state stellar mass black holes; Ponti et al. 2012), then these events 
could be the primary source (or at least contribute) to form the lobes. 

All these processes appear similarly likely to have an impact on the formation 
of the lobes, from an energetic point of view. However, we note that the first two 
mechanisms are powered by a quasi-continuous outflow from \sgras\ or from the central
stellar cluster. In such scenarios, therefore, the sharpness of the edges at the 
extremities of the lobes remains rather puzzling, favouring explosive-outbursting 
scenarios. 

\subsubsection{Are Sgr~A's lobes the SNR of SGR~J1745-2900?}
\label{lobes}

The recent discovery of a young magnetar, SGR~J1745-2900 (Degenaar et al.\ 
2013; Dwelly \& Ponti 2013; Mori et al.\ 2013), most probably in orbit around 
Sgr~A$^\star$ (Rea et al.\ 2013), raises the quest for finding its young SNR. 
SGR~J1745-2900 is estimated to be only about $9\times10^3$~yr old and to be 
located at $\lsimeq0.07-2$~pc from \sgras\ (Rea et al. 2013). 
Therefore, the supernova that generated SGR~J1745-2900 should have exploded 
near the centroid of the lobes and likely inside the inner radius of the CND. If the shock propagated at 
the sound speed ($v_s\sim750$~km~s$^{-1}$), then a present size of $\sim6.5$~pc would 
be expected. This is slightly smaller than the observed size of the lobes ($\sim12$~pc), 
but is consistent with the observed size within the uncertainties in the age and the sound speed.
We also note that the thermal energy content in the lobes ($E_{th}\sim9\times10^{49}$~erg) 
is lower, but of comparable order of magnitude, to the energy released by typical supernova 
explosions. It is therefore plausible that the lobes are indeed the SNR associated with 
SGR~J1745-2900. 

Another viable possibility is that the lobes have been generated by the supernova 
that created the PWN candidate G359.945-0.044, located only $\sim8$" 
from \sgras, with an estimated age of few thousand years (Wang et al.\ 2006). 
It is therefore possible that at least one SN exploded close to \sgras\ within the last 
$\sim10$~kyr. If so, its blast wave likely propagated into a pre-existing 
hot, low density cavity created by \sgras's outflows and the collective 
winds of the central stellar cluster. Given the density and the temperature 
in the lobes ($kT\sim2$~keV; Morris et al.\ 2003), the shock is estimated 
to reach 15~pc in $9\times10^3$~yr (Wang et al.\ 2005), a value consistent 
with the observed size of the lobes. 

As mentioned above, the presence of the sharp edges to the lobes seems to favour 
an explosive mechanism for their creation. A supernova exploding inside the 
CND would expand into the pre-existing, stationary outflow from the center and be
collimated in the same way. Hydrodynamical simulations typically show that SN 
shock fronts are reflected away when encountering the walls of a dense molecular 
cloud, such as the CND (Ferreira \& de Jager 2008). 
In this scenario the sharp edge of the lobes would be due to the SN shock front. 
If the supernova recurrence time is longer than or comparable to the lobe 
expansion time (a few thousand years) then this would not appear as a stationary 
process. Assuming a recurrence time between $1-10\times10^4$~yr (similar to 
the SN recurrence time of the central young cluster, in which $\sim100$ massive 
stars presumably becoming SN over a $\sim10^7$~yr time interval), we estimate a 
time-averaged kinetic power release of $3-30\times10^{38}$~erg~s$^{-1}$.
This indeed suggests that: i) SN explosions of the central star cluster can 
contribute to powering the lobes; ii) the lobes are quasi-stationary 
features; and iii) it is not unlikely that we observe such features created by 
a rare event such as a SN explosion.

Finally, we note that, although the characteristics of the X-ray emission 
from the lobes appear consistent with being the X-ray remnant of a recent 
supernova that exploded within a few parsecs of \sgras, the lack of associated 
nonthermal radio emission from such a young SNR is problematical 
for this hypothesis. 

\subsection{G359.77-0.09 and G359.79-026: a ring from a hot superbubble 
southwest of \sgras}

A series of diffuse, soft features appear to the southwest of \sgras\ 
(see Fig. \ref{RGBhard}), namely G359.77-0.09, G359.79-0.26 
and a newly recognised extended feature, G0.0-0.16. 
If these are physically connected, they form, in projection, a roughly elliptical 
shape whose major axis has an inclination of about $60^{\circ}$ with respect 
to the Galactic plane (see Fig. \ref{RGBhard}, \ref{FC2}, \ref{FCZoom}, 
\ref{RGBSoftLines1}, \ref{RGBSoftLines2}, \ref{SoftLines} and \ref{Terrier}). 
These features show similar colours and strong, soft line 
emissions, indicating a similar thermal origin (see Fig. \ref{RGBhard}, 
\ref{RGBSoftLines1} and \ref{RGBSoftLines2}). 
This elliptical structure appears brightest at the softest energies, however 
it is not observed below $1.5$ keV, suggesting a location near the 
Galactic centre and a low temperature for its plasma, compared to 
the surrounding emission. We note that this feature is characterised by 
a very bright edge with strong Si~{\sc xiii} emission on the outside of 
the ellipse, suggesting a lower temperature of the edge compared to the interior. 

The ellipse center is located at $l\sim359.9^\circ$, $b\sim-0.125^{\circ}$ 
and it has minor and major axes of about $7.8$ and $\sim12$ arcmin,
respectively (corresponding to 18 and 28 pc). 
Both Mori et al. (2009) and Heard \& Warwick (2013) considered 
that these structures/group of structures were physically connected 
and form a superbubble candidate. 
We note that the recognition of such an elliptical ring critically depends 
on the presence of the dark lane running from $l\sim0.02^\circ$, 
$b\sim-0.22^\circ$ to $l\sim0.05^\circ$, $b\sim-0.07^\circ$ 
(see Fig. \ref{RGBhard}, \ref{RGBSoftLines1}, \ref{RGBSoftLines2}, 
\ref{FCZoom} and \ref{Terrier}), which separates G0.0-0.16
 from the emission of G0.1-0.1; this dark lane helps define the quasi-continuous 
 elliptical morphology of the ring. However, the lane might simply be due 
to absorption by foreground material,  in which case G0.0-0.16, forming the eastern 
part of the ring, could simply be connected to G0.1-0.1. 

\subsubsection{S~{\sc xv} emission filling the superbubble}

The S~{\sc xv} emission provides a key piece of information to better
understand the superbubble. We note, in fact, that the S~{\sc xv} emission 
completely fills the region inside the ring with a roughly uniform brightness (several 
times brighter than in the surrounding region) and sharply drops at the ring's edge. 
This indicates that the superbubble is a shell of hot gas that we see 
projected onto the plane of the sky (see middle top panel of 
Fig.\ \ref{SoftLines}). To further corroborate this, we note that
a sharp emissivity drop appears to be located just outside of the ring, 
running all around the ring's external edge. 
Such X-ray depression might be produced by a 
high column density of cold gas pushed away by the superbubble's 
shock front and accumulated in large quantities just outside the shock. 
If so, the observed depression could be indicating that the superbubble 
is located in front of G0.1-0.1. Fig. \ref{FCZoom} 
also shows a small depression in the top part of the northern lobe that 
could easily be explained by absorption associated with the superbubble 
if it is located in front of the lobes.  This situation would then be somewhat
analogous to the colour variation in the south lobe.

Mori et al. (2009) and Heard \& Warwick (2013) estimate a total 
thermal energy contained within the superbubble of 
$E_{th}\sim10^{51} f^{1/2}$~erg (where $f$ is the volume filling factor 
of the emitting plasma). Such a large energy content does, 
indeed, require multiple supernova events. 
Those authors also estimate for G359.77-0.09 and G359.79-0.26 
an ionisation time-scale of $t_{ion}\sim3\times10^4$~yr (assuming 
$f\sim1$). 

\subsubsection{Origin}

The origin of such a superbubble is not clear. We note that many of 
the massive stars that are suggested to have escaped from the Quintuplet 
cluster (Habibi et al. 2013; 2014) are projected inside the superbubble. It is 
therefore possible that explosions of stars lost by the Quintuplet cluster have 
contributed to energising the superbubble. 

A more speculative point is that the estimated age of the 
superbubble is of the same order of magnitude as the recurrence time 
of tidal disruption events by Sgr~A$^\star$: $t_{TDE}\sim1-3\times10^4$~yr 
(Alexander \& Hopman 2003).  While \sgras\ appears, in projection, to be 
located inside the superbubble, it is $\sim6.7$~arcmin off from the superbubble's 
center. Khokhlov \& Melia (1996) suggested that an explosion associated with 
a tidal disruption event would liberate a large 
amount of energy on the order of $E\sim10^{52}$~erg that would propagate
as a powerful shock wave into the local interstellar medium. 
As with the remnant of the SGR~J1745-2900, we expect that the 
shockwaves of a tidal disruption event would interact with the CND. 
However, the unbound part of the tidally disrupted star would be ejected into 
a limited solid angle, producing a strongly elongated and asymmetric remnant 
(Khokhlov et al.\ 1996; Ayal et al.\ 2000). 
Such a remnant would then appear as a very energetic shell of hot gas and 
remain visible for a time comparable to the age of a typical SNR. 
Assuming a shock survival time of $\sim1-10\times10^4$~yr, 
we could potentially observe a few remnants resulting from tidal disruptions. 
We suggest that the superbubble G359.77-0.09 has properties that make 
it a possible candidate. No other feature with properties obviously related to 
a tidal disruption event appear to be observed close to \sgras. 

\subsection{The arc bubble: a second superbubble in the GC}
\label{G0.1-0.1}

Highly enhanced soft X-ray emission is observed east of \sgras\ from 
the region called G0.1-0.1. This feature appears in Figures \ref{RGBSoftLines1} 
and \ref{RGBSoftLines2} as a slightly elliptical feature of enhanced 
emission with center at $l\sim0.09^\circ$, $b\sim-0.09^{\circ}$ and with 
radius of $\sim5$ arcmin (corresponding to $\sim10$ pc). 
The top panel of Fig. \ref{RGBSoftLines2} 
shows that G0.1-0.1 and the Radio Arc regions both show distinct 
red emission. This indicates large equivalent widths of the emission 
lines from this plasma and therefore a thermal origin. However, the top 
panel of Fig. \ref{RGBSoftLines1} and the bottom of Fig. 
\ref{RGBSoftLines2} show strong colour gradients within these regions, 
indicating that they might have different contributions 
from distinct components. 

The PWN candidate G0.13-0.11 ($l=0.131^\circ$, $b=-0.111^\circ$; Wang et 
al. 2002; Heard \& Warwick 2013) stands out from the general thermal 
emission in G0.1-0.1, appearing as a
distinct point-like source characterised by a light blue/white 
colour\footnote{Note that in the top panel of Fig. \ref{RGBSoftLines1} 
despite an enhancement of diffuse emission from the region surrounding 
the core of G0.13-0.11, no point like source is detected in the soft X-ray 
line image. }, indicating its non-thermal origin (dominated by intense soft and 
hard continuum emission with no soft line emission, see the top panel of 
Fig. \ref{RGBSoftLines2})\footnote{Please note that an unrelated point source 
at $l=0.142^\circ$, $b=-0.109^\circ$ is located at the same latitude, but  
$\sim1.5$ pc to the Galactic east of the PWN candidate G0.13-0.11.}. 
The point-like head of G0.13-0.11 
appears to be accompanied by a tail extending to the south for $4.5-5$ pc;
in Fig. \ref{Face}, it appears with a white-violet colour. 

Heard \& Warwick (2013) suggest that the SN that generated G0.13-0.11 might 
be the source of the soft X-ray emission from this region. Those authors
present a spectral study of the X-ray emission from G0.1-0.1 and find a 
gas temperature of $kT=1.1\pm0.1$~keV, and a column density of 
$N_H=5.6\pm0.5\times10^{22}$~cm$^{-2}$, indicating a GC location 
of this emission, and abundances that are about $1.8$ times Solar. Assuming 
that the plasma volume is only 3.5~pc$^3$, corresponding to only 1.5~arcmin 
radius around the PWN (see the red circle in Fig.\ \ref{MIPS}), 
the authors estimated a thermal energy of $E_{th}=3.1\times10^{49}$~erg 
(and a plasma ionisation time-scale of at least $t=1.8\times10^4$~yr), thus 
consistent with being produced by a single supernova explosion. 

\subsubsection{Energetics of the arc bubble}

\begin{figure*}
\hbox{\includegraphics[width=1.0\textwidth,angle=0]{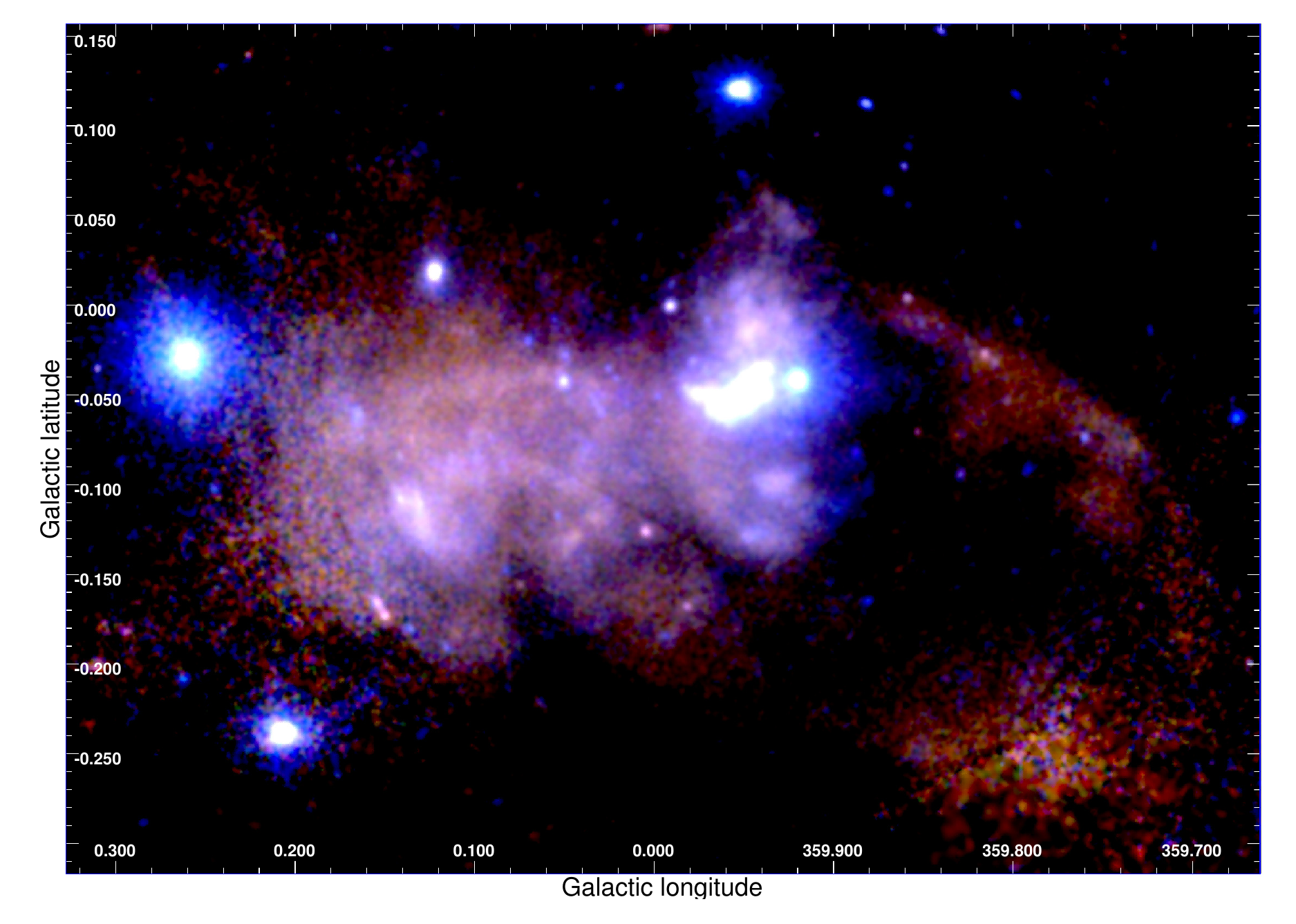}}
\caption{RGB image with colour scale chosen to highlight 
enhancements and depressions in the diffuse emission east of 
Sgr~A$^{\star}$. In red the sum of the Si-S and S~{\sc xv} bands is shown. 
In green the sum of S-Ar plus the S~{\sc xv} and Ar~{\sc xvii} bands is shown. 
The blue image shows the sum of the Ar-Ca plus Blue-Ca and Ca~{\sc xix} bands
(see Tab.\ \ref{Ebands} for the definition of the energy bands). }
\label{Face}
\end{figure*}
\begin{figure*}
\includegraphics[width=0.495\textwidth,angle=0]{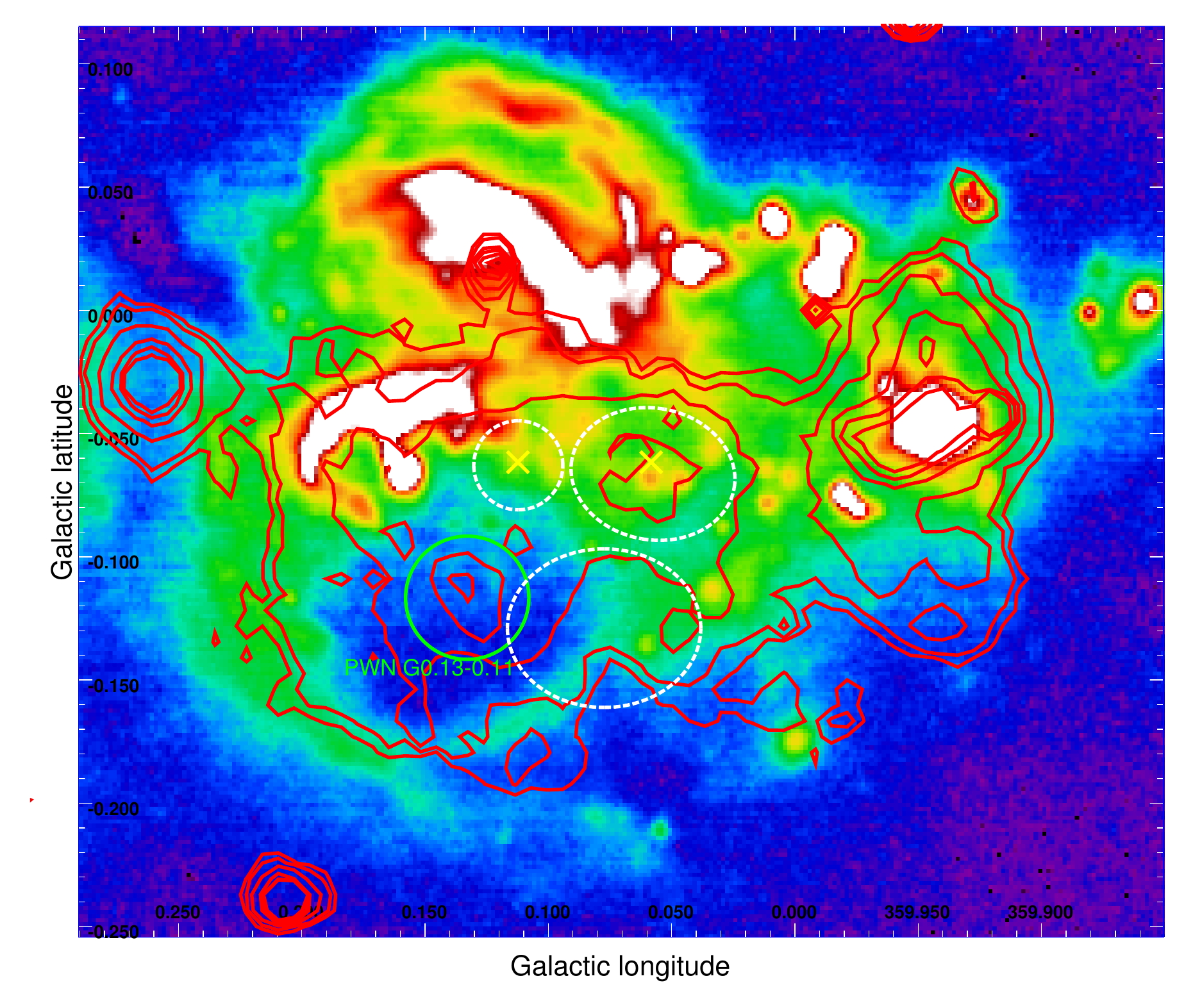}
\includegraphics[width=0.495\textwidth,angle=0]{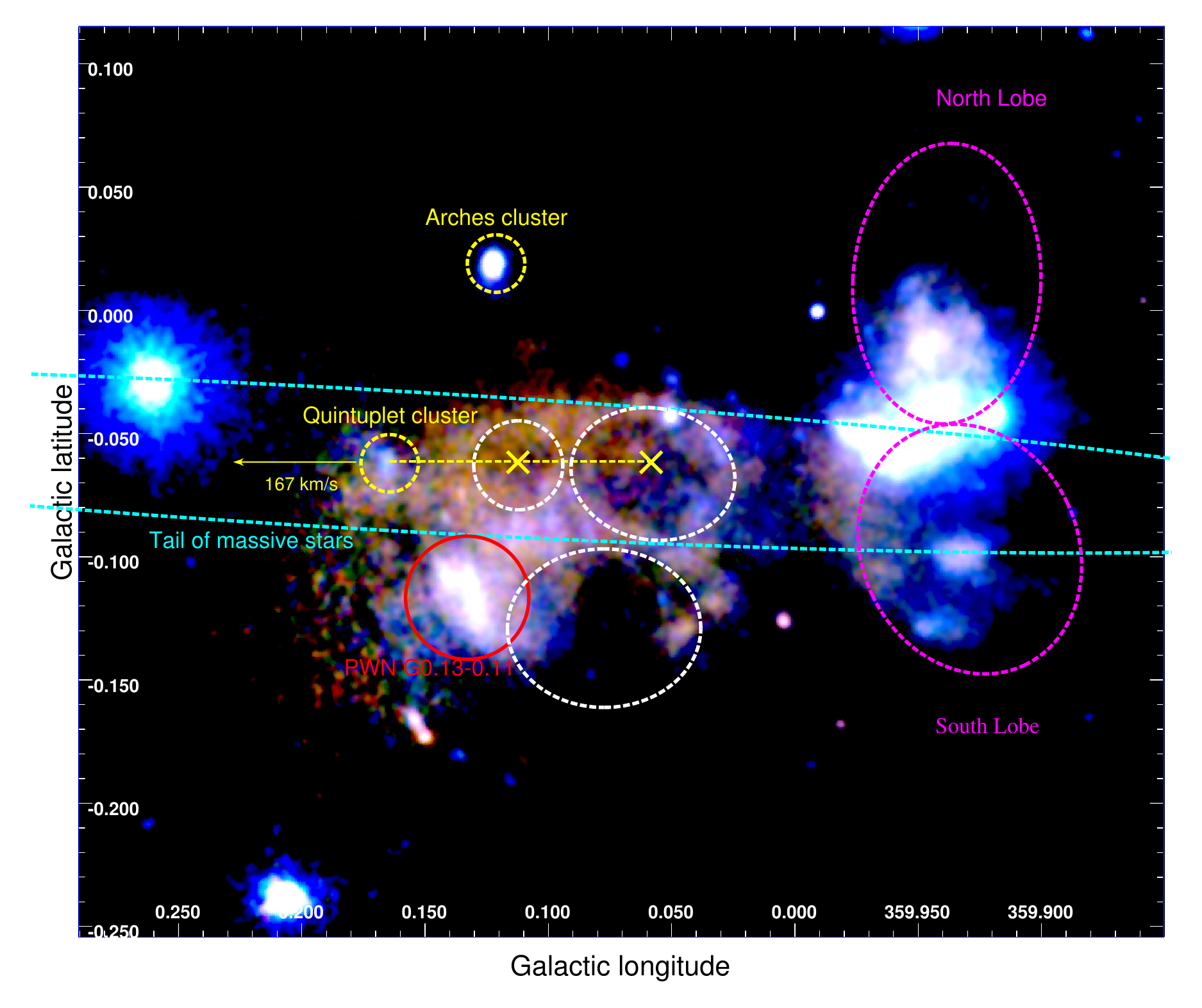}
\caption{{\it (Left panel)} $20~\mu$m MSX map of the GC. 
The contours indicate the intensity of S~{\sc xv} emission. 
Soft X-ray emission fills the arc bubble observed in the mid-IR. 
The green solid circle and the white dashed ellipses indicate the 
position of the PWN G0.13-0.11 and three structures in the soft X-ray 
emission map (see right panel). 
{\it (Right panel)} Soft X-ray map of the GC (the same energy bands used in 
Fig.\ \ref{Face} are displayed). The position of the 
PWN G0.13-0.11 is indicated by a red circle with $1.5$~arcmin 
radius. At least three sub-structures, appearing like holes, are 
observed within G0.1-0.1 (here indicated with white dashed ellipses). 
The positions of the Arches and Quintuplet star clusters are indicated 
by yellow dashed circles. The direction of the supersonic motion of 
the Quintuplet cluster is indicated and its past location is indicated 
by the yellow dashed line. The inferred positions of the Quintuplet 
$4\times10^4$ and $9\times10^4$ years ago are indicated 
with yellow crosses. The cyan dashed ellipse indicates the region in
which many massive stars that might have been expelled by the 
Quintuplet cluster are located.}
\label{MIPS}
\end{figure*}

We note from Figs.~\ref{Face}, \ref{RGBhard}, \ref{RGBSoftLines1}, \ref{RGBSoftLines2}, 
\ref{FCZoom}, \ref{SoftLines} and \ref{Terrier} that G0.1-0.1 extends 
further from G0.13-0.11 than the 1.5~arcmin region size considered by 
Heard \& Warwick (2013), with no clear boundary at $1.5$~arcmin (see Fig.\ \ref{MIPS}).
To illustrate this, Fig. \ref{Face} shows the soft emission lines RGB image 
with colour scales chosen to highlight intensity variations present within 
this region. The left panel of Fig. \ref{MIPS} shows the contours of the 
S~{\sc xv} emission overlaid on the $20~\mu$m MSX image (Price et al.\ 2001). 
Figure~\ref{MIPS} clearly shows that the empty mid-IR bubble (the 
so called arc bubble; Levine et al.\ 1999; Rodriguez-Fernandez et al.\ 
2001; Simpson et al. 2007) is completely filled with warm X-ray emitting plasma 
and that the soft X-ray emission is not confined to within $\sim1.5$~arcmin 
of G0.13-0.11, but it extends much further, for about $7$~arcmin. 
Assuming a uniform surface brightness, if the bubble is 4.5 times larger, we would 
expect a thermal energy of $E_{th}\sim1.5\times10^{51}$ erg, thus most 
probably requiring multiple supernova events, and supporting the notion
that G0.1-0.1 is a second superbubble candidate in the GC. 

Rodriguez Fernandez et al.\ (2001) have noticed that the radio Arc bubble 
is filled with continuum X-ray emission seen by ASCA which they ascribed 
to X-ray sources inside the bubble. Here we find that the bubble is in fact filled 
with diffuse thermal X-rays, most likely originating from SN explosions 
of massive stars associated with the Quintuplet cluster.

We also note that the soft line emission is at least as extended as the arc 
bubble and is highly inhomogeneous (see Figs. 19 and 20). 
Three depressions having roughly circular shapes can be discerned 
in G0.1-0.1 (Fig.\ \ref{MIPS}). Two cavities are located at about the 
same latitude, with centers close to $l=0.057^\circ$, $b=-0.067^\circ$
and to $l=0.116^\circ$, $b=-0.071^\circ$ and with radii 
of $\sim1.6$ arcmin (corresponding to $\sim3.7$ pc) and $\sim1$ arcmin, 
respectively. These cavities appear to be surrounded by a thin rim of brighter 
emission. 
A third depression is centered at  $l=0.083^\circ$, $b=-0.123^\circ$, with 
$\sim1.8$ arcmin radius.  This cavity also seems to be confined by a 
thin shell of brighter material, except for its southern edge, where it appears open 
(see Fig. \ref{Face}), possibly because of the presence of a dark absorbing lane. 

\subsubsection{Association with the Quintuplet cluster?}

Despite its offset position from the centre of the mid-IR arc bubble, the 
Quintuplet cluster is often considered responsible for creating and
maintaining the bubble with a combination of supernovae and strong stellar
winds (e.g., Simpson et al. 2007). 
Johnson et al.\ (2007) invoke a possible non-uniformity of the ambient medium 
as a possible origin of this asymmetry. This might also apply to the X-ray emission. 
However, we note that the Quintuplet cluster is moving supersonically within 
the CMZ (Stolte et al.\ 2014). Given its projected velocity, the cluster would have 
taken $\sim100$ kyr to cross the width of the IR arc bubble, in which case the 
bubble would have been inflated on a time scale much smaller than typical 
superbubble formation times (Castor et al.\ 1975; Weaver et al.\ 1977; Mc~Low \& McCray 1988). 
Several SN explosions in that amount of time would therefore have been required. 
Furthermore, the Quintuplet cluster would have been located in the middle 
of the two northern cavities about $4\times10^4$ and 
$9\times10^4$~yr ago, respectively (the right panel of Fig. \ref{MIPS} 
illustrates the direction of the cluster's motion and the position 
of the cavities). The relatively small cavities observed in the northern part 
of G0.1-0.1 are unlikely to have been generated by multiple cluster stars. 
In fact, a supernova 
exploding in the hot plasma of G0.1-0.1 is expected to undergo a significantly 
different evolution than a typical SNR. In particular, the sound velocity is 
significantly larger than in a typical low pressure medium. Tang \& Wang (2005) 
have shown that the shock velocity follows a Sedov solution but quickly 
deviates from it when it becomes mildly supersonic. This translates into a much 
faster evolution and much larger cavities would be expected if the SN exploded 
$4\times10^4$ and $9\times10^4$~yr ago, when the quintuplet cluster was at 
that location. It is more likely, therefore, that the two cavities to the Galactic west 
of the Quintuplet might have been generated by supernova explosions of massive 
stars either stripped from the Quintuplet cluster (Habibi et al. 2013) or having 
no association with it. 

The large thermal energy filling the arc superbubble could have been produced 
by some combination of winds from the young stars and by multiple supernova 
explosions, including the supernova 
explosion associated with the PWN G0.13-0.11\footnote{We note that the 
PWN~G0.13-0.11 is located right in the middle of the mid-IR arc bubble.}. 

\subsection{A hot atmosphere around the GC - A link to the GC lobe? 
And to the \fermi\ bubbles?}
\label{Fermi}

\subsubsection{General morphology}

As observed in all soft X-ray maps (Figs.~\ref{RGBhard}, \ref{RGBSoftLines1} and 
\ref{RGBSoftLines2}) and confirmed by the soft plasma 
intensity map (obtained through spectral-images decomposition, Fig.~\ref{Terrier}), 
the regions at higher Galactic latitudes are significantly brighter in soft X-rays 
than the regions closer to the disc, presumably in part because of the smaller 
extinction at the higher latitudes. 

The western border of this enhanced emission is defined by a relatively sharp edge 
between $l=359.63^\circ$, $b=0.06^\circ$ and $l=359.55^\circ$, 
$b=0.46^\circ$ (Fig.\ \ref{Terrier}, see also Figs. \ref{RGBSoftLines1} and 
\ref{RGBSoftLines2}). The soft X-ray emission peaks above the location of the 
GC Radio Arc, appearing as a continuation of the Radio Arc itself. 
Further west the soft X-ray emission appears to fade with increasing Galactic longitude. 
In particular, the spectral decomposition provides hints for the presence of an 
eastern edge, fainter but similar to the western edge, of this hot GC atmosphere. 
However the presence of an edge to the west is less obvious because of the 
soft X-ray extinction likely caused by a series of molecular clouds (e.g., the Brick) 
present at that location and because of the partial high-latitude coverage of this region. 

\subsubsection{Radiative process}

The presence of intense, soft X-ray emission lines (see Figs.\ \ref{RGBSoftLines1} 
and \ref{RGBSoftLines2}) and, in particular, the good fit of a spectral 
decomposition based on a thermally emitting gas (see Fig.\ \ref{Terrier}) indicate 
that most of the high-latitude emission is generated by a thermal radiative process
in a warm plasma. To demonstrate this, we accumulate the EPIC-pn spectrum 
from a circular region of 8.28~arcmin radius centered at $l=0.181^\circ$, 
$b=0.359^\circ$.  The resulting spectrum is well fitted with an absorbed thermal 
emission component ({\sc apec}) with $kT=0.96\pm0.1$~keV, $N_H=(2.3\pm0.2)\times10^{22}$~cm$^{-2}$ 
and A$_{\rm apec}=(1.5\pm0.4)\times10^{-2}$~cm$^{-5}$. 

\subsubsection{Eastern edge}

The eastern edge of the high-latitude emission rises from the position of the Radio Arc. 
This raises the interesting question of whether the soft X-ray emission 
at the location of the Radio Arc might have two contributions, one associated with 
the G0.1-0.1 superbubble (filling the mid-IR arc bubble;  see \S~\ref{G0.1-0.1}), 
while the second is associated with enhanced soft X-ray emission due to the 
presence of the Radio Arc and its polarized radio plumes at higher latitudes
(Seiradakis et al. 1985; Tsuboi et al. 1986; Yusef-Zadeh \& Morris 1988). 
If that is indeed the case and if the two structures have different X-ray colours 
(e.g. the superbubble produces lower temperature thermal X-ray lines, while the 
Radio Arc has a larger continuum to lines ratio), then we should observe 
variations in the X-ray colour distribution. In particular, we would expect 
a whiter colour and a green-yellow colour (similar to the one characterising the 
lobes of Sgr~A) at the location of the Radio Arc region compared to G0.1-0.1, 
in the top panel of Fig. \ref{RGBSoftLines1} and in Fig. \ref{RGBSoftLines2}, respectively.  
This idea is, indeed, in agreement with the colour variations and the evolution of the 
line intensities observed between the G0.1-0.1 and Radio Arc complexes 
(Fig.\ \ref{RGBSoftLines1} and \ref{RGBSoftLines2}). 

\subsubsection{Western edge, the Chimney and AFGL5376} 

Running almost parallel to the western edge of the high-latitude plasma 
is another region of enhanced soft X-ray emission, located near the Galactic plane, 
the so called Chimney (l~$=359.45^\circ$; Tsuru et al. 2009). 
The Chimney appears as a column of soft X-ray emitting plasma 
extending all the way between the core of Sgr~C and the northern limit of 
the \xmm\ scan ($b\sim0.15^{\circ}$; see Fig. \ref{RGBSoftLines1}, 
\ref{RGBSoftLines2} and \ref{RGBhardNH}). 
Tsuru et al. (2009) suggested that the Chimney is an outflow emanating from 
the supernova remnant candidate G359.41-0.12. 
They estimated a thermal energy and dynamical time for G359.41-0.12 and 
the Chimney of 
$E_{th}=5.9\times10^{49}$~erg, $t_{dy}=2.4\times10^4$~yr and 
$E_{th}=7.6\times10^{49}$~erg, $t_{dy}=4\times10^4$~yr, respectively. 
The energetics and time-scales are consistent with typical GC supernova 
remnants and Tsuru et al. suggested that the very peculiar morphology 
of the outflow producing the Chimney might be due to a peculiar distribution 
of molecular clouds that block the plasma expansion in the other 
directions (Tsuru et al. 2009).  We note that the morphology of the Chimney 
resembles that of the Radio Arc. It originates near the Galactic plane (where 
dense and massive molecular clouds are located) and extends almost 
perpendicular to the Galactic plane.
Within the gap between the Chimney and the western edge (see Fig. 6), a bright 
non-thermal radio filament with an X-ray counterpart is observed: G359.54+0.18 
-- the Ripple filament, with a radio length of $0.08^\circ$
(Lu et al. 2003; Yusef-Zadeh et al.\ 2005). It is oriented parallel to the edge 
of the soft X-ray plasma distribution (Bally et al.\ 1989; 
Yusef-Zadeh et al.\ 1997; 2004; Staguhn et al.\ 1998; see also Yamauchi et al. 2014). 
Similar to the Radio Arc, other X-ray and  non-thermal radio filaments 
are observed at the base of the Chimney. The high concentration of non-thermal 
filaments indicates the importance of magnetic structures in this region 
(e.g., Morris 2014). 

The bright IR source AFGL5376 is located further to the northwest, along the 
continuation of the sharp X-ray edge and the Chimney (unfortunately just off 
the \xmm\ map; see Uchida et al.\ 1990; 1994). It is associated with high-velocity 
CO emission and defines the most prominent portion of a strong large scale 
($\sim90$~pc) shock front that extends all the way down to Sgr C (Uchida et al.\ 1994). 
Because the Chimney appears to be associated with a shock, and because it
is spatially coincident with magnetic filaments along its length (Yusef-Zadeh
et al. 2004), we suggest that it is not a simple supernova remnant, but is a 
phenomenon associated with a footpoint of a larger scale structure, the GCL. 

\subsubsection{Is the outflow confined inside the GCL? Inside the \fermi\ bubbles?}

The GC is considered a mini-starburst environment, producing intense 
outflows (Crocker 2012; Yoast-Hull et al.\ 2014). 
The warm plasma detected at high-latitudes is therefore, most probably, 
associated with intense star formation and it can be a pervasive 
atmosphere above the entire CMZ. Even in the absence of another confining
force, the gravitational potential (e.g., that of Breitschwerdt et al.\ 1991; 
Launhardt et al. 2002) would bind the $\sim1$~keV plasma (having a sound 
speed of $\sim500$~km~s$^{-1}$; Muno et al.\ 2004) to the Galaxy, but in hydrostatic 
equilibrium, would allow it to extend to heights of several hundred parsecs. 
If so, it would require a large average star formation rate and concomitant energy input 
to generate and maintain it.  

However, the detection of edges in its distribution suggests that such 
plasma might be confined within known structures.  As noted by Blanton (2008), 
the locations of the eastern and western footpoints of the GCL (Law 2011) 
correspond to the positions of the Radio Arc and the Chimney, respectively. 
Indeed, the GCL and its possible magnetic nature might confine 
the warm plasma observed in soft X-rays. 
This opens the exciting possibility that the observed high-latitude 
enhanced X-ray emission from the GC "atmosphere" is indeed 
the warm plasma filling the GCL. 
Based on the spectral fit, we deduce a density 
of $n_e=0.06$~cm$^{-3}$ inside the GCL.  Assuming uniform physical 
conditions inside the GCL (a cylinder of 45~pc radius and 160~pc height) 
and extrapolating over the entire GCL, we estimate a mass of 
$\sim4\times10^3$~M$_\odot$ filling the GCL with a total thermal energy 
of $\sim10^{52}$~erg. This value is of the same order of magnitude 
as the energy required to inflate the GCL as estimated by Law (2011).

Just after the first detection of the GCL, 
Uchida et al.\ (1985) noted its similarity with the lobes in nearby radio 
galaxies (although smaller in size and strength). The authors 
interpreted the lobes as created by a magneto-dynamic acceleration 
mechanism where the magnetic twist is produced by the rotation 
of a contracting disc of gas in the Galactic plane. Under such 
conditions, the plasma is accelerated into a conical cylinder with 
a helical velocity field (Uchida et al.\ 1985). Alternatively, 
Bland-Hawthorn \& Cohen (2003) suggested that the GCL could 
be produced by a large-scale bipolar Galactic wind, that would be 
the result of a powerful ($E=10^{54-55}$~erg) nuclear starburst 
that took place a few $10^6$~yr ago. These authors show that dust is 
associated with the entire GCL structure and they suggest that the GC 
(and the centers of many Galaxies) would drive large-scale winds 
into the halo with a recurrence time of about $10$~Myr (Bland-Hawthorn 
\& Cohen 2003). Other, alternative, scenarios for the origin of the GCL 
involve outflows associated with enhanced activity of Sgr~A$^\star$  
(Ponti et al.\ 2013) or intense star formation (Crocker et al.\ 2011; 2012).

It is not excluded that the GCL could be simply one part of an even larger 
scale feature extending over a physical scale of several kiloparsecs above 
and below the Galaxy, the so called \fermi\ bubbles (Su et al.\ 2010). 
These gamma-ray bubbles, detected with \fermi, are interpreted as 
produced by highly relativistic particles emitting brightly at GeV energies and 
beyond and they appear to 
contain and confine soft X-ray emitting plasma traced by the ROSAT all 
sky survey, from very large scales down to the Milky Way's center (Su et al.\ 2010). 
However, close to the Galactic plane, the bubbles' edges start to become 
confused. Whatever their origin might be, the \fermi\ bubbles appear to originate 
(and be collimated) from the CMZ, within the region that \xmm\ scanned 
here.

Additional \xmm\ observations at high Galactic latitudes, in particular, inside 
and at the border of the GCL, covering the AFGL~5376 source and the edges 
of the base of the \fermi\ bubbles will be needed to measure the extent of this
high-latitude emission and to help disentangle the hypotheses for its origin. 
Furthermore, higher spatial resolution observations (such as provided by \chandra) 
at high latitudes would allow one to pin down what fraction of the extended,
high-latitude X-ray emission is associated with faint point sources that are 
relatively less subject to extinction than sources near the plane. 

\subsection{Soft X-ray emission from the Sgr~D and Sgr~E regions}

Intense, soft, diffuse X-ray emission is observed from G359.12-0.05, the region 
around 1E1740.7-2942. A radio SNR (G359.07-0.02) is observed at about 
the same position (LaRosa et al.\ 2000). G359.12-0.05 has an emission spectrum 
typical of an SNR (Nakashima et al.\ 2010). In particular, the high extinction suggests 
it is located at the GC. Nakashima et al.\ (2010) suggest that G359.12-0.05 
might be associated with the great annihilator and therefore be the second system 
(such as SS433 and the radio SNR~W~50) where a BH is associated with its SNR.  

The core of the Sgr D complex is also observed to show enhanced medium 
energy emission (see Fig.\ \ref{RGBhard}). In radio, a SNR southwest of 
Sgr~D's core and H II regions are clearly observed (Fig.\ \ref{FC2}; LaRosa et 
al.\ 2000). Sawada et al.\ (2009) analysed X-ray data from the \xmm\ and \suzaku\ 
satellites and observed soft X-ray emission from two diffuse X-ray sources, DS1 and 
one associated with the core of Sgr~D. They suggest that DS1 is a new 
SNR at the GC.

\subsection{Star formation estimate from counts of supernova remnants  }

We observe a total of $\sim10-12$ supernova remnant candidates in the CMZ 
(plus $\sim5$ independent radio SNR; see Tab. \ref{TabSNR}) plus two superbubbles, 
each likely created by many (3-10) supernova events. 
These remnants have typical estimated ages of a few tens of thousands of years 
and temperatures of $kT\sim0.4-1.5$~keV. Due to the presence of the two 
superbubbles requiring multiple SNe and the high absorption towards the GC 
(such as in the star forming region Sgr~B) that hampers us from observing a 
potentially larger population of remnants characterised by lower temperatures, 
the number of SNR observed in the GC is most probably under-estimated. 
However, assuming lifetimes of 10-40 kyr, the observed number of SNR yields 
a rate, averaged over the past several thousands of years, of 
$r_{\rm SN}\sim 3.5-15\times10^{-4}$~yr$^{-1}$, consistent with other 
estimates (Crocker 2012). This implies a kinetic energy input higher 
than $1.1\times10^{40}$~erg~s$^{-1}$. 
To estimate the star formation rate, we assume that all stars with masses greater 
than 8~M$_\odot$ produce supernovae and that all SNR are observable.
Therefore, we multiply the supernova rate by the integral of the initial mass 
function (IMF) over all masses divided by the integral of the IMF above 
8~M$_{\odot}$. To reflect the GC IMF, we assume the Kroupa (2002) formulation.
The star formation rate then results to be: $r_{\rm SFR}\sim0.035-0.15$~M$_\odot$~yr$^{-1}$.  
If the IMF in the CMZ is top-heavy, as some have argued, then a smaller star formation 
rate is implied. 

As noted also by Mori et al. (2008; 2009) and Heard \& Warwick (2013), 
we observe that the two superbubbles have far hotter temperatures 
(higher density and smaller size) than all the ones observed 
in the Galactic plane or in the Large Magellanic Cloud (typically with 
temperatures $kT\sim0.1-0.3$~keV, densities $n_e\sim0.01-0.03$~cm$^{-2}$, 
sizes $l\sim140-450$~pc; but see also Sasaki et al. 2011; Kavanagh 
et al. 2012). This could simply be the consequence of the high extinction
towards the GC, hiding a population of normal and very soft 
superbubbles, or it could be a characteristic feature of GC superbubbles, 
inducing a different evolution because of the interaction with the peculiar 
GC environment. Further investigation is required to solve this problem. 

\section{Conclusions}

We have systematically analysed more than 100 \xmm\ observations pointed 
within one degree of Sgr~A$^\star$ and have created the deepest, few arcsec resolution, 
X-ray images of the CMZ. This includes a total of about $1.5$~Ms of EPIC-pn cleaned 
exposure in the central $15$ arcmin and about $200$~ks at all other points of 
the Central Molecular Zone (CMZ). 
We present here, for the first time, not only broad-band X-ray continuum maps, but 
also mosaicked maps of both soft line intensities and inter-line emission from 
the entire CMZ region. 

\begin{itemize}

\item{} The remarkably similar distributions of both the soft line emitting plasma 
(Si~{\sc xiii}, S~{\sc xv}, Ar~{\sc xvii} and Ca~{\sc xix}) and the soft continuum 
(intra-line bands) indicate that most of the diffuse soft X-ray emission arises from
a thermal process generating both continuum and lines.

\item{} Starting from the mosaic maps of the different narrow energy bands 
and assuming the GC emission is produced by three different components, 
we fit the maps at different energies and derive the integrated intensity map 
of the thermal plasma emission. Integrating over the entire 
CMZ, the total observed (un-absorbed) flux is:
$F_{2-4.5~keV}=4.2\times10^{-11}$~erg~cm$^{-2}$~s$^{-1}$, 
$F_{4.5-12~keV}=1.2\times10^{-10}$~erg~cm$^{-2}$~s$^{-1}$, 
corresponding to a luminosity of $L_{2-12}=3.4\times10^{36}$~erg~s$^{-1}$ 
at an assumed 8~kpc distance.

\item{} Counting the number of supernova remnants in the CMZ, we 
estimate a supernova rate between $r_{\rm SN}\sim3.5-15\times10^{-4}$~yr$^{-1}$, 
consistent with other estimates (Crocker 2012), that corresponds to a star 
formation rate of $r_{\rm SFR}\sim0.035-0.15$~M$_\odot$~yr$^{-1}$ 
over the past several thousand years. This implies a kinetic energy input 
greater than $1.5\times10^{40}$~erg~s$^{-1}$. 

\item{} We report the discovery of a new X-ray filament XMM~J0.173-0.413 
perpendicular to the 
Galactic plane and south of the GC Radio Arc spatially corresponding to a 
non-thermal radio filament. XMM~J0.173-0.413 is the first of the four cases 
known where the X-ray emission is not at or near a location where the radio 
filaments show unusually strong curvature. 

\item{} The soft GC X-ray emission is absorbed not only by high-column-density 
foreground clouds located in the Galactic disk, but also by some clouds located
on the near side of the CMZ, such as the core and 
envelope of Sgr~B2, M0.25+0.01 (the "Brick"), and even a few clouds at higher 
Galactic latitudes, M0.18+0.126 and M0.20-0.48. 
However, the majority of the observed variations in the soft X-ray emission 
are true emissivity modulations and not a product of absorption. 

\item{} Several SNR candidates are identified by their soft X-ray emission that appears 
to fill holes in the column density distribution of gas-and-dust derived from 
submillimeter maps. 

\item{} Our data shed new light on two quasi-symmetric lobes situated to Galactic 
north and south of Sgr~A$^\star$. The Northern lobe shows a bright and sharp transition 
at its edge, suggesting the presence of a shock.
Such features are possibly the remnant of the SN that generated SGR~J1745-2900 
or the PWN candidate G359.945-0.044. 
Alternatively, the lobes might constitute a long-lived bipolar structure produced by 
an isotropic outflow produced by either 1) the cumulative winds from the young 
stars of the central cluster, 2) a wind associated with the accretion flow onto \sgras, 
or 3) the same process that generated the X-ray reflection nebulae (if such 
activity has been recurrent over the past millennia). 

\item{} The uniform X-ray colour of the superbubble G359.9-0.125, its sharp 
external edge and its being filled with S~{\sc xv} emitting plasma suggest that 
the soft X-ray features southwest of Sgr~A$^\star$ form a unique shell-like 
structure with total energy $E_{th}\sim10^{51}$~erg, therefore making it a 
superbubble candidate in the GC (high absorption indicates that 
G359.9-0.125 is located at the GC). Alternatively, it might be the remnant 
of a very energetic event at the GC, such as a tidal disruption event. 

\item{} We discover new evidence for the GC superbubble G0.1-0.1, also known 
as the arc-bubble from mid-IR observations: its soft X-ray (e.g. S~{\sc xv}) emission 
completely fills the mid-IR bubble, and indicates a thermal energy as large as 
$E_{th}\sim1.5\times10^{51}$~erg.
At present the Quintuplet cluster, which is moving at very high speed through 
the CMZ, is located at the border of the superbubble. 
However, it was more centrally located a few $10^4$~yr ago and it could 
have, at least in part, energised it. 
We do not observe similar soft X-ray emission trailing the Arches cluster, 
but this might be ascribed to its younger age. 

\item{} We suggest that the Galactic Center Lobe might be a magnetic 
structure filled with warm, soft, X-ray-emitting plasma. In fact, we observe: 
i) enhanced soft X-ray emission at high Galactic latitudes; 
ii) enhanced soft X-ray emission at, and between, the longitudes of the 
Radio Arc and the Chimney associated with Sgr C, corresponding 
to the east and west foot-points of the GCL; iii) a sharp edge 
(at $l=359.63^\circ, b=0.06^\circ$ and $l=359.55^\circ, b=0.46^\circ$), 
running parallel to the nonthermal ripple filament (G359.54+0.18) and Sgr~C thread, 
defining the western border of the enhanced soft X-ray emission. 
The GCL could be the relatively small base of an even larger structure, the 
so-called \fermi\ Bubbles. Additional observations will be needed to clarify this.

\item{} A new very faint X-ray transient, XMMU~J17450.3-291445, 
has been discovered during the new \xmm\ campaign to reach a peak luminosity 
of $L_X\sim10^{35}$~erg~s$^{-1}$ for $\sim2$~hr (Soldi et al. 2014). 

\end{itemize}

\section*{Acknowledgments}

This research has made use both of data obtained with \xmm, an ESA 
science mission with instruments and contributions directly funded by ESA Member 
States and NASA, and data obtained from the Chandra Data Archive.
We kindly acknowledge Sergio Molinari for providing the Herschel map, Casey Law 
for the GBT images and Namir Kassim for the VLA 90-cm map.
GP acknowledges Roland Crocker, Barbara De Marco and Pierre Maggi for useful 
discussions. GP also acknowledges Frederick Baganoff and Nanda Rea for discussions 
about the origin of the lobes and the association with the SNR of SGR~J1745-2900. 
We thank the referee for a careful reading of the paper. 
GP acknowledges support via an EU Marie Curie Intra-European Fellowship under 
contract no. FP7-PEOPLE-2012-IEF-331095. 
The GC \xmm\ monitoring project is partially supported by the Bundesministerium 
f\"{u}r Wirtschaft und Technologie/Deutsches Zentrum f\"{u}r Luft- und Raumfahrt 
(BMWI/DLR, FKZ 50 OR 1408) and the Max Planck Society. 
Partial support through the COST action MP0905 Black Holes in a Violent 
Universe is acknowledged. The authors thank the ISSI in Bern. 

\begin{table*}
\begin{center}
\tiny
\begin{tabular}{ | c c c c c c c c c c c c c c c c c }
\hline
                      &          & \multicolumn{2}{c}{EPIC-pn} & \multicolumn{2}{c}{EPIC-MOS1}& \multicolumn{2}{c}{EPIC-MOS2} & \multicolumn{2}{c}{EPIC-pn} & \multicolumn{2}{c}{EPIC-MOS1}& \multicolumn{2}{c}{EPIC-MOS2} &  \multicolumn{3}{c}{Threshold} \\ 
OBSID           & rev    & Exp & Exp & Exp & Exp & Exp & Exp &mode& filter&mode& filter&mode& filter & pn & M1 & M2 \\
                     &           & Mod & ID & Mod & ID & Mod & ID &                  &         &          &         &           &         & c/s & c/s & c/s \\
\hline
\multicolumn{14}{c}{{\bf NEW CMZ \xmm\ scan}} \\
0694640101 & 2335 & S & 003 & S & 001 & S & 002 &  FF & med &   FF & med &   FF & med & 6.0 & 2.0 & 2.0 	 \\  
0694640201 & 2335 & S & 003 & S & 001 & S & 002 &  FF & med &   FF & med &   FF & med & 6.0 & 2.0 & 2.0 	 \\   
0694640301 & 2335 & S & 003 & S & 001 & S & 002 &  FF & med &   FF & med &   FF & med & 6.0 & 2.0 & 2.0   \\   
0694640401 & 2335 & S & 003 & S & 001 & S & 002 &  FF & med &   FF & med &   FF & med & 8.0 & 2.5 & 2.5   \\   
0694640501 & 2335 & S & 003 & S & 001 & S & 002 &  FF & med &   FF & med &   FF & med & 6.0 & 2.0 & 2.0   \\   
0694640601 & 2335 & S & 003 & S & 001 & S & 002 &  FF & med &   FF & med &   FF & med & 6.0 & 2.0 & 2.0   \\   
0694640701 & 2335 & S & 003 & S & 001 & S & 002 &  FF & med &   FF & med &   FF & med & 6.0 & 2.0 & 2.0   \\   
0694640801 & 2335 & S & 003 & S & 001 & S & 002 &  FF & med &   FF & med &   FF & med & 6.0 & 2.0 & 2.0   \\   
0694640901 & 2335 & S & 003 & S & 001 & S & 002 &  FF & med &   FF & med &   FF & med & 6.0 & 2.0 & 2.0  \\
0694641001 & 2335 & S & 003 & S & 001 & S & 002 &  FF & med &   FF & med &   FF & med & 6.0 & 2.0 & 2.0   \\
0694641101 & 2335 & S & 003 & S & 001 & S & 002 &  FF & med &   FF & med &   FF & med & 6.0 & 2.0 & 2.0   \\
0694641201 & 2335 & S & 003 & S & 001 & S & 002 &  FF & med &   FF & med &   FF & med & 6.0 & 2.0 & 2.0   \\
0694641301 & 2335 & S & 003 & S & 001 & S & 002 &  FF & med &   FF & med &   FF & med & 6.0 & 2.0 & 2.0   \\
0694641401 & 2335 & S & 003 & S & 001 & S & 002 &  FF & med &   FF & med &   FF & med & 6.0 & 2.0 & 2.0   \\
0694641501 & 2335 & S & 003 & S & 001 & S & 002 &  FF & med &   FF & med &   FF & med & 6.0 & 2.0 & 2.0   \\
0694641601 & 2335 & S & 003 & S & 001 & S & 002 &  FF & med &   FF & med &   FF & med & 6.0 & 2.0 & 2.0   \\
\hline
\multicolumn{14}{c}{{\bf OLD CMZ \xmm\ scan}} \\
 0112970101 & 0145 & U & 002 & U & 002 & U & 002 &  FF  & med &  FF & med &   FF & med & 6.0 & 2.0 & 2.0  \\   
 0112970201 & 0145 & S & 003  & S & 001 & S & 002 &  eFF & med &  FF & med &   FF & med & 6.0&  2.0&  2.0  \\   
 0112970401 & 0143 & S & 003 & S & 001 & S & 002 &  eFF & med &  FF & med &   FF & med & 6.0 & 2.0 & 2.0 \\   
 0112970501 & 0144 & S & 003 & S & 001 & S & 002 &  eFF & med &  FF & med &   FF & med & 6.0 & 2.0 & 2.0 \\   
 0112970701 & 0139 & S & 003 & S & 001 & S & 002 &  eFF & med &  FF & med &   FF & med & 6.0 & 2.0 & 2.0 \\   
 0112970801 & 0144 & S & 003 & S & 001 & S & 002 &  eFF & med &  FF & med &   FF & med & 6.0 & 1.5 & 1.5 \\   
 0112971001 & 0145 & S & 003 & S & 001 & S & 002 &  FF  & tck &  FF & med &   FF & med & 5.0 & 1.5 & 1.5   \\   
 0112971301 & 0143 & S & 003 & S & 001 & S & 408 &  SW  & med &  TU & med &   RF & med & 6.0 & 1.5 & 1.5 \\   
 0112971501 & 0240 & S & 003 & S & 001 & S & 002 &  eFF & med &  FF & med &   FF & med & 6.0 & 2.0 & 2.0 \\   
 0112971601 & 0240 & S & 003 & S & 001 & S & 002 &  eFF & med &  FF & med &   FF & med & 6.0 & 2.0 & 2.0 \\   
 0112971701 & 0240 & S & 003 & S & 011 & S & 010 &  SW  & med &  TU & med &   SW & med & 6.0 & 2.0 & 2.0 \\   
 0112971801 & 0240 & S & 003 & S & 001 & S & 002 &  eFF & med &  FF & med &   FF & med & 6.0 & 2.0 & 2.0 \\   
 0112971901 & 0240 & S & 003 & S & 001 & S & 002 &  eFF & med &  FF & med &   FF & med & 5.0 & 1.5 & 1.5 \\   
 0112972101 & 0318 & S & 003 & S & 001 & S & 002 &  eFF & med &  FF & med &   FF & med & 6.0 & 1.5 & 1.5 \\   
										       
\hline
\multicolumn{14}{c}{{\bf Pointing toward Sgr~A$^{\star}$}} \\
\multicolumn{14}{c}{{\it 2002}} \\
 0111350101 & 0406 & U & 002 & S & 006 & S & 005 &  FF &  tck &  FF & med  &  FF & med  &  6.0  & 2.0  & 2.0	 \\   
 0111350301 & 0516 & S & 001 & S & 006 & S & 005 &  FF &  tck &  FF & med  &  FF & med  &  6.0  & 2.0  & 2.0	 \\   
\multicolumn{14}{c}{{\it 2004}} \\					     	                              
 0202670501 & 0788 & U & 002 & U & 003 & U & 003 &  eF &F med &  FF & med  &  FF & med  &  6.0  & 2.0  & 2.0	 \\   
 0202670601 & 0789 & S & 003 & S & 001 & S & 002 &  eF &F med &  FF & med  &  FF & med  &  6.0  & 2.0  & 2.0	 \\   
 0202670701 & 0866 & S & 003 & S & 001 & S & 002 &  FF &  med &  FF & med  &  FF & med  &  6.0  & 2.0  & 2.0	 \\   
 0202670801 & 0867 & S & 003 & S & 001 & S & 002 &  FF &  med &  FF & med  &  FF & med  &  6.0  & 1.5  & 1.5	 \\   
\multicolumn{14}{c}{{\it 2006}} \\					     	                              
 0302882601 & 1139 & S & 003 & S & 001 & S & 002 &  FF &  med &  FF & med  &  FF & med  &  6.0  & 2.0  & 2.0	 \\   
 0302884001 & 1236 & S & 003 & S & 001 & S & 002 &  FF &  med &  FF & med  &  FF & med  &  7.0  & 2.0  & 2.0	 \\   
\multicolumn{14}{c}{{\it 2007}} \\					     	                              
 0402430301 & 1339 & S & 001 & S & 002 & S & 003 &  FF &  med &  FF & med  &  FF & med  &  8.0  & 2.0  & 2.0   \\   
 0402430401 & 1340 & U & 002 & U & 002 & U & 002 &  FF &  med &  FF & med  &  FF & med  &  8.0  & 2.0  & 2.0   \\   
 0402430701 & 1338 & S & 001 & S & 002 & S & 003 &  FF &  med &  FF & med  &  FF & med  &  8.0  & 2.0  & 2.0   \\   
 0504940201 & 1418 & S & 003 & S & 001 & S & 002 &  FF &  med &  FF & med  &  FF & med  &  8.0  & 2.0  & 2.0   \\   
\multicolumn{14}{c}{{\it 2008}} \\					     	                              
 0511000301 & 1508 & S & 003 & S & 001 & S & 002 &  FF &  thn &  FF & thn  &  FF & thn  &  7.0  & 2.0  & 2.0   \\   
 0511000401 & 1610 & S & 003 & U & 002 & U & 002 &  FF &  thn &  FF & thn  &  FF & thn  &  7.0  & 2.5  & 2.5   \\   
 0505670101 & 1518 & U & 002 & U & 002 & U & 002 &  FF &  med &  FF & med  &  FF & med  &  8.0  & 2.0  & 2.0   \\   
\multicolumn{14}{c}{{\it 2009}} \\					     	                              
 0554750401 & 1705 & S & 003 & S & 001 & S & 002 &  FF &  med &  FF & med  &  FF & med  &  8.0  & 2.5  & 2.5   \\   
 0554750501 & 1706 & S & 003 & S & 001 & S & 002 &  FF &  med &  FF & med  &  FF & med  &  8.0  & 2.5  & 2.5   \\   
 0554750601 & 1707 & U & 002 & S & 001 & S & 002 &  FF &  med &  FF & med  &  FF & med  &  8.0  & 2.5  & 2.5   \\   
\multicolumn{14}{c}{{\it 2011}} \\					     	                              
 0604300601 & 2069 & S & 003 & S & 001 & S & 002 &  FF &  med &  FF & med  &  FF & med  &  7.0  & 2.0  & 2.0   \\   
 0604300701 & 2070 & U & 002 & S & 001 & S & 002 &  FF &  med &  FF & med  &  FF & med  &  7.0  & 2.0  & 2.0   \\   
 0604300801 & 2071 & U & 002 & U & 002 & U & 002 &  FF &  med &  FF & med  &  FF & med  &  8.0  & 2.0  & 2.0  \\   
 0604300901 & 2072 & S & 003 & S & 001 & S & 002 &  FF &  med &  FF & med  &  FF & med  &  7.0  & 2.0  & 2.0   \\   
 0604301001 & 2073 & S & 003 & S & 001 & S & 002 &  FF &  med &  FF & med  &  FF & med  &  7.0  & 2.0  & 2.0   \\   
 0658600101 & 2148 & S & 001 & S & 002 & S & 003 &  FF &  med &  FF & med  &  FF & med  &  8.0  & 2.5  & 2.5   \\   
 0658600201 & 2148 & S & 001 & S & 002 & S & 003 &  FF &  med &  FF & med  &  FF & med  &  7.0  & 1.8  & 1.8   \\   
\multicolumn{14}{c}{{\it 2012}} \\					     	                              
 0674600601 & 2245 & S & 003 & S & 001 & S & 002 &  FF &  med &  FF & med  &  FF & med  &  7.0  & 2.0  & 2.0   \\   
 0674600701 & 2246 & S & 003 & S & 001 & S & 002 &  FF &  med &  FF & med  &  FF & med  &  7.0  & 2.0  & 2.0   \\   
 0674600801 & 2248 & S & 003 & S & 001 & S & 002 &  FF &  med &  FF & med  &  FF & med  &  6.0  & 1.8  & 1.8   \\   
 0674601001 & 2249 & S & 003 & S & 001 & S & 002 &  FF &  med &  FF & med  &  FF & med  &  7.0  & 1.8  & 1.8   \\   
 0674601101 & 2247 & S & 003 & U & 002 & U & 002 &  FF &  med &  FF & med  &  FF & med  &  6.0  & 2.0  & 2.0   \\   
\hline
\end{tabular}
\caption{List of all \xmm\ observations considered in this work. 
Exposure Mode: U, S stand for unscheduled and scheduled, respectively. 
Filters: Med, thn, tck stand for medium, thin and thick filters, respectively. 
FF, eFF, SW, Ti, TU stand for full frame, extended full frame, small window, timing and 
time uncompressed, respectively.}
\label{TabObs1}
\end{center}
\end{table*} 

\begin{table*}
\begin{center}
\tiny
\begin{tabular}{ | c c c c c c c c c c c c c c c c c }
\hline
                      &          & \multicolumn{2}{c}{EPIC-pn} & \multicolumn{2}{c}{EPIC-MOS1}& \multicolumn{2}{c}{EPIC-MOS2} & \multicolumn{2}{c}{EPIC-pn} & \multicolumn{2}{c}{EPIC-MOS1}& \multicolumn{2}{c}{EPIC-MOS2} &  \multicolumn{3}{c}{Threshold} \\ 
OBSID           & rev    & Exp & Exp & Exp & Exp & Exp & Exp &mode& filter&mode& filter&mode& filter & pn & M1 & M2 \\
                     &           & Mod & ID & Mod & ID & Mod & ID &                  &         &          &         &           &         & c/s & c/s & c/s \\
\hline
\multicolumn{14}{c}{{\bf Other observations of the CMZ}} \\
 0030540101 & 0504 & S & 003 & S & 001 & S & 002 &  SW &  tck &  SW & tck &   SW & tck  &  6.0  & 2.0  & 2.0 \\  
 0144220101 & 0596 & U & 002 & U & 002 & U & 002 &  SW &  med &  FF & med &   FF & med  &  6.0  & 1.5  & 1.5 \\  
 0152920101 & 0607 & S & 003 & S & 001 & S & 002 &  FF &  tck &  FF & tck &   FF & tck  &  6.0  & 1.5  & 1.5 \\  
 0144630101 & 0688 & S & 003 & S & 001 & S & 002 &  SW &  med &  TU & med &   SW & med  &       &      &     \\  
 0203930101 & 0868 & S & 003 & S & 001 & S & 002 &  eF &F med &  FF & med &   FF & med  &  6.0  & 1.5  & 1.5 \\  
 0205240101 & 0956 & S & 003 & S & 001 & S & 002 &  FF &  med &  FF & med &   FF & med  &  6.0  & 1.5  & 1.5 \\  
 0304220301 & 1048 & S & 004 & S & 002 & S & 003 &  SW &  med &  FF & med &   FF & med  &  6.0  & 1.5  & 1.5 \\  
 0304220101 & 1063 & S & 003 & S & 001 & S & 002 &  SW &  med &  FF & med &   FF & med  &  6.0  & 1.5  & 1.5 \\  
 0303210201 & 1065 & S & 003 & S & 001 & S & 002 &  SW &  med &  TU & med &   TU & med  &  6.0  & 1.5  & 1.5 \\  
 0302882501 & 1139 & S & 003 & S & 001 & S & 002 &  FF &  med &  FF & med &   FF & med  &  6.0  & 2.0  & 2.0 \\  
 0302882701 & 1139 & S & 003 & S & 001 & S & 002 &  FF &  med &  FF & med &   FF & med  &  6.0  & 2.0  & 2.0 \\  
 0302882801 & 1139 & S & 003 & S & 001 & S & 002 &  FF &  med &  FF & med &   FF & med  &  6.0  & 2.0  & 2.0 \\  
 0302882901 & 1139 & S & 003 & S & 001 & S & 002 &  FF &  med &  FF & med &   FF & med  &  6.0  & 2.0  & 2.0 \\  
 0302883001 & 1139 & S & 003 & S & 001 & S & 002 &  FF &  med &  FF & med &   FF & med  &  6.0  & 2.0  & 2.0 \\  
 0302883101 & 1139 & S & 003 & S & 001 & S & 002 &  FF &  med &  FF & med &   FF & med  &  6.0  & 2.0  & 2.0 \\  
 0302883201 & 1139 & S & 003 & S & 001 & S & 002 &  FF &  med &  FF & med &   FF & med  &  6.0  & 2.0  & 2.0 \\  
 0305830701 & 1157 & S & 003 & S & 001 & S & 002 &  FF &  med &  FF & med &   FF & med  &  6.0  & 1.5  & 1.5 \\  
 0302883901 & 1236 & S & 003 & S & 001 & S & 002 &  FF &  med &  FF & med &   FF & med  &  6.0  & 2.0  & 2.0 \\  
0302884101 & 1236 & S & 003 & S & 001 & S & 002 &  FF &  med &  FF & med &   FF & med  &  6.0  & 2.0  & 2.0 \\  
0302884201 & 1236 & S & 003 & S & 001 & S & 002 &  FF &  med &  FF & med &   FF & med  &  6.0  & 2.0  & 2.0 \\  
0302884301 & 1236 & S & 003 & S & 001 & S & 002 &  FF &  med &  FF & med &   FF & med  &  6.0  & 2.5  & 2.5 \\  
0302884401 & 1236 & S & 003 & S & 001 & S & 002 &  FF &  med &  FF & med &   FF & med  &  6.0  & 2.0  & 2.0 \\  
0302884501 & 1236 & S & 003 & S & 001 & S & 002 &  FF &  med &  FF & med &   FF & med  &  6.0  & 2.0  & 2.0 \\  
0406580201 & 1241 & S & 003 & S & 001 & S & 002 &  FF &  med &  FF & med &   FF & med  &  6.0  & 2.0  & 2.0 \\  
0410580401 & 1243 & N & 000 & S & 002 & S & 003 &  Ti &  tck &  FF & tck &   TU & tck  &  6.0  & 2.0  & 2.0 \\  
0410580501 & 1245 & N & 000 & S & 002 & S & 003 &  Ti &  tck &  FF & tck &   TU & tck  &  6.0  & 2.0  & 2.0 \\  
0400340101 & 1244 & S & 003 & S & 001 & S & 002 &  FF &  med &  FF & med &   FF & med  &  6.0  & 2.0  & 2.0 \\  
0506291201 & 1322 & N & 000 & S & 001 & S & 002 &  Ti &  med &  FF & med &   FF & med  &  6.0  & 2.0  & 2.0 \\  
0504940101 & 1418 & S & 003 & S & 001 & S & 002 &  FF &  med &  FF & med &   FF & med  &  6.0  & 2.0  & 2.0 \\  
0504940401 & 1418 & S & 003 & S & 001 & S & 002 &  FF &  med &  FF & med &   FF & med  &  6.0  & 2.0  & 2.0 \\  
0504940501 & 1418 & S & 003 & S & 001 & S & 002 &  FF &  med &  FF & med &   FF & med  &  6.0  & 2.0  & 2.0 \\  
0504940601 & 1418 & S & 003 & S & 001 & S & 002 &  FF &  med &  FF & med &   FF & med  &  7.0  & 2.0  & 2.0 \\  
0504940701 & 1418 & S & 003 & S & 001 & S & 002 &  FF &  med &  FF & med &   FF & med  &  6.0  & 2.0  & 2.0 \\  
0511010701 & 1505 & S & 003 & S & 001 & S & 002 &  FF &  med &  FF & med &   FF & med  &  6.0  & 2.5  & 2.5 \\  
0511000101 & 1508 & S & 003 & S & 001 & S & 002 &  FF &  thn &  FF & thn &   FF & thn  &  6.0  & 2.0  & 2.0 \\  
0511000501 & 1508 & S & 003 & S & 001 & S & 002 &  FF &  thn &  FF & thn &   FF & thn  &  6.0  & 2.0  & 2.0 \\  
0511000701 & 1508 & S & 003 & S & 001 & S & 002 &  FF &  thn &  FF & thn &   FF & thn  &  6.0  & 2.0  & 2.0 \\  
0511000901 & 1508 & S & 003 & S & 001 & S & 002 &  FF &  thn &  FF & thn &   FF & thn  &  6.0  & 2.5  & 2.5 \\  
0511001101 & 1508 & S & 003 & S & 001 & S & 002 &  FF &  thn &  FF & thn &   FF & thn  &  6.0  & 2.5  & 2.5 \\  
0511001301 & 1508 & S & 003 & S & 001 & S & 002 &  FF &  thn &  FF & thn &   FF & thn  &  6.0  & 2.0  & 2.0 \\  
0511000201 & 1510 & S & 003 & S & 001 & S & 002 &  FF &  thn &  FF & thn &   FF & thn  &  6.0  & 2.5  & 2.5 \\  
0511000601 & 1510 & S & 003 & S & 001 & S & 002 &  FF &  thn &  FF & thn &   FF & thn  &  6.0  & 2.5  & 2.5 \\  
0511000801 & 1512 & S & 003 & S & 001 & S & 002 &  FF &  thn &  FF & thn &   FF & thn  &  6.0  & 2.5  & 2.5 \\  
0511001001 & 1512 & S & 003 & S & 001 & S & 002 &  FF &  thn &  FF & thn &   FF & thn  &  6.0  & 2.5  & 2.5 \\  
0511001201 & 1512 & S & 003 & S & 001 & S & 002 &  FF &  thn &  FF & thn &   FF & thn  &  8.0  & 2.5  & 2.5 \\  
0511001401 & 1512 & S & 003 & S & 001 & S & 002 &  FF &  thn &  FF & thn &   FF & thn  &  6.0  & 2.5  & 2.5 \\  
0505870301 & 1511 & S & 003 & S & 001 & S & 002 &  FF &  med &  FF & med &   FF & med  &  6.0  & 2.0  & 2.0 \\  
0603850201 & 1891 & S & 003 & S & 001 & S & 002 &  FF &  med &  FF & med &   FF & med  &  6.0  & 2.0  & 2.0 \\  
0655670101 & 2065 & N & 000 & S & 001 & S & 002 &  Ti &  med &  FF & med &   FF & med  &  6.0  & 2.0  & 2.0 \\  
\hline
\end{tabular}
\caption{List of all \xmm\ observations considered in this work. 
Exposure Mode: U, S stand for unscheduled and scheduled, respectively. 
Filters: Med, thn, tck stand for medium, thin and thick filters, respectively. 
FF, eFF, SW, Ti, TU stand for full frame, extended full frame, small window, timing and 
time uncompressed, respectively.}
\label{TabObs2}
\end{center}
\end{table*}

\begin{table*}
\begin{center}
\tiny
\begin{tabular}{ | c c c c c c c c c c c c c c c c c }
\hline
OBSID           & obs date & Exp pn & Exp M1 & Exp M2 & Exp pn & Exp M1 & Exp M2 \\
\hline
\multicolumn{8}{c}{{\bf NEW CMZ \xmm\ scan}} \\
0694640101 &  2012-09-07  &  41978  &  43452 &   43605 &   38739 &   38980  &  38985 \\
0694640201 &  2012-08-30  &  45035  &  46616 &   46619 &   45038 &   46616  &  46619 \\
0694640301 &  2012-08-31  &  40041  &  41616 &   41619 &   40041 &   41616  &  41619 \\
0694640401 &  2012-09-02  &  52954  &  51442 &   51460 &   38736 &   40073  &  40075 \\
0694640501 &  2012-09-05  &  44976  &  46606 &   46621 &   32935 &   33180  &  33185 \\
0694640601 &  2012-09-06  &  40042  &  41614 &   41621 &   40042 &   41614  &  41621 \\
0694640701 &  2012-10-02  &  42539  &  44099 &   44120 &   42539 &   44117  &  44120 \\
0694640801 &  2012-10-06  &  40041  &  41616 &   41619 &   40041 &   41616  &  41619 \\
0694640901 &  2012-09-12  &  43031  &  44617 &   44604 &   42202 &   43784  &  43786 \\
0694641001 &  2012-09-23  &  46021  &  47607 &   47620 &   46041 &   47614  &  47620 \\
0694641101 &  2012-09-24  &  40041  &  41616 &   41619 &   40041 &   41616  &  41619 \\
0694641201 &  2012-09-26  &  40008  &  41559 &   41577 &   40008 &   41588  &  41598 \\
0694641301 &  2012-09-26  &  53842  &  56260 &   56348 &   46667 &   48012  &  48018 \\
0694641401 &  2012-09-30  &  45816  &  46751 &   46920 &   32466 &   33767  &  33770 \\
0694641501 &  2012-10-06  &  49746  &  51483 &   51486 &   39167 &   40518  &  40507 \\
0694641601 &  2012-10-08  &  40005  &  41585 &   41585 &   27250 &   27795  &  27803 \\
\hline
\multicolumn{8}{c}{{\bf OLD CMZ \xmm\ scan}} \\
0112970101 &  2000-09-23  &  12870  &  15806 &   15611 &   12252 &   14679  &  14637 \\
0112970201 &  2000-09-23  &  13499  &  17394 &   17392 &   12999 &   16894  &  16892 \\
0112970401 &  2000-09-19  &  25411  &  29365 &   29391 &   21880 &   23849  &  23847 \\
0112970501 &  2000-09-21  &  21119  &  24914 &   24911 &   10289 &   14084  &  14081 \\
0112970701 &  2000-09-11  &  19518  &  23419 &   23413 &   19383 &   23221  &  23218 \\
0112970801 &  2000-09-21  &  19969  &  23892 &   23892 &   13462 &   17198  &  17198 \\
0112971001 &  2000-09-24  &  12599  &  16492 &   16482 &    8774 &   12529  &  12529 \\
0112971301 &  2000-09-19  &  12800  &      0 &   13091 &       0 &       0  &      0 \\
0112971501 &  2001-04-01  &  20293  &  25020 &   25017 &    6752 &    7017  &   7017 \\
0112971601 &  2001-03-31  &      0  &   3996 &    3949 &       0 &       0  &      0 \\
0112971701 &  2001-03-31  &  11000  &      0 &   11799 &       0 &       0  &      0 \\
0112971801 &  2001-04-01  &   9927  &  14513 &   14542 &    1900 &    2069  &   2069 \\
0112971901 &  2001-04-01  &   4698  &   9191 &    9191 &    4147 &    8379  &   8379 \\
0112972101 &  2001-09-04  &  21687  &  26039 &   26055 &   20130 &   23515 2&3517 \\
\hline
\multicolumn{8}{c}{{\bf Pointing toward Sgr~A$^{\star}$}} \\
0111350101 &  2002-02-26 &   40030 &   52105 &   52120 &   40030 &   52118 &   52120 \\ 
0111350301 &  2002-10-03 &   15377 &   16960 &   16996 &    8261 &    9877 &    9880 \\  
\multicolumn{8}{c}{{\it 2004}} \\					     	                              
0202670501 &  2004-03-28 &  110170 &    5733 &    6087 &   45847 &       0 &       0 \\
0202670501 &  2004-03-28 &       0 &  107784 &  108572 &       0 &       0 &       0 \\
0202670501 &  2004-03-30 &       0 &     650 &     848 &       0 &       0 &       0 \\
0202670601 &  2004-03-30 &  112204 &     585 &     538 &   56926 &       0 &       0 \\
0202670601 &  2004-03-30 &       0 &  120863 &  122251 &       0 &       0 &       0 \\
0202670701 &  2004-08-31 &  127470 &  132469 &  132503 &   78857 &   78921 &   78915 \\
0202670801 &  2004-09-02 &  130951 &  132997 &  133036 &   91795 &   93131 &   93126 \\
\multicolumn{8}{c}{{\it 2006}} \\					     	                              
0302882601 &  2006-02-27 &    4937 &    6563 &    6568 &    1700 &    3160 &    3163 \\
0302884001 &  2006-09-08 &    4987 &    6563 &    6570 &    4787 &    6365 &    6370 \\
\multicolumn{8}{c}{{\it 2007}} \\					     	                              
0402430301 &  2007-04-01 &  101319 &   93947 &   94022 &   50962 &   50955 &   50958 \\
0402430401 &  2007-04-03 &   93594 &   97566 &   96461 &   36886 &   36892 &   36876 \\ 
0402430701 &  2007-03-30 &   32338 &   33912 &   33917 &   21240 &   22820 &   22825 \\
0504940201 &  2007-09-06 &   11092 &   12649 &   12652 &    7392 &    8949 &    8960 \\
\multicolumn{8}{c}{{\it 2008}} \\					     	                              
0511000301 &  2008-03-03 &    5057 &    6615 &    6620 &    3305 &    4863 &    4868 \\
0511000401 &  2008-09-23 &    5058 &    4358 &    4342 &    5058 &    4358 &    4342 \\
0505670101 &  2008-03-23 &   96601 &   97787 &   97787 &   64200 &   65143 &   65153 \\
\multicolumn{8}{c}{{\it 2009}} \\					     	                              
0554750401 &  2009-04-01 &   38034 &   39614 &   39619 &   31934 &   33358 &   33363 \\
0554750501 &  2009-04-03 &   42434 &   44016 &   44018 &   38634 &   40216 &   40218 \\
0554750601 &  2009-04-05 &   32837 &   38816 &   38818 &   31485 &   37464 &   37466 \\
\multicolumn{8}{c}{{\it 2011}} \\					     	                              
0604300601 &  2011-03-28 &   45306 &   48467 &   48491 &   28768 &   30121 &   30119 \\
0604300701 &  2011-03-30 &   42305 &   48579 &   48584 &   32872 &   39149 &   39156 \\
0604300801 &  2011-04-01 &   37321 &   38642 &   38494 &   33771 &   36149 &   36129 \\
0604300901 &  2011-04-03 &   36568 &   37589 &   37573 &   19941 &   21140 &   21143 \\
0604301001 &  2011-04-05 &   48210 &   47757 &   47646 &   32571 &   33917 &   33914 \\
0658600101 &  2011-08-31 &   47585 &   49169 &   49159 &   47653 &   49169 &   49177 \\
0658600201 &  2011-09-01 &   51324 &   52903 &   52908 &   39634 &   41109 &   41115 \\
\multicolumn{8}{c}{{\it 2012}} \\					     	                              
0674600601 &  2012-03-13 &   19594 &   21167 &   21172 &    8594 &    9296 &    9301 \\
0674600701 &  2012-03-15 &   14040 &   15616 &   15618 &    6802 &    8209 &    8212 \\
0674600801 &  2012-03-19 &   21041 &   22615 &   22618 &   16784 &   18358 &   18358 \\
0674601001 &  2012-03-21 &   22034 &   23616 &   23619 &   19841 &   21416 &   21419 \\
0674601101 &  2012-03-17 &   25682 &   24638 &   24628 &    8956 &    8173 &    8178 \\
\hline
\end{tabular}
\caption{List of all \xmm\ observations considered in this work. 
Total and cleaned exposure time (in seconds) for each camera, 
respectively. }
\label{Exp1}
\end{center}
\end{table*}

\begin{table*}
\begin{center}
\tiny
\begin{tabular}{ | c c c c c c c c c c c c c c c c c }
\hline
OBSID           & obs date & Exp pn & Exp M1 & Exp M2 & Exp pn & Exp M1 & Exp M2 \\
\hline
\multicolumn{8}{c}{{\bf Other observations of the CMZ}} \\
0030540101 &  2002-09-09 &   27689 &   27842 &   27844 &   27339 &   27495 &   27494 \\
0144220101 &  2003-03-12 &   46746 &   49905 &   49843 &   28820 &   31550 &   31461 \\
0152920101 &  2003-04-02 &   50182 &   51639 &   51774 &   48486 &   50082 &   50097 \\
0144630101 &  2003-09-11 &    8469 &       0 &    8661 &       0 &     316 &     311 \\
0203930101 &  2004-09-04 &   46544 &   50438 &   50446 &   39078 &   43003 &   43013 \\
0205240101 &  2005-02-26 &   46919 &   50625 &   50604 &   14946 &   15251 &   15243 \\
0304220301 &  2005-08-29 &   20031 &   20213 &   20226 &    6417 &    6615 &    6620 \\
0304220101 &  2005-09-29 &    8051 &    8237 &    8250 &    5621 &    5816 &    5821 \\
0303210201 &  2005-10-02 &   23472 &       0 &       0 &      23 &     314 &     315 \\
0302882501 &  2006-02-27 &    7561 &    9176 &    9178 &    6364 &    7999 &    8002 \\
0302882701 &  2006-02-27 &    5237 &    6851 &    6869 &    2937 &    4564 &    4569 \\
0302882801 &  2006-02-27 &    5937 &    7558 &    7571 &    5537 &    7164 &    7171 \\
0302882901 &  2006-02-27 &    5936 &    7566 &    7569 &    4437 &    6066 &    6069 \\
0302883001 &  2006-02-27 &    5937 &    7540 &    7558 &    3137 &    4766 &    4771 \\
0302883101 &  2006-02-27 &    9814 &   11432 &   11448 &    8614 &   10248 &   10258 \\
0302883201 &  2006-03-29 &    4896 &    6518 &    6526 &    3898 &    5547 &    5539 \\
0305830701 &  2006-04-04 &    6399 &   11266 &   11256 &       0 &    1028 &    1028 \\
0302883901 &  2006-09-08 &    4987 &    6565 &    6568 &    4787 &    6365 &    6368 \\
0302884101 &  2006-09-08 &    4987 &    6565 &    6570 &    4000 &    5578 &    5583 \\
0302884201 &  2006-09-08 &    4987 &    6565 &    6570 &    4987 &    6565 &    6570 \\
0302884301 &  2006-09-09 &    4987 &    6565 &    6568 &    4987 &    6565 &    6568 \\
0302884401 &  2006-09-09 &    4036 &    5616 &    5621 &    4037 &    5616 &    5621 \\
0302884501 &  2006-09-09 &    6787 &    8364 &    8370 &    6787 &    8364 &    8369 \\
0406580201 &  2006-09-18 &   28034 &   29607 &   29609 &   13896 &   14809 &   14814 \\
0410580401 &  2006-09-22 &       0 &   32558 &       0 &       0 &   32367 &   32326 \\
0410580501 &  2006-09-26 &       0 &   32116 &       0 &       0 &   30108 &   30096 \\
0400340101 &  2006-09-24 &   40001 &   41575 &   41580 &   16312 &   17481 &   17486 \\
0506291201 &  2007-02-27 &       0 &   38616 &   38621 &       0 &   30937 &   30937 \\
0504940101 &  2007-09-06 &    5058 &    6615 &    6620 &    4958 &    6515 &    6520 \\
0504940401 &  2007-09-06 &    5058 &    6615 &    6620 &    5058 &    6615 &    6620 \\
0504940501 &  2007-09-06 &    5057 &    6615 &    6620 &    5006 &    6563 &    6568 \\
0504940601 &  2007-09-06 &    5058 &    6615 &    6620 &    1720 &    3175 &    3181 \\
0504940701 &  2007-09-06 &    5058 &    6615 &    6620 &    4558 &    6115 &    6120 \\
0511010701 &  2008-02-27 &    7455 &    9004 &    9004 &    5803 &    7362 &    7368 \\
0511000101 &  2008-03-03 &    6943 &    8500 &    8500 &     546 &     796 &     800 \\
0511000501 &  2008-03-04 &    5058 &    6615 &    6620 &    4658 &    6215 &    6220 \\ 
0511000701 &  2008-03-04 &    5058 &    6615 &    6620 &    4506 &    6063 &    6068 \\
0511000901 &  2008-03-04 &    5058 &    6615 &    6620 &    5058 &    6514 &    6518 \\
0511001101 &  2008-03-04 &    5057 &    6615 &    6620 &    5057 &    6615 &    6620 \\
0511001301 &  2008-03-04 &    5058 &    6615 &    6620 &    3800 &    5132 &    5137 \\
0511000201 &  2008-09-23 &    5058 &    6615 &    6620 &    5058 &    6615 &    6620 \\
0511000601 &  2008-09-23 &    5058 &    6615 &    6620 &    5058 &    6615 &    6620 \\
0511000801 &  2008-09-27 &    5035 &    6602 &    6620 &    5035 &    6615 &    6620 \\
0511001001 &  2008-09-27 &    5034 &    6615 &    6620 &    5034 &    6615 &    6620 \\
0511001201 &  2008-09-27 &    5034 &    6615 &    6620 &    5034 &    6615 &    6620 \\
0511001401 &  2008-09-27 &    5034 &    6615 &    6620 &    5034 &    6615 &    6620 \\
0505870301 &  2008-03-10 &   29885 &   31614 &   31494 &    7249 &    7250 &    7255 \\
0603850201 &  2010-04-07 &   22503 &   21643 &   21663 &   16919 &   18271 &   18266 \\
0655670101 &  2011-03-19 &       0 &  103934 &  103954 &       0 &   80716 &   80729 \\
\hline
\end{tabular}
\caption{List of all \xmm\ observations considered in this work. 
Total and cleaned exposure time (in seconds) for each camera, 
respectively. }
\label{Exp2}
\end{center}
\end{table*}

\end{document}